\documentclass[traditabstract]{aa} 
\usepackage{array}
\usepackage{graphicx}
\usepackage{natbib}
\usepackage{lscape}
\usepackage{rotating}
\usepackage{txfonts}
\def\l{$\lambda$}

\newcommand{\kms}{km\ s$^{-1}$}

\newcommand{\xmm}{{\sc XMM}\emph{-Newton}}

\newcommand{\lxlb}{$L_{\rm X}/L_{\rm BOL}$}
\newcommand{\loglxlb}{$\log[L_{\rm X}/L_{\rm BOL}]$}
\begin{document}
  \title{X-ray properties of the young open clusters HM1 and IC\,2944/2948\footnote{Tables \ref{hm1det}, \ref{icdet}, and \ref{iccorrel} are available at CDS.} \thanks{Based on observations collected with XMM-Newton, an ESA Science Mission with instruments and contributions directly funded by ESA Member States and the USA (NASA).}}

   \author{Ya\"el Naz\'e\inst{1}\fnmsep\thanks{Research Associate FRS-FNRS}, Gregor Rauw\inst{1}, Hugues Sana\inst{2}, Michael F. Corcoran\inst{3}.
          }

   \institute{GAPHE, D\'epartement AGO, Universit\'e de Li\`ege, All\'ee du 6 Ao\^ut 17, Bat. B5C, B4000-Li\`ege, Belgium\\
              \email{naze@astro.ulg.ac.be}
\and Astronomical Institute Anton Pannekoek, Amsterdam University, Science Park 904, 1098~XH, Amsterdam, The Netherlands
\and CRESST/NASA Goddard Space Flight Center, Code 662, Greenbelt, MD 20771; Univ. Space Res. Ass., Columbia, MD, USA
             }
\authorrunning{Naz\'e et al.}
\titlerunning{X-rays from HM1 and IC\,2944/2948}


 
  \abstract
{Using \xmm\ data, we study for the first time the X-ray emission of HM1 and IC\, 2944/2948. Low-mass, pre-main-sequence objects with an age of a few Myr are detected, as well as a few background or foreground objects. Most massive stars in both clusters display the usual high-energy properties of that type of objects, though with \loglxlb\ apparently lower in HM1 than in IC\,2944/2948. Compared with studies of other clusters, it seems that a low signal-to-noise at soft energies, due to the high extinction, may be the main cause of this difference. In HM1, the two Wolf-Rayet stars show contrasting behaviors: WR89 is extremely bright, but much softer than WR87. It remains to be seen whether wind-wind collisions or magnetically confined winds can explain these emissions. In IC\,2944/2948, the X-ray sources concentrate around HD\,101205; a group of massive stars to the north of this object is isolated, suggesting that there exist two subclusters in the field-of-view.} 


   \maketitle
%

\section{Introduction}

X-ray observations of OB associations and young open clusters reveal emission from the early-type stars as well as from a population of low-mass pre-main-sequence stars (e.g.\ NGC\,6231 \citealt{Sana}; Carina OB1 \citealt{Igor}, \citealt{naz11}; Cyg\,OB2 \citealt{AC}, \citealt{Rauw}). X-ray emission is a ubiquitous characteristic of massive stars of spectral type earlier than about mid-B. This emission is usually attributed to a distribution of wind-embedded shocks produced by the so-called line deshadowing instability (LDI, \citealt{Feldmeier}) in the radiatively-driven winds of these objects. The corresponding pockets of shock-heated gas at characteristic temperatures of a few million degrees are believed to be scattered throughout the wind volume, extending to very near the stellar photosphere. The X-ray emission from this distribution of pockets of hot plasma is usually not expected or seen to vary \citep{zetaPup2}. 

Ever since the discovery of the X-ray emission of early-type stars, it has been found that the X-ray luminosity of O-type stars linearly scales with their bolometric luminosity (e.g. \citealt{naz09}). The situation is quite different for Wolf-Rayet stars, where no clear dependence between X-ray and bolometric luminosities has been observed  \citep{Wessolowski}. From the theoretical point of view, the origin of the empirical \lxlb\ relation of O-type stars remains unknown \citep{OC}, although an adequate mixing of the hot and cool material could explain this relation \citep{Owocki}. From the observational point of view, it is important to study this relation for a variety of open clusters and associations to probe the various environmental parameters that could impact on this relation and produce the scatter around the relation that is observed when considering large samples of O-stars \citep{naz09}.  

Additional X-ray emission can arise in massive stars' winds from large-scale shocks associated with wind interactions in massive (non-compact) binaries (e.g.\ \citealt{SBP}), or magnetic confinement (e.g.\ \citealt{gag05}). In general, wind-wind collisions and magnetically confined wind shocks occur at much higher Mach numbers than LDI shocks and are observed to (generally) produce stronger and harder X-ray emission up to 10\,keV. However, the observations do not reveal a systematic overluminosity of massive binaries with respect to single O-type stars (e.g.\ \citealt{Sana,naz09}) and some magnetic O-type stars also fail to comply with the theoretical picture  \citep{Ofp}.

In this context, we study for the first time the X-ray emission of HM\,1 and IC\,2944/2948, two clusters with different massive star populations. Section 2 presents the observation and data reduction, sections 3 and 4 show the results for HM1 and IC\,2944/2948, respectively, section 5 discusses the derived \lxlb\ and section 6 summarizes the results and concludes.

\begin{figure*}
\includegraphics[width=9cm]{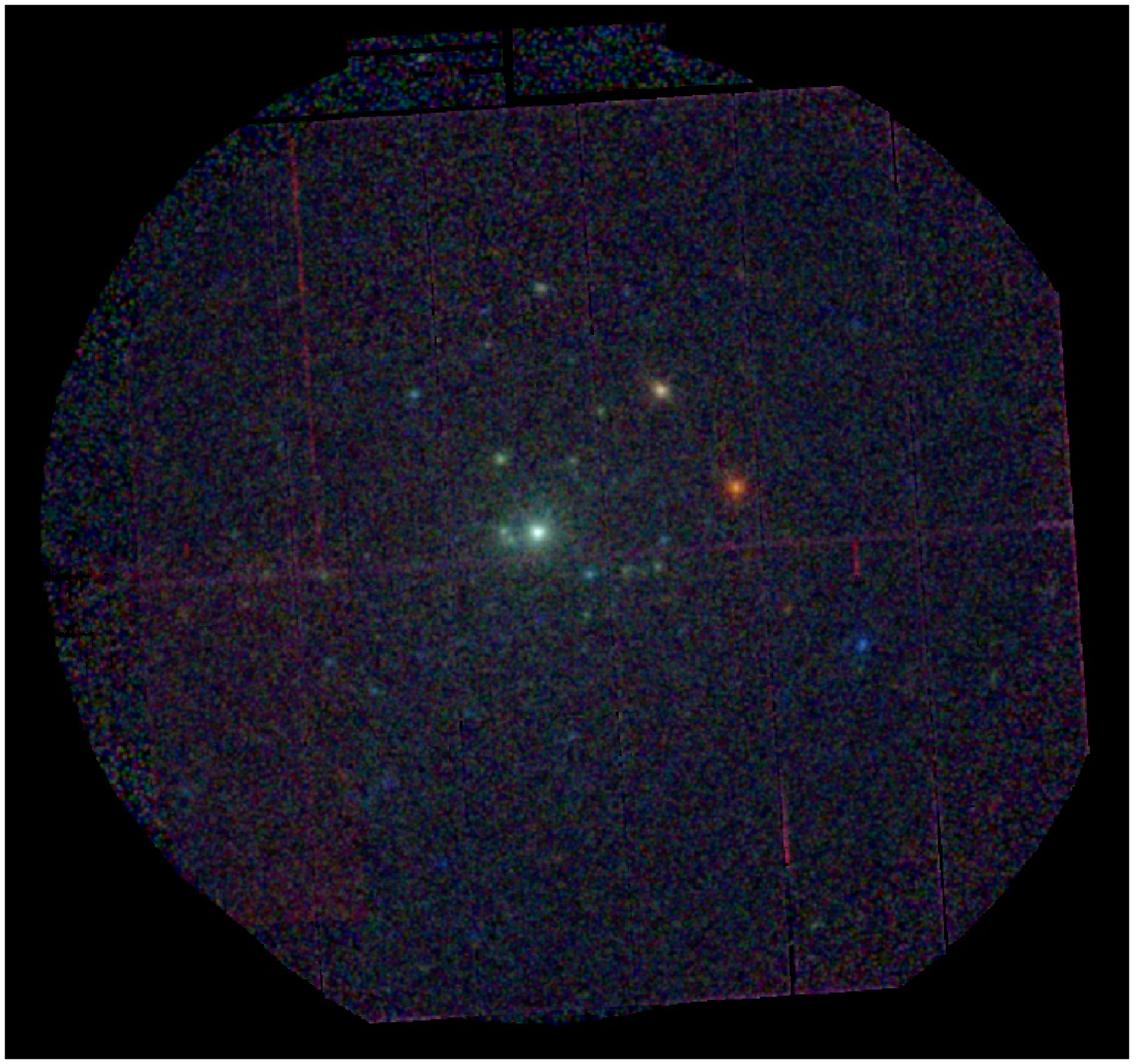}
\includegraphics[width=9cm, bb=35 185 550 675, clip]{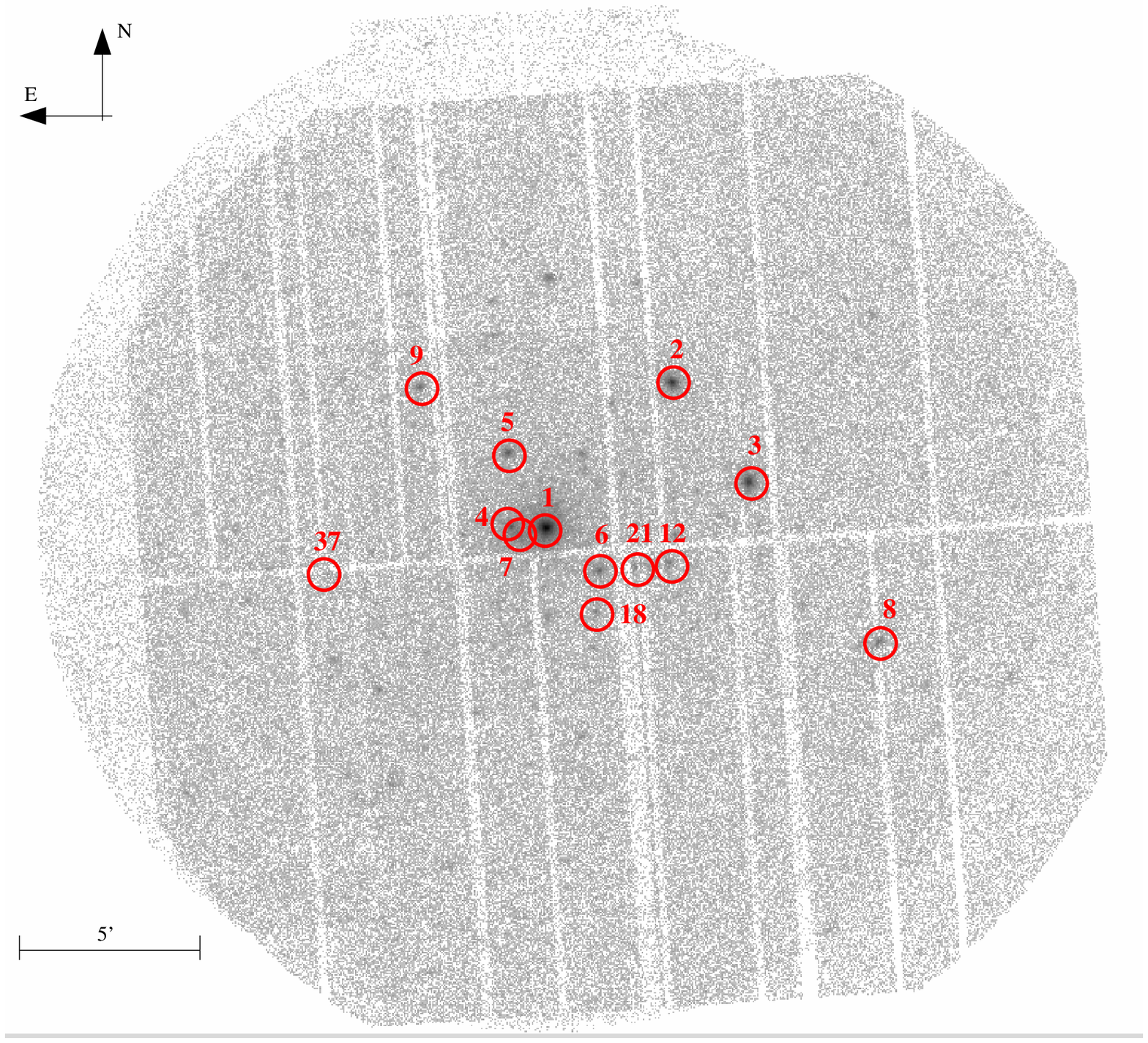}
\caption{Left: Three-color image of HM1, made using the SAS ``images'' script. Red, green, blue correspond respectively to the soft, medium, and hard energy bands (see text for definition). The greenish source at the field's center is WR89. Right: Grayscale image of HM1, with the objects discussed in the text identified by their X-ray source number (see Table \ref{hm1det}). In both images, orientation and scale are the same. The color version of this figure is available online.}
\label{hm1col}
\end{figure*}

\section{\xmm\ observations}

\xmm\ has observed HM1 for 25\,ks on Rev. 1877 (with MEDIUM filter) and IC\,2944/2948 for 40\,ks on Rev. 2209 (with THICK filter). No background flare affected the observations, and no source is bright enough to suffer from pile-up. The data were reduced with SAS v12.0.0 using calibration files available on July 1, 2012 and following the recommendations of the \xmm\ team\footnote{SAS threads, see \\ http://xmm.esac.esa.int/sas/current/documentation/threads/ Note the recent inclusion of the CTI correction for pn data (task {\it epspatialcti}), which was applied here.}. 

The source detection was performed using the task {\it edetect\_chain} on the three EPIC datasets and in three energy bands (soft=S=0.3--1.0\,keV  energy band, medium=M=1.0--2.0\,keV energy band, hard=H=2.0--10.0\,keV energy band). This task searches for sources by using a sliding box and finds the final source parameters from point-spread-function (PSF) fitting. The detection likelihood was adjusted to detect as many sources as possible with few spurious detections (see below). For the brightest sources (i.e., with more than 500 EPIC counts), we also extracted EPIC spectra and light curves in circular regions, centered on the detection position for the sources and as close as possible to the targets for the backgrounds. The background positions as well as the extraction radii were adapted taking into account the crowding near the source, and elliptical regions were used for sources close to field-of-view (FOV) edges since the PSF is highly distorted at large off-axis angles. After extraction and calibration with response matrices, EPIC spectra were grouped, using {\it specgroup}, to obtain an oversampling factor of five and to ensure that a minimum signal-to-noise ratio of three was reached in each spectral bin of the background-corrected spectra. These spectra were fitted within Xspec v12.7.0. The EPIC light curves of the source and background were extracted for 1\,ks and 5\,ks time bins and in the total (0.3--10.\,keV) energy band using the SAS task {\it epiclccorr}, which provides equivalent on-axis, full PSF count rates. Light curves were then tested using $\chi^2$ tests for constancy, linear trend, and quadratic trend hypotheses, and we furthermore compared the improvement of the $\chi^2$ when increasing the number of parameters in the model (e.g., linear trend vs constancy) thanks to using F-tests (see Sect. 12.2.5 in \citealt{lin68}). Similar tests have been applied to stars in NGC6231 \citep[see section 5.1 in ][]{san04} and to $\zeta$\,Puppis \citep{zetaPup2}

\section{HM1}
The open cluster HM1 was first described by \citet{hav77}. It is relatively nearby ($DM=12.6$, or 3.3\,kpc), tight (3'), and significantly reddened ($E(B-V)=1.84\pm0.07$\,mag, \citealt{vaz01}). This high extinction renders its study in the optical range difficult, explaining the small number of studies that were devoted to this cluster. 

However, HM1 displays an interesting massive star population. \citet{hav77} identified 15 members and 9 non-member stars, thanks to UBV photometry. They also obtained low-resolution spectroscopy for a few objects, leading to the identification of six hot stars amongst the member objects: two Wolf-Rayet (WR87 and WR89), two Of stars, and two O8 objects. Further photometric work was performed by \citet{the82} and \citet{vaz01}, who extended the previous dataset to fainter stars and other photometric bands: e.g., \citet{vaz01} provided UBVRI photometry for 802 objects, among which 39 are members and 36 are probable members\footnote{Compared with DSS and 2MASS data, two objects (HM1 VB 14 and 52) are actually spurious (misidentifications due to proximity of the brighter sources HM1 VB 1 and 4). }. In parallel, \citet{mas01} obtained spectra for eight stars. They confirmed the presence of the two Of stars, identified two additional O-type objects, and revised the O8 classification to earlier spectral types. Their final classification for the two WRs is WN7 and that for the O-stars reveals O4If+, O5If+, O5V, O6If, O7V((f)), and O9.5V\footnote{or III, since text and table do not agree.} stars. \citet{gam08} revised the classification of the main-sequence O5 to O5III(f) and mentioned that it is actually a binary with a 5.9\,d period and minimum masses of 31 and 15\,M$_{\odot}$.

\begin{figure*}
\includegraphics[width=6.cm]{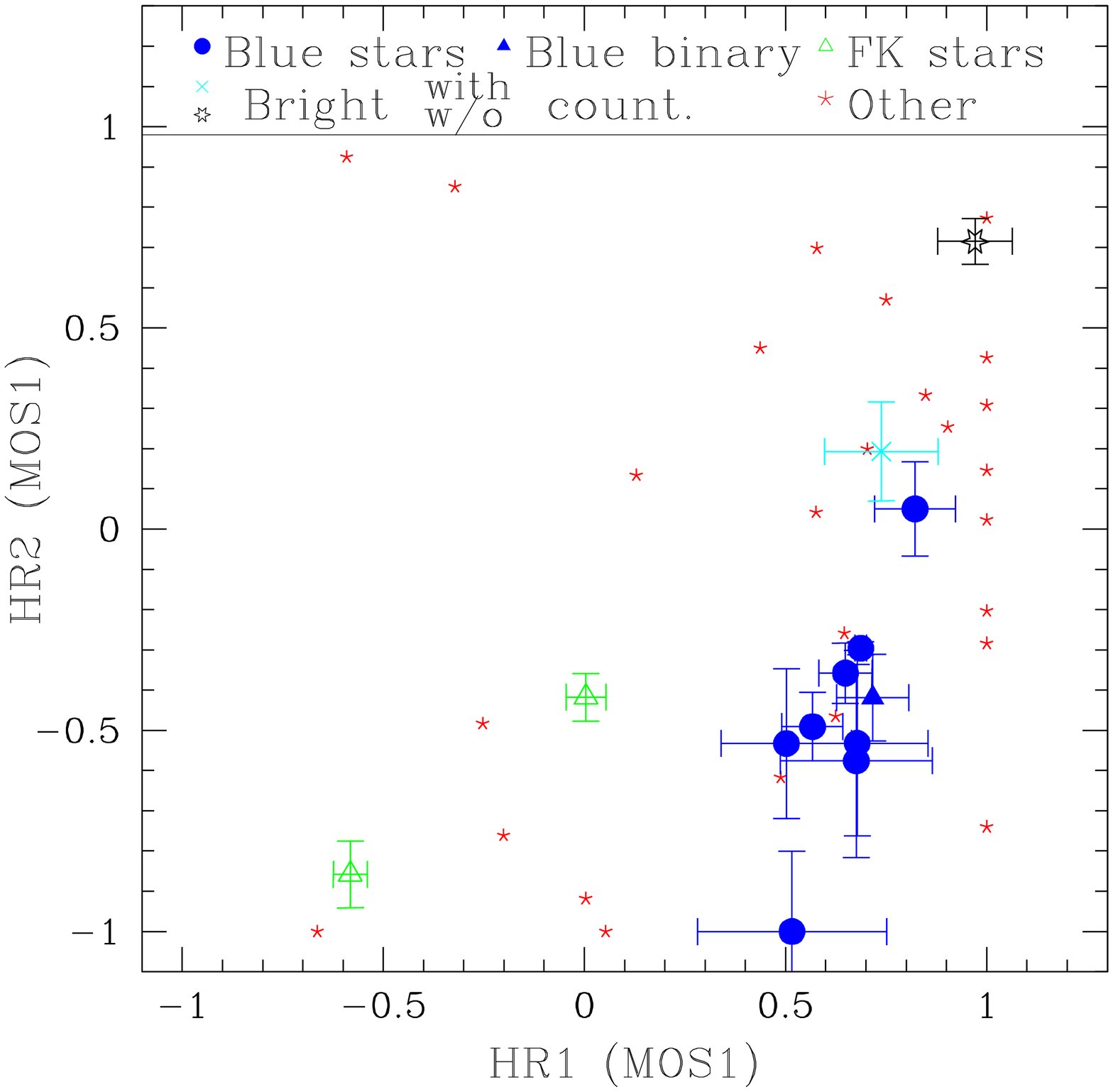}
\includegraphics[width=6.cm]{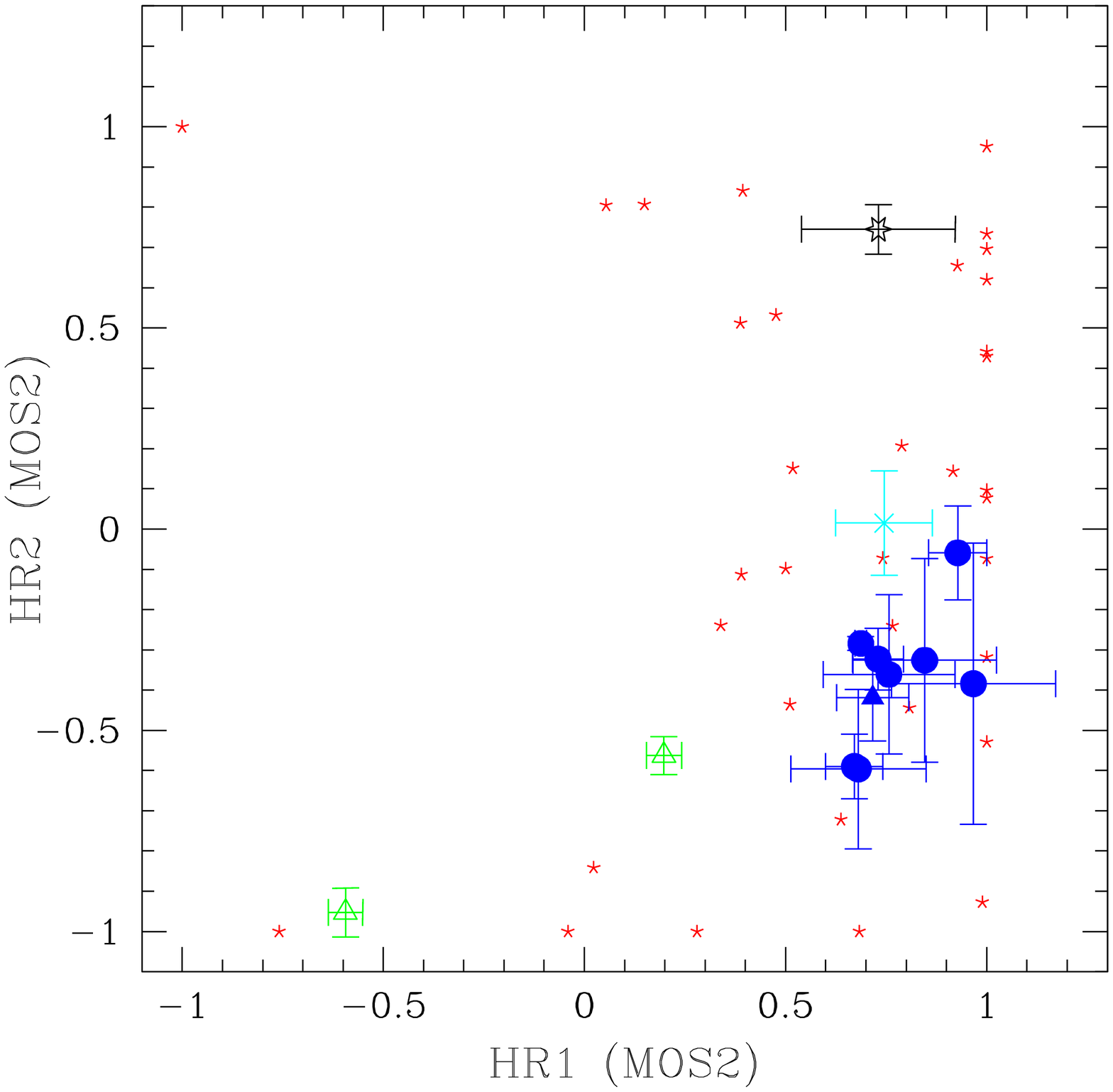}
\includegraphics[width=6.cm]{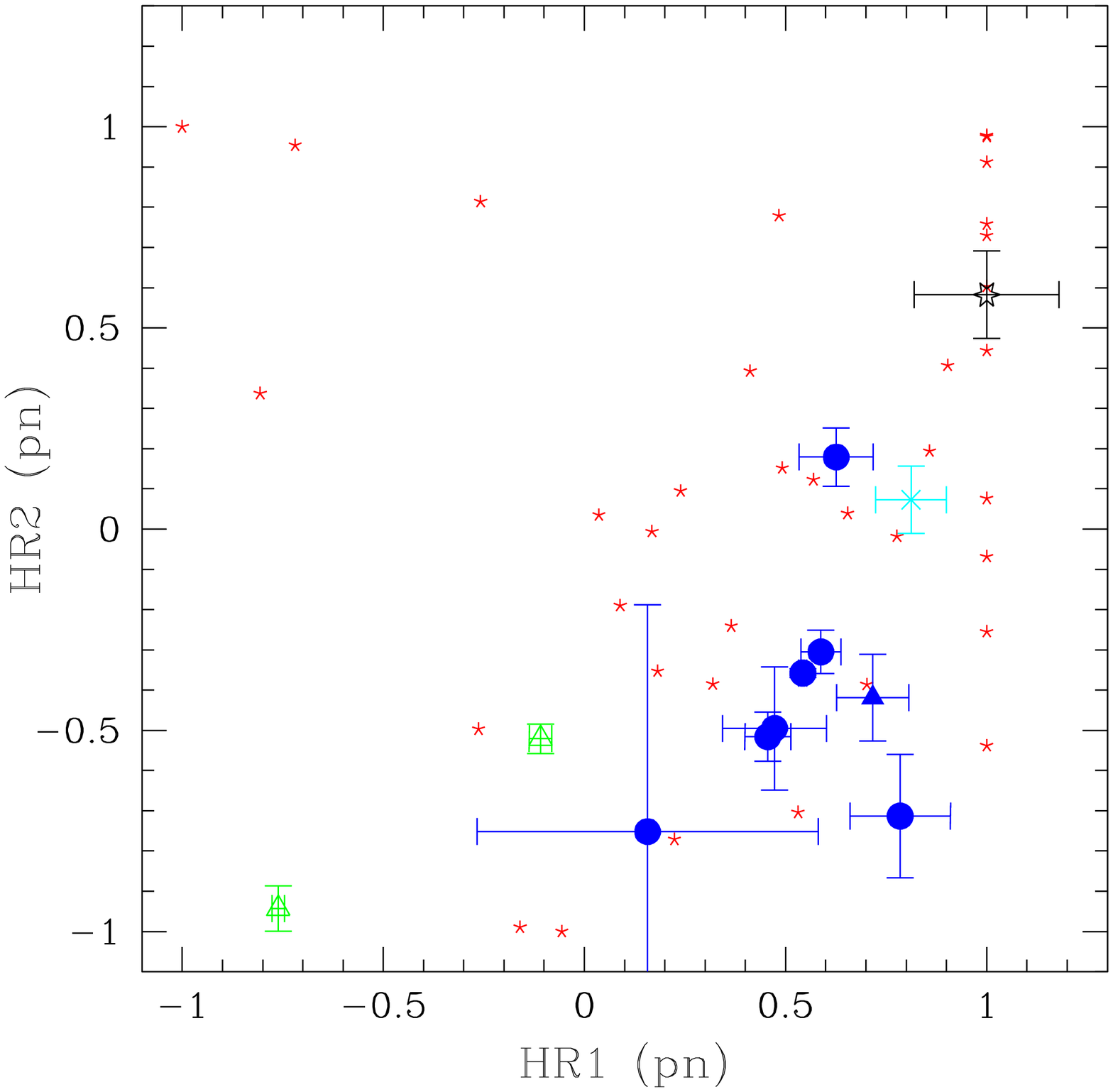}
\caption{Hardness ratios of sources in HM1 for MOS1 (left), MOS2 (middle), and pn (right).  Different types of sources are distinguished by symbols. The massive, blue stars are shown with blue filled symbols (triangle for the known binary, circles for the others). The low-mass FK stars are shown with green open triangles. The other X-ray bright objects (i.e., with at least 500 EPIC counts) with and without 2MASS counterparts are displayed as cyan crosses and black stars, respectively. The remaining, faint sources are represented by red asterisks (see top of the figure - error bars are not shown for the other sources), and potentially spurious sources are not shown. WR87 is the hardest massive star. The color version of this figure is available online.}
\label{hm1HR}
\end{figure*}

HM1 now appears as a young (1-2\,Myr, \citealt{mas01}, or 2-4\,Myr, \citealt{vaz01}) cluster containing a number of rare, very early objects and several transition stars (WR and Of). Although only eight members of the cluster were studied spectroscopically, the list of cluster members \citep[table 4 in ][]{vaz01} contains 24 objects that are very bright ($M_V<-3.8$) and blue ($[B-V]_0<-0.2$, $[U-B]_0<-0.9$, taking into account the error bars), hence are O or early B-type candidates. Indeed, masses over 20\,M$_{\odot}$ have been suggested for some of these candidates by \citet{mas01}. This would make HM1 one of the richest clusters in massive stars in our Galactic neighborhood.

To date, this cluster has not been studied in X-rays. Our \xmm\ observation reveals only 58 sources, of which 7 may be spurious as they are not even clearly detected by eye (Fig. \ref{hm1col} and Table \ref{hm1det}). Because of the high extinction toward this direction, the hardness ratio $HR_1=(M-S)/(M+S)$, where $S$ and $M$ are the soft and medium-energy count rates, is always close to one (Fig. \ref{hm1HR}). A second hardness ratio relying on the medium and hard bands, $HR_2=(H-M)/(H+M)$, instead shows a continuum of properties, with massive and F stars generally soft, and the bright X-ray source without near-infrared (NIR) counterpart very hard (see below). 

We have cross-correlated this list of sources with stars measured by \citet{vaz01}, as well as with the Simbad database and 2MASS catalog (Table \ref{hm1det}). To find the best correlation radius in the latter case, we used the technique outlined by \citet{jef97} and used by \citet{rau02}: the best radius, ensuring as many good identifications as possible while reducing the number of spurious correlations is 3''. Within such a radius, 33 of the 58 sources possess a 2MASS counterpart. Using the same radius, we found that nine X-ray sources correlate with known cluster members or probable members, and three additional ones correlate with non-member objects. Simbad provides two additional identifications for XID 2 and 3. Table \ref{hm1det} summarizes these identifications. Finally, we analyzed the light curves and spectra of the nine sources with at least 500 EPIC counts (Table \ref{hm1fit}). Details for each of these sources, and for groups of fainter sources, are given below.

  \begin{sidewaystable*}[htb]
  \tiny
  \centering
  \caption{Parameters of the detected X-ray sources in HM1.}
  \label{hm1det} 
  \begin{tabular}{>{\footnotesize}l >{\footnotesize}c >{\footnotesize}c | >{\footnotesize}c >{\footnotesize}c >{\footnotesize}c >{\footnotesize}c >{\footnotesize}c | >{\footnotesize}c >{\footnotesize}c | >{\footnotesize}c >{\footnotesize}c | >{\footnotesize}l}
  \hline
XID & RA & DEC & CR(MOS1) & CR(MOS2) & CR(pn) & $HR_1$(pn) & $HR_2$(pn) & d('') & VB\# & d('') & 2MASS & Add. info\\
  \hline
  1 & 17:19:00.576 & -38:48:50.58 & $214.95\pm3.45$ &  $ 209.52\pm3.39$ & $624.25\pm6.48$ &  $ 0.54\pm0.01$ &  $-0.36\pm0.01$ & 1.6 & 1 &  1.0 &  17190052-3848513 & WR89, WN7\\ 
  2 & 17:18:42.379 & -38:44:46.04 & $ 33.71\pm1.57$ &  $  37.04\pm1.58$ & $117.59\pm3.20$ &  $-0.11\pm0.03$ &  $-0.52\pm0.04$ &     &   &  0.4 &  17184240-3844462 & V504 Sco, eclip. bin.\\ 
  3 & 17:18:31.353 & -38:47:33.68 & $ 30.58\pm1.53$ &  $  28.83\pm1.52$ & $148.29\pm3.66$ &  $-0.76\pm0.02$ &  $-0.94\pm0.06$ &     &   &  0.4 &  17183138-3847336 & HD\,156301, F5V\\ 
  4 & 17:19:05.529 & -38:48:48.98 & $ 15.14\pm1.10$ &  $  14.55\pm1.10$ & $ 41.37\pm2.09$ &  $ 0.59\pm0.05$ &  $-0.31\pm0.05$ & 1.5 & 4 &  0.7 &  17190554-3848496 & VB4, O4If+\\ 
  5 & 17:19:06.016 & -38:46:43.97 & $  9.42\pm0.78$ &  $   8.55\pm0.74$ & $ 25.51\pm1.47$ &  $ 0.46\pm0.06$ &  $-0.52\pm0.06$ & 1.8 & 5 &  0.8 &  17190602-3846448 & VB5, O5If+\\ 
  6 & 17:18:52.879 & -38:50:03.13 & $  7.14\pm0.82$ &  $   6.52\pm0.76$ & $ 19.77\pm1.42$ &  $ 0.63\pm0.09$ &  $ 0.18\pm0.07$ & 1.1 & 9 &  0.4 &  17185287-3850035 & WR87, WN7\\ 
  7 & 17:19:04.528 & -38:49:04.53 & $  8.94\pm0.95$ &  $   7.61\pm0.91$ & $ 24.85\pm1.84$ &  $ 0.64\pm0.07$ &  $-0.37\pm0.08$ & 1.9 & 10&  1.1 &  17190443-3849048 & VB10, O5III(f)+OB\\ 
  8 & 17:18:12.677 & -38:52:04.19 & $ 22.15\pm1.89$ &  $  18.24\pm1.76$ & $ 57.81\pm8.52$:&  $ 1.00\pm0.18$ &  $ 0.58\pm0.11$ &     &   &      &                   & \\ 
  9 & 17:19:18.701 & -38:44:52.68 & $  5.13\pm0.64$ &  $   4.99\pm0.63$ & $ 15.20\pm1.28$ &  $ 0.81\pm0.09$ &  $ 0.07\pm0.08$ &     &   &  1.9 &  17191871-3844545 & \\ 
 10 & 17:19:00.093 & -38:41:49.41 &                 &  $   6.51\pm0.81$ & $ 19.95\pm1.57$ &  $ 0.32\pm0.08$ &  $-0.39\pm0.10$ &     &   &      &                   & \\
 11?& 17:19:52.359 & -38:49:17.53 & $  0.00\pm0.16$ &  $   0.06\pm0.21$ & $  8.93\pm1.10$ &  $-1.00\pm0.04$ &                 &     &   &      &                   & \\
 12 & 17:18:42.800 & -38:49:48.78 & $  3.08\pm0.54$ &  $   2.70\pm0.52$ & $  8.00\pm1.05$ &  $ 0.47\pm0.13$ &  $-0.50\pm0.15$ & 1.8 & 11&  1.6 &  17184281-3849503 & VB11, O6If\\ 
 13 & 17:18:41.940 & -38:49:03.87 & $  2.73\pm0.51$ &  $   3.14\pm0.53$ & $  7.18\pm0.98$ &  $ 0.57\pm0.19$ &  $ 0.12\pm0.14$ &     &   &      &                   & \\
 14?& 17:18:59.252 & -38:48:22.19 & $  1.64\pm0.56$ &  $   1.85\pm0.62$ & $  4.11\pm1.09$ &  $ 0.33\pm0.29$ &  $-0.28\pm0.31$ &     &   &  2.4 &  17185938-3848203 & \\ 
 15?& 17:19:00.819 & -38:49:21.22 & $  1.68\pm0.54$ &  $   1.30\pm0.55$ & $  5.47\pm1.17$ &  $ 0.61\pm0.19$ &  $-0.57\pm0.22$ &     &   &  1.9 &  17190080-3849231 & \\ 
 16?& 17:18:30.185 & -38:47:32.24 & $  1.02\pm0.56$ &  $   2.86\pm0.66$ & $  0.00\pm0.47$ &                 &                 &     &   &      &                   & \\
 17 & 17:18:55.371 & -38:46:46.49 & $  2.69\pm0.50$ &  $   2.75\pm0.48$ & $  6.35\pm0.90$ &  $ 0.18\pm0.15$ &  $-0.35\pm0.18$ & 1.1 &354&  1.1 &  17185544-3846471 & \\ 
 18 & 17:18:53.402 & -38:51:11.91 & $  1.82\pm0.42$ &  $   1.02\pm0.35$ & $  6.04\pm0.92$ &  $ 0.79\pm0.12$ &  $-0.71\pm0.15$ & 2.3 & 20&  1.4 &  17185337-3851132 & VB20, blue, bright, massive\\ 
 19 & 17:19:24.640 & -38:53:22.54 & $  3.12\pm0.56$ &  $   2.23\pm0.52$ & $  5.03\pm0.95$ &  $ 0.65\pm0.23$ &  $ 0.04\pm0.19$ &     &   &  2.9 &  17192452-3853250 & \\ 
 20 & 17:18:59.126 & -38:48:11.95 & $  1.77\pm0.52$ &  $   1.22\pm0.45$ & $  5.30\pm1.00$ &  $ 0.78\pm0.23$ &  $-0.02\pm0.19$ &     &   &      &                   & \\
 21 & 17:18:47.634 & -38:49:55.96 & $  2.52\pm0.54$ &  $   2.51\pm0.49$ &                 &                 &                 & 2.3 & 13&  1.9 &  17184764-3849578 & VB13, O7V((f))\\ 
 22 & 17:18:51.278 & -38:45:21.53 & $  2.45\pm0.47$ &  $   1.64\pm0.39$ & $  5.77\pm1.67$:&  $ 1.00\pm0.29$ &  $-0.26\pm0.28$ & 1.1 & 23&  1.0 &  17185124-3845224 & \\ 
 23 & 17:19:08.380 & -38:42:28.26 &                 &  $   1.63\pm0.48$ & $  5.93\pm0.99$ &  $ 0.48\pm0.54$ &  $ 0.78\pm0.12$ &     &   &  1.7 &  17190852-3842283 & \\ 
 24 & 17:19:22.691 & -38:56:02.91 & $  4.55\pm0.80$ &  $   2.67\pm0.70$ & $  6.58\pm1.22$ &  $ 1.00\pm2.04$ &  $ 0.91\pm0.11$ &     &   &      &                   & \\
 25 & 17:19:17.446 & -38:35:13.37 &                 &  $   6.34\pm1.17$ &                 &                 &                 &     &   &      &                   & \\
 26 & 17:19:07.966 & -38:43:26.32 & $  2.05\pm0.47$ &  $   1.55\pm0.48$ & $  5.00\pm0.91$ &  $ 0.17\pm0.20$ &  $-0.01\pm0.22$ &     &   &      &                   & \\
 27 & 17:19:52.526 & -38:56:16.53 & $  4.22\pm0.95$ &  $   4.04\pm1.01$ & $  4.77\pm1.48$ &  $ 1.00\pm0.19$ &  $ 0.08\pm0.30$ &     &   &      &                   & \\
 28 & 17:19:31.202 & -38:52:32.25 & $  2.07\pm0.51$ &  $   2.72\pm0.59$ & $  4.33\pm0.90$ &  $ 1.00\pm0.13$ &  $-0.07\pm0.21$ &     &   &      &                   & \\
 29 & 17:19:00.369 & -38:51:21.44 & $  1.48\pm0.37$ &  $   1.35\pm0.42$ & $  5.01\pm0.81$ &  $ 1.00\pm3.58$ &  $ 0.97\pm0.08$ &     &   &  2.1 &  17190045-3851196 & \\ 
 30 & 17:18:55.457 & -38:54:42.49 & $  1.63\pm0.47$ &  $   2.75\pm0.63$ & $  4.17\pm0.93$ &  $ 0.41\pm0.37$ &  $ 0.39\pm0.21$ &     &   &  2.2 &  17185554-3854444 & \\ 
 31 & 17:18:23.595 & -38:51:02.13 & $  1.21\pm0.43$ &  $   2.23\pm0.60$ & $  5.97\pm1.14$ &  $-0.06\pm0.17$ &  $-1.00\pm0.39$ &     &   &  0.5 &  17182355-3851024 & \\ 
 32 & 17:18:13.552 & -38:42:50.39 & $  4.52\pm1.03$ &  $   4.56\pm0.96$ & $  6.90\pm1.67$ &  $-0.26\pm0.53$ &  $ 0.81\pm0.16$ &     &   &      &                   & \\
 33 & 17:18:53.837 & -38:52:12.90 & $  1.18\pm0.36$ &  $   0.53\pm0.29$ & $  3.32\pm0.77$ &  $ 0.09\pm0.25$ &  $-0.19\pm0.30$ &     &   &  1.7 &  17185369-3852133 & \\ 
 34?& 17:18:47.037 & -38:49:53.06 & $  1.25\pm0.46$ &  $   0.63\pm0.40$ & $ 17.44\pm3.95$ &  $ 0.14\pm0.23$ &  $-0.33\pm0.31$ &     &   &      &                   & \\
 35 & 17:18:47.581 & -38:41:59.05 &                 &  $   1.59\pm0.46$ & $  7.22\pm1.17$ &  $ 0.86\pm0.19$ &  $ 0.19\pm0.15$ &     &   &  2.2 &  17184747-3842008 & \\ 
 36?& 17:18:56.872 & -38:48:54.77 & $  0.47\pm0.34$ &  $   1.39\pm0.47$ & $  3.60\pm0.85$ &  $ 1.00\pm0.20$ &  $-0.87\pm0.19$ &     &   &  2.9 &  17185709-3848561 & \\ 
 37 & 17:19:32.030 & -38:50:06.33 & $  1.55\pm0.40$ &  $   1.98\pm0.49$ & $  2.29\pm1.01$:&  $ 0.16\pm0.42$ &  $-0.75\pm0.56$ & 2.6 & 26&  1.4 &  17193203-3850076 & VB26, blue, bright\\ 
 38 & 17:18:19.541 & -39:00:06.65 &                 &  $   4.23\pm1.22$ & $ 12.95\pm2.58$ &  $ 0.37\pm0.19$ &  $-0.24\pm0.25$ &     &   &  2.5 &  17181974-3900073 & \\ 
 39 & 17:19:14.575 & -38:58:20.55 & $  2.51\pm0.70$ &  $   2.57\pm0.77$ & $  7.99\pm1.54$ &  $ 1.00\pm0.57$ &  $ 0.73\pm0.13$ &     &   &      &                   & \\
 40 & 17:19:29.358 & -38:55:47.64 & $  0.63\pm0.39$ &  $   1.26\pm0.50$ & $  4.80\pm0.98$ &  $-0.16\pm0.19$ &  $-0.99\pm0.34$ &     &   &  1.3 &  17192925-3855481 & \\ 
 41 & 17:19:18.006 & -38:52:21.42 & $  1.13\pm0.52$ &  $   0.69\pm0.34$ & $  3.69\pm0.92$ &  $ 0.90\pm0.27$ &  $ 0.41\pm0.26$ & 2.8 &440&  1.6 &  17191806-3852229 & \\ 
 42 & 17:19:40.569 & -38:49:58.87 & $  2.23\pm0.49$ &  $   0.82\pm0.39$ & $  2.56\pm0.99$:&  $ 1.00\pm0.25$ &  $-0.54\pm0.43$ &     &   &  1.1 &  17194049-3849596 & \\ 
 43?& 17:19:02.825 & -38:48:15.37 & $  1.72\pm0.49$ &  $   0.89\pm0.47$ & $  2.65\pm0.78$ &  $ 1.00\pm0.28$ &  $ 0.01\pm0.29$ &     &   &      &                   & \\
 44 & 17:19:43.636 & -38:50:53.17 & $  1.62\pm0.48$ &  $   1.14\pm0.43$ & $  3.13\pm0.88$ &  $-0.26\pm0.25$ &  $-0.50\pm0.57$ &     &   &  0.9 &  17194361-3850540 & \\ 
 45 & 17:19:38.387 & -38:46:20.90 & $  1.23\pm0.41$ &  $   1.46\pm0.45$ & $  2.96\pm1.14$:&  $ 0.49\pm0.52$ &  $ 0.15\pm0.39$ &     &   &  1.6 &  17193836-3846224 & \\ 
 46 & 17:19:22.554 & -38:38:44.29 &                 &  $   1.71\pm0.56$ & $  4.75\pm1.24$ &  $ 1.00\pm0.54$ &  $ 0.76\pm0.18$ &     &   &      &                   & \\
 47 & 17:18:39.484 & -38:51:41.31 & $  0.96\pm0.44$ &  $   0.98\pm0.38$ & $  3.26\pm0.91$ &  $ 1.00\pm0.93$ &  $ 0.60\pm0.22$ &     &   &      &                   & \\
 48 & 17:18:55.565 & -38:43:53.22 & $  1.13\pm0.47$ &  $   1.16\pm0.36$ & $  1.80\pm0.67$ &  $ 0.04\pm0.43$ &  $ 0.03\pm0.45$ &     &   &      &                   & \\
 49 & 17:17:44.591 & -38:50:30.92 &                 &                   & $ 11.68\pm2.69$ &  $ 0.24\pm0.24$ &  $ 0.09\pm0.26$ &     &   &      &                   & \\
 50 & 17:19:12.466 & -38:49:01.98 & $  1.00\pm0.33$ &  $   0.72\pm0.32$ & $  2.51\pm0.68$ &  $-0.72\pm0.84$ &  $ 0.95\pm0.15$ &     &   &  2.2 &  17191253-3849040 & \\ 
 51 & 17:19:19.508 & -38:45:58.09 & $  0.75\pm0.30$ &  $   0.78\pm0.32$ & $  1.83\pm0.58$ &                 &  $ 1.00\pm0.22$ &     &   &      &                   & \\
 52 & 17:17:53.257 & -38:56:41.63 &                 &                   & $  9.55\pm2.45$ &  $-0.81\pm0.20$ &  $ 0.34\pm0.69$ &     &   &      &                   & \\
 53 & 17:19:45.343 & -38:48:49.15 & $  0.93\pm0.51$ &  $   1.53\pm0.56$ & $  2.71\pm0.99$ &  $ 0.70\pm0.41$ &  $-0.39\pm0.41$ &     &   &      &                   & \\
 54 & 17:18:30.451 & -38:44:07.41 & $  0.29\pm0.30$ &  $   0.63\pm0.32$ & $  3.29\pm0.86$ &  $ 0.22\pm0.24$ &  $-0.77\pm0.37$ &     &   &      &                   & \\
 55 & 17:18:06.276 & -38:47:25.28 & $  1.03\pm0.58$ &  $   2.98\pm0.82$ & $  3.81\pm1.38$ &  $-1.00\pm1.07$ &  $ 1.00\pm0.34$ &     &   &      &                   & \\
 56 & 17:19:25.874 & -38:56:33.59 & $  1.61\pm1.11$:&  $   1.60\pm0.61$ & $  5.36\pm1.21$ &  $ 1.00\pm0.23$ &  $ 0.44\pm0.19$ &     &   &  2.0 &  17192574-3856323 & \\ 
 57 & 17:18:17.310 & -38:42:26.67 & $  2.48\pm0.88$ &  $   0.89\pm0.51$:& $  5.74\pm1.40$ &  $ 1.00\pm10.1$ &  $ 0.98\pm0.08$ &     &   &      &                   & \\
 58 & 17:17:53.448 & -38:52:57.90 &                 &  $   1.41\pm0.98$ & $  8.76\pm2.20$ &  $ 0.53\pm0.21$ &  $-0.70\pm0.35$ &     &   &  1.1 &  17175335-3852574 & \\ 
  \hline
  \end{tabular}
\tablefoot{
A ``?'' after the XID indicates a potentially spurious source. Sources 34 and 21 actually refer to the same object, whereas sources 14, 15, 36 and 43 are in the wings of source 1 and source 16 appears also as a wing of source 3. The count rates are given in cts\,ks$^{-1}$ for the total band (0.3--10.\,keV energy band) ; uncertain values are indicated by a ``:'' (source partially in a gap) ; missing values could not be calculated (e.g., the source is in a gap, over a bad column, or out of the field-of-view for that instrument). The hardness ratios $HR_1$ and $HR_2$ are computed as $(M-S)/(M+S)$ and $(H-M)/(H+M)$, respectively, where $S$, $M$, and $H$ are the count rates recorded in the soft (0.3--1.0\,keV) energy band, medium (1.0--2.0\,keV) energy band, and hard (2.0--10.0\,keV) energy band, respectively.}
  \end{sidewaystable*}

  \begin{sidewaystable*}[htb]
  \tiny
  \centering
  \caption{Spectral parameters of the best-fit thermal models to EPIC data of HM1.}
  \label{hm1fit}
  \begin{tabular}{l c c c c c c c c c c c c }
  \hline
XID & $N_{\rm H}$ & $kT_1$ & $norm_1$ & $kT_2$ & $norm_2$ & $kT_3$ & $norm_3$ & $\chi^2$/dof (dof) & $F_{\rm X}^{obs}$ & $L_{\rm X}^{abscor}$ & $\log(L_{\rm BOL})$ & $\log(L_{\rm X}^{abscor}/L_{\rm BOL})$ \\
& $10^{22}$\,cm$^{-2}$ & keV & $10^{-3}$\,cm$^{-5}$ & keV & $10^{-3}$\,cm$^{-5}$ & keV & $10^{-3}$\,cm$^{-5}$ & & 10$^{-13}$\,erg\,cm$^{-2}$\,s$^{-1}$ & 10$^{32}$\,erg\,s$^{-1}$ & (L$_{\odot}$) & \\
  \hline
1 & $1.1 + (0.066\pm0.007)$ & $0.140\pm0.009$ & $51\pm29$   & $0.57\pm0.03$ & $0.97\pm0.10$ & $1.99\pm0.08$ & $0.44\pm0.03$  & 1.41 (414) & 15.5 & 154.7$\pm$1.2 & 6.32 & $-5.713\pm0.003$\\
2 & $0.26\pm0.13$         & $0.20\pm0.05$ & $0.06\pm0.17$   & $0.95\pm0.04$ & $0.05\pm0.02$ & $2.74\pm0.35$ & $0.108\pm0.009$& 1.15 (136) & 2.11 & & & \\
3 & $0.32\pm0.06$         & $0.03\pm0.02$ & $37543\pm74131$ & $0.32\pm0.02$ & $0.29\pm0.07$ &               &                & 1.79 (77)  & 1.54 & & & \\
4 & $1.1 + (0.51\pm0.07)$ & $0.90\pm0.06$ & $0.24\pm0.03$   & & & & & 1.15 (56) & 0.59 & 2.78$\pm$0.10 & 6.23 & $-7.37\pm0.02$ \\
5 & $1.1 + (0.53\pm0.12)$ & $0.62\pm0.14$ & $0.35\pm0.11$   & & & & & 1.24 (34) & 0.47 & 3.00$\pm$0.13 & 6.13 & $-7.24\pm0.02$ \\
6 & $1.1 + (0.10\pm0.10)$ & $1.98\pm0.54$ & $0.065\pm0.012$ & & & & & 1.67 (29) & 0.68 & 1.58$\pm$0.09 & 6.22 & $-7.61\pm0.02$ \\
7 & $1.1 + (0.40\pm0.08)$ & $0.93\pm0.07$ & $0.18\pm0.02$   & & & & & 1.12 (43) & 0.49 & 2.53$\pm$0.14 & 5.82 & $-7.00\pm0.02$ \\
8 & $1.1 + (1.41\pm1.05)$ & $>7.0$        & $0.44\pm0.06$   & & & & & 0.69 (21) & 5.37 & & & \\
9 & $1.1 + (0.\pm0.16)$   & $2.58\pm0.46$ & $0.064\pm0.005$ & & & & & 0.93 (17) & 0.49 & & & \\
  \hline
  \end{tabular}
\tablefoot{ The fitted model has the form $wabs\times wabs\times \sum  apec$, where the first, interstellar absorption was fixed to $1.1\times 10^{22}$\,cm$^{-2}$ for all but the known foreground sources (XID 2 and 3). Fluxes refer to the 0.5--10.\,keV energy band, luminosities were corrected for the interstellar absorption only, and abundances of the thermal emission component and additional absorption are solar \citep{and89}, except for the two WRs (XID 1 and 6, see text). For massive stars, approximate errors on X-ray luminosities and \loglxlb, shown on Fig. \ref{hm1lxlb}, were calculated using the relative errors on the total count rates (Tables \ref{hm1det} - this agrees well with results of the ``flux err'' command and ``cflux'' model within xspec for both observed fluxed and absorption-corrected ones) - other uncertainties in the absolute calibration, the spectral models, and the extinction or bolometric luminosities of the individual stars are not included in these errors. }
  \end{sidewaystable*}

\subsection{Massive stars}

\subsubsection{WR89 (XID 1)}
WR89 was identified as a possible non-thermal radio emitter by \citet{cap04}. Such an emission is thought to be linked to relativistic electrons accelerated in the shocks of a wind-wind collision occurring inside a massive binary system \citep{van06}. However, the radio emission of WR89 was found to be thermal by several other authors, notably \citet{lei95}, \citet{dou00} and \citet{mon09}. Accordingly, the star was assumed to be single by \citet{ham06}. The same authors have fitted the visible spectrum of WR89 with their model atmosphere code, deriving the physical properties of the star, e.g., temperature, luminosity (in Table \ref{hm1fit}, the quoted luminosity value was corrected for the different distance used here), and abundances. These abundances were used for the Xspec fitting procedure: their mass fractions of 0.20, 0.78, 0.015, 0.0001, and 0.0014 for hydrogen, helium, nitrogen, carbon, and iron, respectively, correspond to elemental abundances compared to hydrogen with respect to the solar values of \citet{and89} of 10, 48, 0.1, 2.7 and 3.5 for helium, nitrogen, carbon, iron, and other elements, respectively. 

In our \xmm\ data, WR89 clearly is the most luminous object of the field. It presents a high \loglxlb\ ($-$5.7), which is typical of wind-wind collisions in WR+O systems (see \citealt{gue09} and references therein). Moreover, its spectrum requires one to fit a non-negligible  hot (2\,keV) component. Finally, while the source is constant during the 7h exposure (Fig. \ref{hm1lc}), this may not be the case on other timescales. Indeed, \citet{pol95} quoted a PSPC count rate of $8.8\pm5.6$\,cts\,ks$^{-1}$ for WR89\footnote{The Rosat Faint Source Catalog, also based on the survey data, lists a source at 1.9' from WR89: 1RXS\,J171909.6$-$384926, with a count rate of $20.8\pm9.8$\,cts\,ks$^{-1}$. The reason of the discrepancy with Pollock's value is unknown, but it is still much fainter than predicted from the observed \xmm\ data.}. Our fit would result in a PSPC count rate of about 59\,cts\,ks$^{-1}$, about an order of magnitude brighter. Therefore, the X-ray flux of WR89 is most probably variable with time, which is again typical of colliding-wind systems. In fact, it is quite possible that the \xmm\ data were taken close to periastron time, which would naturally explain the strong brightening (cf. the case of Cyg OB2 \#9, \citealt{naz12}). 

To find clues of binarity, we have examined the archival optical spectra of WR89: 24 spectra taken with FEROS on four consecutive nights in mid-2005, one EMMI/NTT spectrum from March 2002, three ESO1.52m/Boller \& Chivens spectra from May 1996, and one CTIO 1.5m/Cspec dataset from May 1999. To the limit of the data quality and taking into account the broad line widths, no obvious shift is detected\footnote{For the narrow lines of N\,{\sc iv}\,\l\,7103.24 and 7109.35\,\AA, we found velocities of $\sim-45$\,\kms\ and $-10$\,\kms, respectively.}. Either the runs were, by chance, performed at similar orbital phases, or the system has a very long period (decades), or the system's period is shorter (a few years) but the binary is highly eccentric, with velocities changing only during a short phase interval (e.g., the case of Cyg OB2 \#9, \citealt{naz12}). The latter scenario would also be consistent with other X-ray bright WR+O colliding-wind binaries (e.g., WR25, \citealt{gam06}). We therefore recommend a long-term multiwavelength monitoring of WR89, to search for the companion. Only after this can a modeling be undertaken and our understanding of the collision properties improved.

\begin{figure}
\includegraphics[width=9cm]{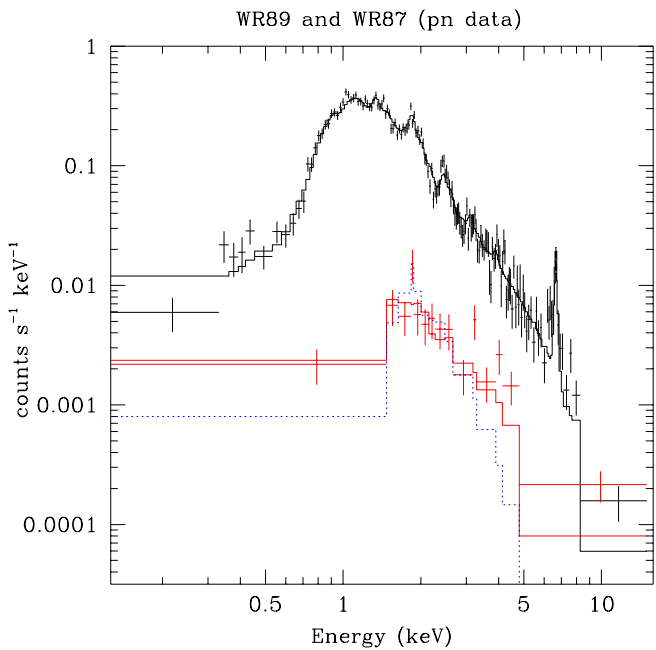}
\caption{The pn spectra of WR89 (top) and WR87 (bottom), along with their best-fit spectral models shown as solid lines (see Table \ref{hm1fit}). While both stars are subject to similar interstellar absorptions \citep{vaz01}, the spectrum of WR87 is obviously much harder and fainter. A dotted blue line shows the best-fit to WR87 spectra when $kT$ is fixed to 0.6\,keV (see text). The color version of this figure is available online.}
\label{specwr}
\end{figure}

\subsubsection{WR87 (XID 6)}
As for WR89, \citet{ham06} estimated the physical properties of WR87 by fitting atmosphere models to the stellar spectrum. Their mass fractions are 0.4 and 0.58 for hydrogen and helium, respectively (mass fractions of other elements like in WR89), which translates into elemental abundances compared to hydrogen with respect to the solar values of \citet{and89} of 3.7, 24, 0.06, 1.3, and 1.8 for helium, nitrogen, carbon, iron, and other elements, respectively. These values were used for our Xspec fitting (Table \ref{hm1fit}). 

Both WRs, which share similar reddening and spectral type, display similar properties at optical wavelengths \citep{mas01,ham06} - the wind density (as traced by $\dot M / [4 \pi R_*^2 v_{\infty}]$) in WR87 is even lower than twice that of WR89. This is in stark contrast with the high-energy properties: the observed X-ray flux of WR87 is about 20 times lower than that of WR89. When correcting for the average interstellar absorption of the cluster \citep[suitable for WR87, see][]{vaz01}, the difference in flux jumps to two orders of magnitude. This is due to the plasma temperature, which is high for WR87 (2\,keV, see also Figs. \ref{hm1HR} and \ref{specwr} - it should be noted that forcing the temperature to be 0.6\,keV, thereby increasing the intrinsic absorption, never leads to a good fit, because $\chi^2$ remains well above 2 in that case (see Fig. \ref{specwr}): WR87 truly seems dominated by very hot plasma). WR87 therefore is a paradoxical case: while its X-ray flux is low, suggesting the star to be single, its high plasma temperature is typical of colliding-wind systems. Wind magnetic confinement can produce hot plasma in single objects, but it would then be associated with an X-ray overluminosity (e.g., the case of $\theta^1$\,Ori\,C, \citealt{gag05}), which is not observed for WR87.

\begin{figure}
\includegraphics[width=9.3cm, clip]{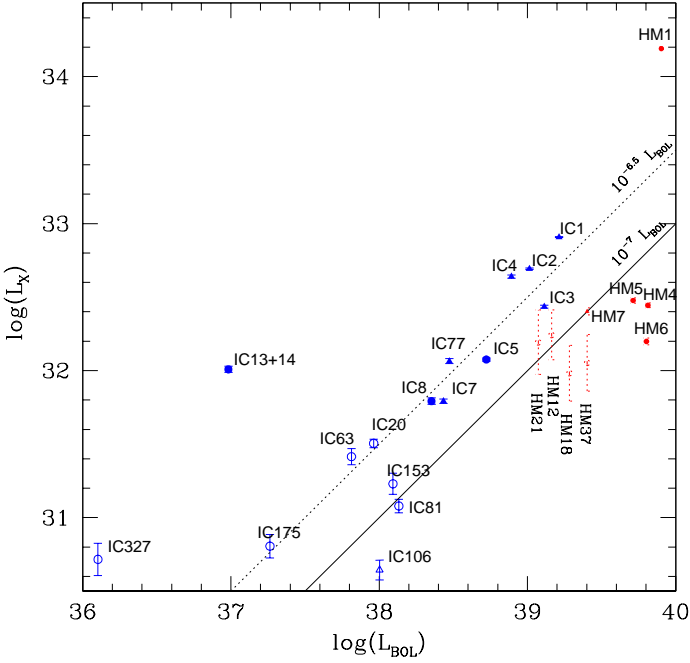}
\caption{X-ray luminosity (in 0.5--10.0\,keV energy band) as a function of bolometric luminosity for massive stars in HM1 (red points) and IC\,2944/2948 (blue points). The associated XID is quoted next to each point, with the prefix HM for sources in HM1 and IC for sources in IC\,2944/2948: except for WR89 (XID 1 in HM1), all stars in HM1 display lower \lxlb\ than those of IC\,2944/2948. Filled symbols correspond to bright objects whose X-ray luminosities were derived from a spectral fit (Table \ref{hm1fit}); open symbols correspond to fainter objects whose count rates were converted into X-ray luminosities; triangles indicate the known binary systems. Approximate error bars were calculated using the relative errors on the total count rates (Tables \ref{hm1det} and \ref{hm1fit}), except for faint objects in HM1, where they indicate the range in X-ray luminosity (see text).}
\label{hm1lxlb}
\end{figure}

Single WC stars typically have low X-ray luminosities, $<10^{31}$\,erg\,s$^{-1}$ with \loglxlb $< -8$, with WR\,135 even having $<7\times 10^{29}$\,erg\,s$^{-1}$ with \loglxlb $< -9$ \citep[ and references therein]{gue09}. However, the situation is less clear for WN stars, which show a large scatter in their X-ray luminosities \citep[ and references therein]{gue09}. Some of them are similar to WC stars: for example, WR40 has a luminosity $< 4\times 10^{31}$\,erg\,s$^{-1}$ and \loglxlb $<-7.6$ \citep{gos05}. Others follow the \loglxlb $=-7$ of O-type stars \citep{osk05,ski12}, but in contrast, a few are highly luminous: for example, WR20b has a luminosity of $ 2\times 10^{33}$\,erg\,s$^{-1}$ and \loglxlb $=-6.1$ \citep{naz08}. A relation between X-ray luminosity and wind kinetic luminosity ($=0.5 \dot M v_{\infty}^2$) of WNL stars has been proposed \citep{ski12}, and WR87 fits into this scheme well, but this relation displays a large scatter (and may not apply to all cases, e.g., WR40 or WR136). Usually, the main temperature associated to the X-ray emission of WR stars is low (0.6\,keV), but a non-negligible hard component may exist, especially for binaries or binary candidates \citep[ and references therein]{gue09}. The closest analog to the unusual case of WR87 may be the WN6 star WR136, which is both underluminous in X-rays and dominated by high-temperature plasma in the high-energy range \citep{ski10}. The nature of the high-energy emission of this latter object is still unknown, however. An additional monitoring of WR87 and WR136, both in X-ray and visible ranges, may shed some light on their intriguing high-energy properties.

\onlfig{5}{
\begin{figure*}
\includegraphics[width=9.cm, bb=50 190 570 650, clip]{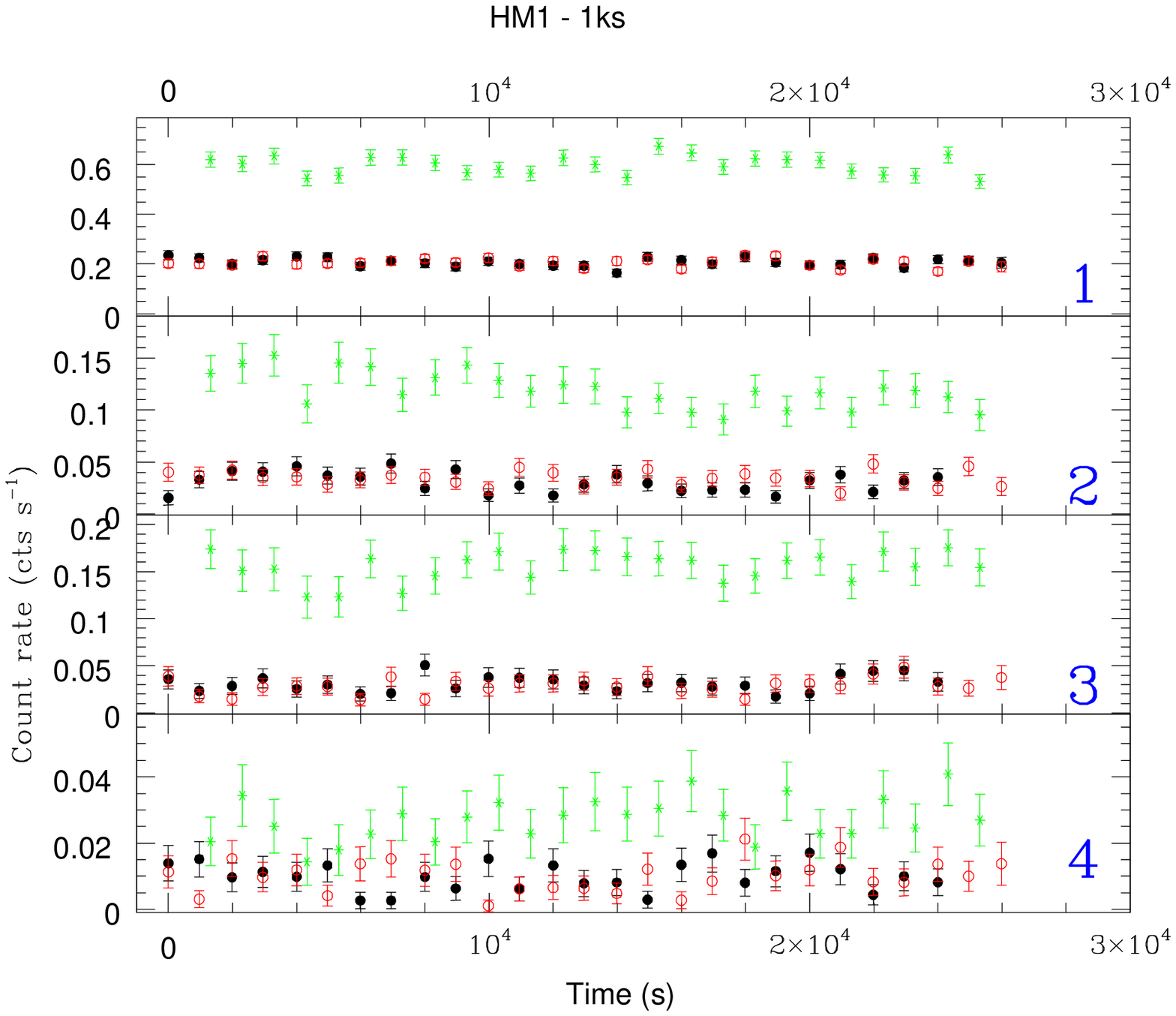}
\includegraphics[width=9.cm, bb=50 190 570 690, clip]{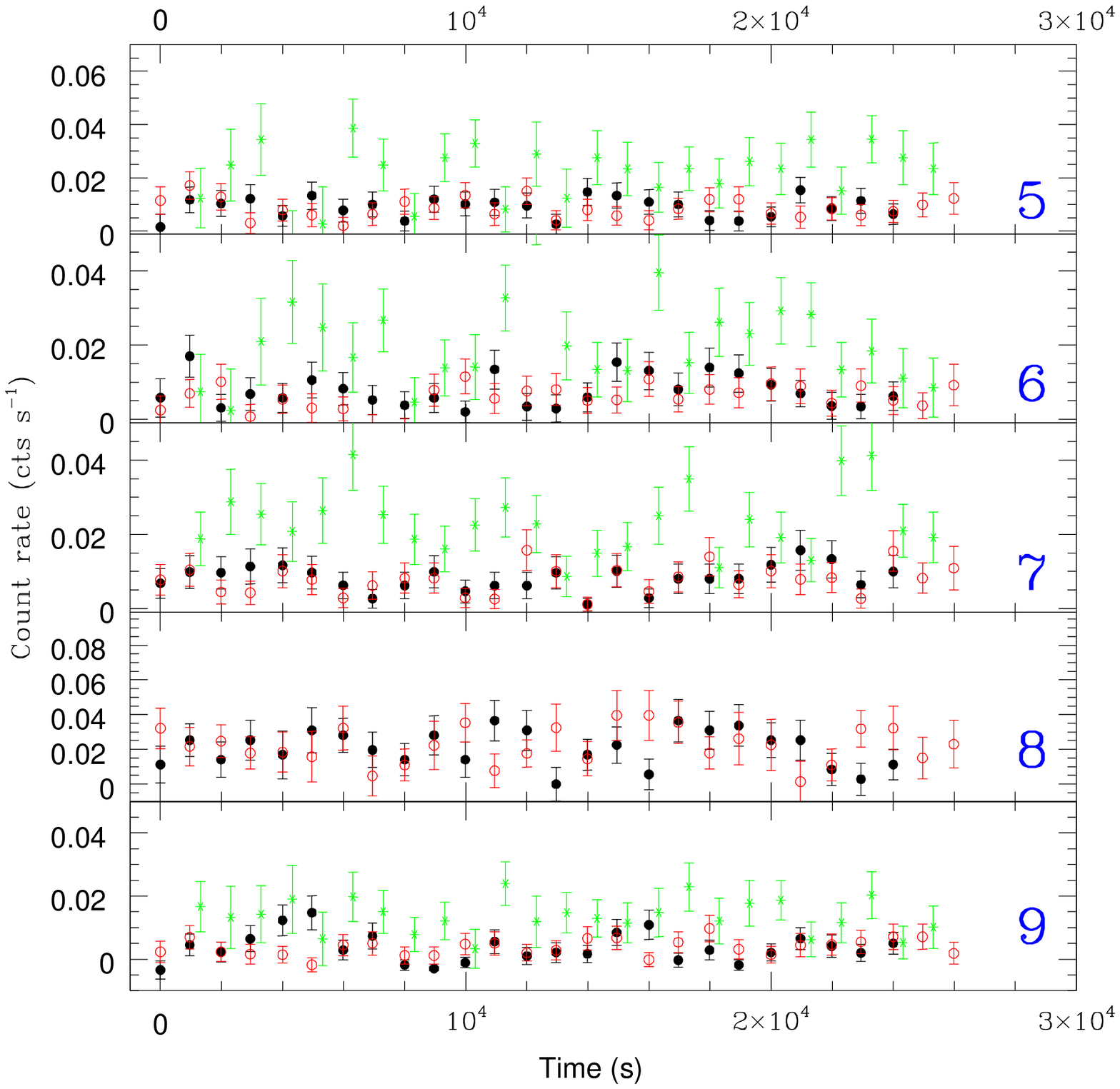}
\includegraphics[width=9.cm, bb=50 190 570 650, clip]{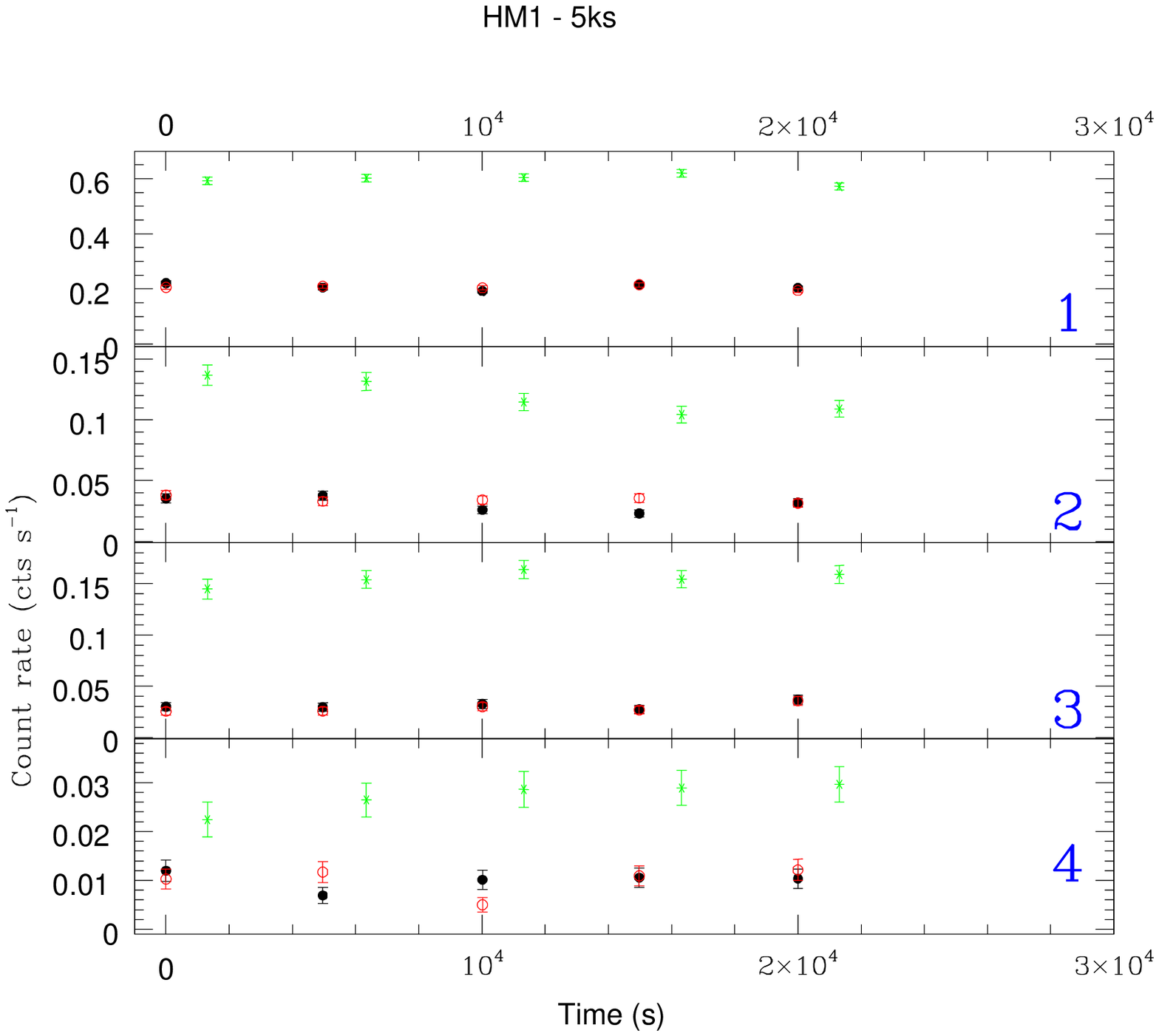}
\includegraphics[width=9.cm, bb=50 190 570 690, clip]{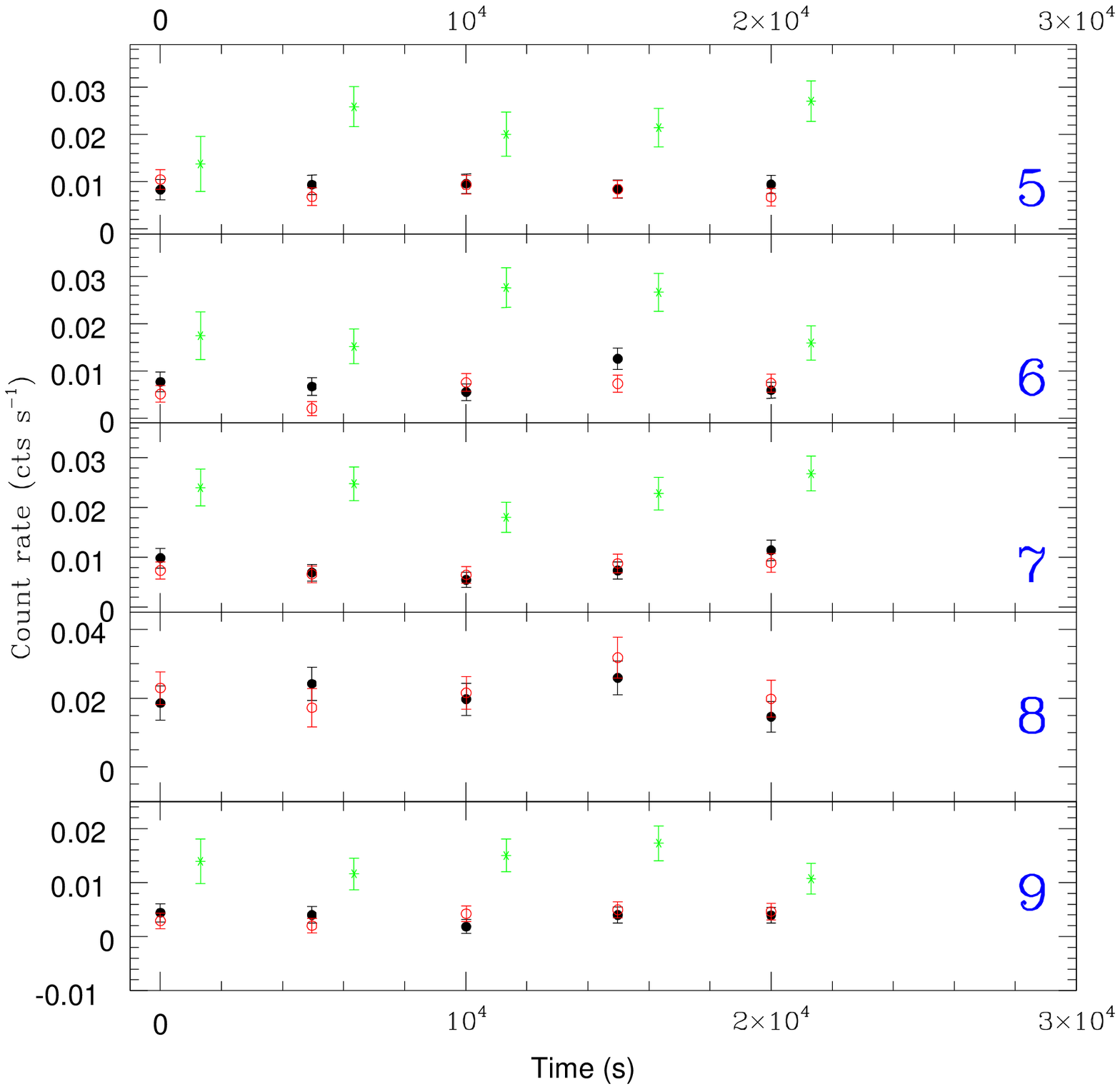}
\caption{The EPIC light curves of the nine brightest X-ray sources (pn=green asterisks, MOS1=black dots, MOS2=red circles) with 1\,ks time bins (top) and 5\,ks time bins (bottom). The x-axis is the time elapsed, in seconds, since the beginning of the MOS1 observation, and the XID of each source is quoted to the right of each lightcurve. $\chi^2$ tests do not find significant (at the 1\% level) and coherent variations in these light curves (XID 6 is found to vary at the 10\% level, and XID 2 is slightly better fitted by a trend than by a constant line). The color version of this figure is available online.}
\label{hm1lc}
\end{figure*}
}

\subsubsection{Detected O-stars}
The bolometric luminosities of O-type stars were estimated from the absolute V-magnitude found by \citet[ see $M_V$ in their Table 4]{vaz01}, using the bolometric corrections of \citet{mar05} for the spectral types, when known \citep{mas01}, or from approximate spectral types corresponding to the colors and V-magnitudes for the other blue and bright objects (see above). We now discuss each star in turn.

For the three O-type stars with enough counts, we performed a fitting within Xspec using a thermal plasma model (Table \ref{hm1fit}). 

With an O4If+ type, VB4 (XID 4) is the earliest star of the cluster. As for all O-stars in HM1, it was never detected in X-rays before and shows no significant (i.e., significance level $SL<1$\%) and coherent (i.e., similar in all instruments) short-term variations of its \xmm\ light curves (Fig. \ref{hm1lc}). Its \loglxlb\ amounts to $-7.4$ (Table \ref{hm1fit}), when a value of $-6.5$ to $-7$ is expected. It is possible that the O-stars in HM1 are truly hotter and fainter, though there is no known cluster property (e.g., abundances,etc.) that would explain this peculiarity. Another, more probable possibility is that the low signal-to-noise ratio at low energies, due to high absorption and short exposure, leads to an underestimate of the X-ray luminosity and an overestimate of the plasma temperature (0.2--0.6\,keV being more typical for such objects, \citealt{Sana,gue09,naz09,naz11}) - note that the additional absorption is typical of O-type stars \citep{naz09,naz11}, however. To check this possibility, we modeled an XMM observation, with a duration (25\,ks) similar to ours, of the ``typical'' 2XMM spectrum of O-stars. The flux ($5\times 10^{-14}$\,erg\,s$^{-1}$) and absorption ($N_{\rm H}=1.1\times10^{22}$\,cm$^{-2}$) were chosen to be typical of the cluster objects. We then fitted this fake X-ray spectrum with a single thermal component, as was done here for the faint O-stars: the deduced temperature is high ($\sim$1\,keV) and the deduced absorption-corrected flux is half (or 0.3\,dex below) the actual input value, possibly explaining the above results. 
 
VB5 (XID 5) displays an O5If+ type, with a \loglxlb\ of $-7.3$ (Table \ref{hm1fit}) and its emission displays the same characteristics as VB4. From the X-ray emission alone, there is no evidence for colliding winds for either of these two objects.

VB10 (XID 7) was classified as O5V by \citet{mas01}, but its nature was recently revised: it is a binary of type O5III(f)+OB, with a period of 5.9d, and minimum masses of 31+15\,M$_{\odot}$ \citep{gam08}. Amongst the three X-ray bright O-stars, VB10 is the hardest and brightest (\loglxlb $= -7.0$). This may hint at a contribution from X-ray bright colliding winds, but additional observations are needed to secure this conclusion, notably searching for the phase-locked variations that are typical of this phenomenon.

Four other massive stars have been detected by \xmm, but they are too faint for spectral analysis. We converted their count rates into intrinsic fluxes using (1) the interstellar absorption and a thermal plasma with temperature of 0.6\,keV, (2) the interstellar absorption and the typical O-star spectrum found in \citet{naz09}, and (3) the average spectral properties of the brightest O-stars (i.e., interstellar absorption plus additional absorption of $4\times10^{21}$\,cm$^{-2}$ and a plasma temperature of 0.85\,keV). Different model assumptions yield different results, explaining the luminosity ranges given below. 

VB11 (XID 12) was classified as O6If by \citet{mas01}, hence a bolometric luminosity of $\log(L_{\rm BOL}/L_{\odot})=5.58$ can be derived from its photometry \citep{vaz01}. The \xmm\ detection leads to an X-ray luminosity of $1.1-2.4\times10^{32}$\,erg\,s$^{-1}$ and a \loglxlb\ of $-6.8$ to $-7.1$. 

VB13 (XID 21) was classified as O7V((f)) by \citet{mas01}, hence a bolometric luminosity of $\log(L_{\rm BOL}/L_{\odot})=5.49$ can be derived from its photometry \citep{vaz01}. The \xmm\ detection leads to a similar X-ray luminosity as for VB11 and a \loglxlb\ of $-6.7$ to $-7.1$.

\begin{figure*}
\includegraphics[width=6cm]{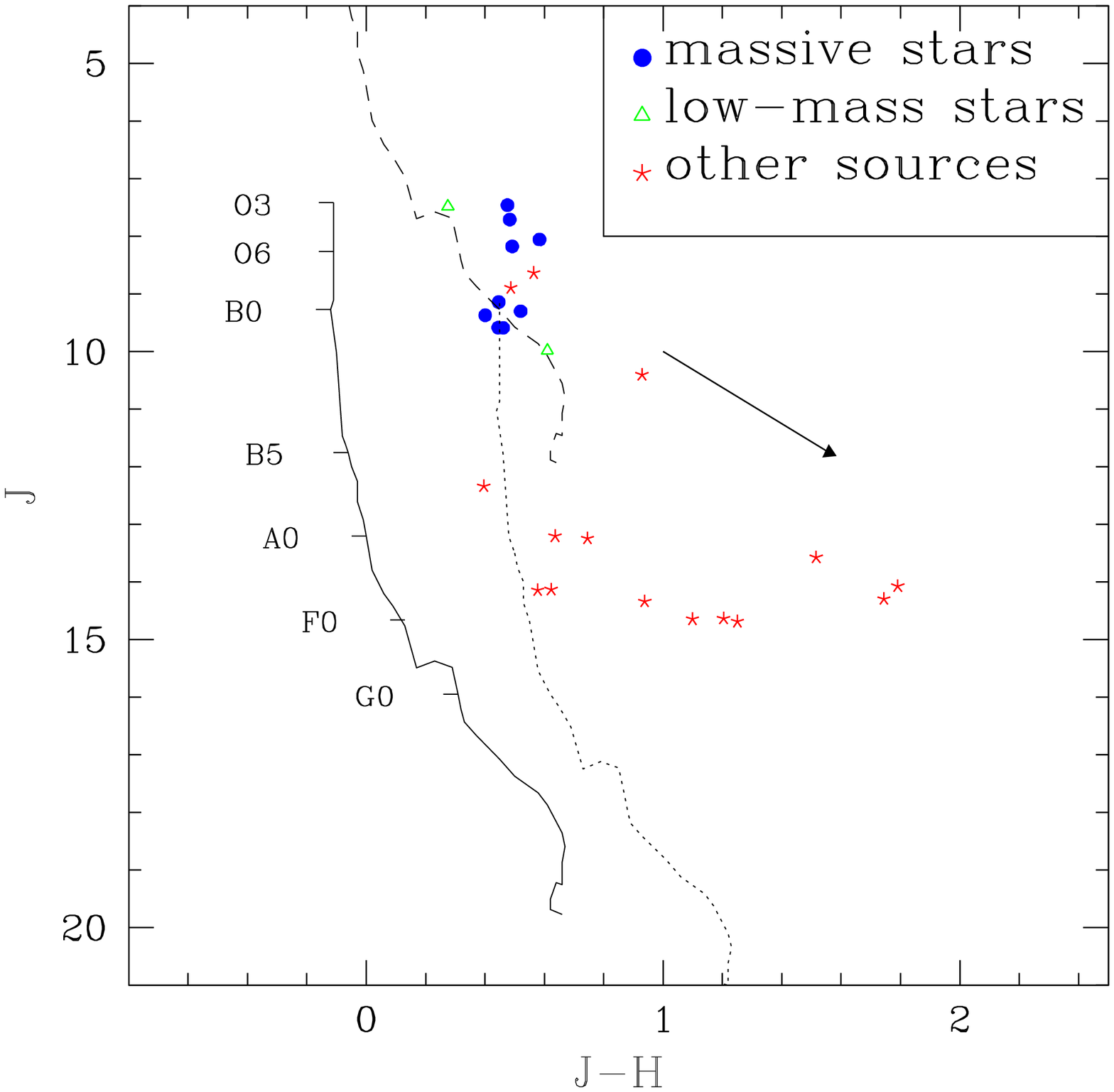}
\includegraphics[width=6cm]{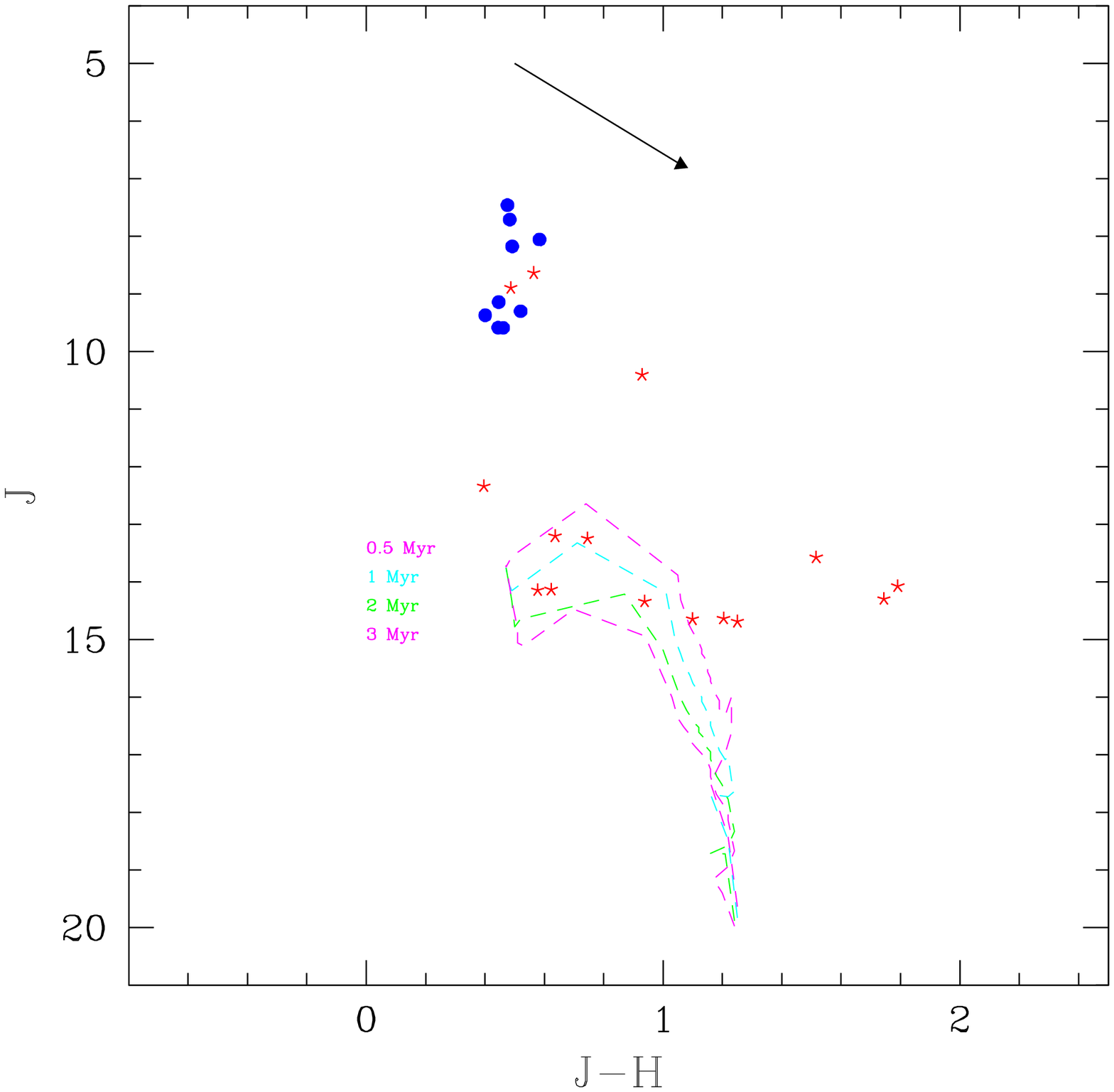}
\includegraphics[width=6cm]{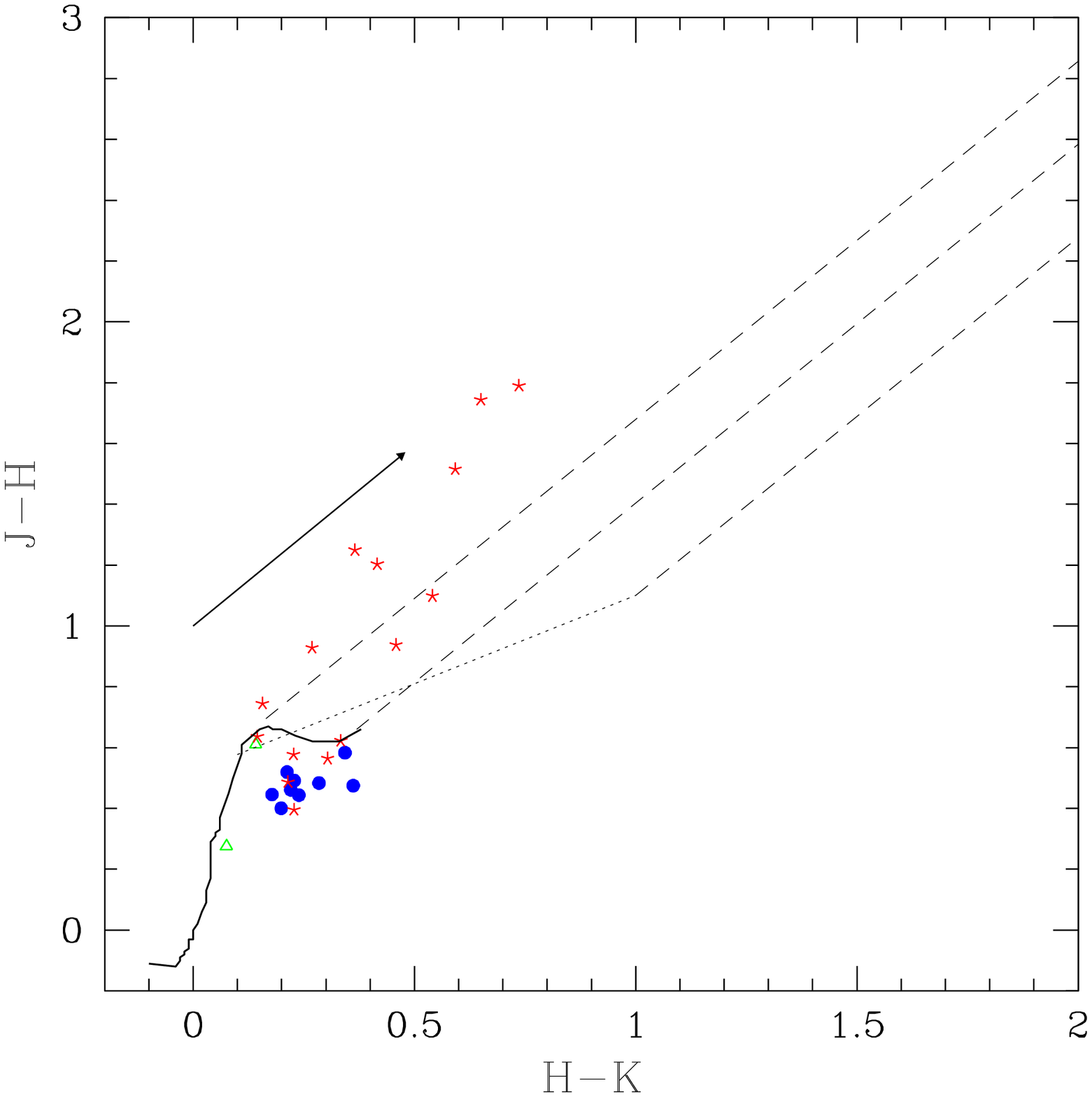}
\caption{{\it Left:} Color-magnitude diagram for non-spurious X-ray sources in HM1 having reliable 2MASS counterparts. The dashed line corresponds to the main sequence shifted for a distance of 92\,pc (that of HD\,156301), the solid and dotted lines correspond to the main sequences for a $DM=12.6$\,mag, typical of the cluster \citep{vaz01}, without and with correction for the cluster's absorption \citep[$E(B-V)=1.84\,mag, R_V=3.3$, see ][]{vaz01}, respectively. The main sequence magnitudes and colors are taken from \citet{mar06} for O-stars (masses $>15$\,M$_{\odot}$) and \citet{tok00} for the other spectral types (masses in the range 0.1--15\,M$_{\odot}$). The reddening vector corresponds to the reddening expected for the cluster. {\it Middle:} The same color-magnitude diagram with \citet{sie00} isochrones, for $DM=12.6$\,mag and $E(B-V)=1.84$\,mag, superimposed as dashed lines. The displayed tracks correspond to masses in the range 0.1--7.0\,M$_{\odot}$ (spectral types M6 to B3). {\it Right:} Color-color diagram for the same objects. The dotted line shows the intrinsic (i.e., dereddened) colors of classical T Tauri stars \citep{mey97}, the dashed lines correspond to increasing values of absorption (using $R_V=3.3$ and \citet{car89} extinction law) for the blue and red limits of the main sequence for low-mass stars and of the TTs sequence. Note that for these figures the 2MASS JHK photometry was transformed into the Bessel \& Brett system (see http://www.astro.caltech.edu/$\sim$jmc/2mass/v3/transformations/ ). The color version of this figure is available online.}
\label{hm12mass}
\end{figure*}

Though its spectral type is unknown, VB20 (XID 18) is both blue and bright \citep[see table 4 in ][]{vaz01} and massive ($>20$\,M$_{\odot}$, \citealt{mas01}). Using V{\'a}zquez \& Baume photometry, we estimated an O9.5I or O3III type for this star, leading to a bolometric luminosity of $\log(L_{\rm BOL}/L_{\odot})=5.44$ or 5.96. Its EPIC count rate (Table \ref{hm1det}) leads to an X-ray luminosity of about $0.5-1.4\times10^{32}$\,erg\,s$^{-1}$, or a \loglxlb\ of $-6.9$ to $-7.8$. 

Also of unknown spectral type, VB26 (XID 37) is a cluster member that is both blue and bright in \citet{vaz01}. We estimated a spectral type of O4III, a bolometric luminosity of $\log(L_{\rm BOL}/L_{\odot})=5.82$, an X-ray luminosity similar to VB20, and a \loglxlb\ of $-7.2$ to $-7.6$.

For the three brighter massive stars, we had found \loglxlb\ of $-7.20$ with a dispersion ($=\sqrt(\sum (X_i-mean)^2/(N-1))$) of 0.19\,dex (or $-7.31$ with a dispersion of 0.09\,dex if excluding the known binary). This agrees well with the values found for the fainter objects. 

\subsubsection{Non-detected massive stars}
Only one O-type star with a known spectral type has not been detected: VB25 (O9.5V). It is faint in the visible, however, with a V-magnitude of 8.2\,mag and an estimated bolometric luminosity of $\log(L_{\rm BOL}/L_{\odot})=4.8$. With a  \loglxlb\ of $-7$, this would yield count rates of $0.9-2.5$\,cts\,ks$^{-1}$ for pn and $0.3-0.7$\,cts\,ks$^{-1}$ for MOS, considering the three spectral models quoted above. Our faintest detections (Table \ref{hm1det}) have count rates of about 3\,cts\,ks$^{-1}$ for pn and 1\,cts\,ks$^{-1}$ for MOS.

Additional candidate-OB stars also exist. Three additional objects are bright and blue in V{\'a}zquez \& Baume photometry and massive ($>20$\,M$_{\odot}$) following the Massey et al. estimates: VB16, VB24, VB37. Eleven others are only bright and blue members of HM1 in V{\'a}zquez \& Baume photometry: VB14, VB32, VB35, VB47, VB52, VB55, VB57, VB65, VB70, VB95, and VB96. We used the available photometry to estimate a spectral type and derive a bolometric luminosity for these cases. Values range from $\log(L_{\rm BOL}/L_{\odot})=4.8$ to 5.4, with only VB14 showing a $\log(L_{\rm BOL}/L_{\odot})=5.7$. The faintest cases should remain undetected owing to our sensitivity, as already seen for VB25. The brightest objects in this sample should have been detected, however, whatever the underlying assumption for the X-ray spectral model. Either the photometry, hence the spectral type estimate, has been overestimated for these objects, or the \loglxlb\ in HM1 is below $-7$, which is possible in view of the properties of some of the detected objects (e.g., $-7.4$ for VB4, see also discussion below).

\subsection{Other X-ray bright objects}

\subsubsection{XID 2 = V504 Sco}
V504 Sco is a known eclipsing variable with a period of 2.59d, a maximum magnitude of $b=13.7$\,mag, and a primary eclipse depth of 0.8\,mag \citep{mal06}. Unfortunately, the spectral types of the components are not known, but the 2MASS photometry suggests an integrated type of about K5V and a distance of about 100\,pc (see Fig. \ref{hm12mass}). In our observations, this bright X-ray source is only surpassed in apparent flux by the overluminous WR89. Its X-ray spectrum is dominated by the thermal component with $kT=2$\,keV and little absorption, consistent with a foreground source (Table \ref{hm1fit}). With the above distance, the X-ray luminosity amounts to $L_X\sim2.5\times10^{29}$\,erg\,s$^{-1}$ in the 0.5--10.0\,keV energy band. The absence of flares in our observation (Fig. \ref{hm1lc}) and the inferred spectral type, plasma temperature, and X-ray luminosity suggest that this object corresponds to a coronal source in quiescence, but the binarity may also point to a lowly-active RS\,CVn system.

\subsubsection{XID 3 = HD156301}
HD156301 is a nearby (92\,pc, \citealt{van07}) low-mass star of type F5V and magnitude $V=8.4$\,mag. It possesses a close (0.6'') and faint ($V=10.7$\,mag) companion \citep{har96,fab00}. In our observations, it is bright, non-flaring, soft ($kT$ of 0.03 and 0.3\,keV) and only lightly absorbed. Its X-ray luminosity corrected for a reddening of $E(B-V)=0.11$\,mag amounts to $7\times 10^{29}$\,erg\,s$^{-1}$, resulting in \loglxlb\ of about $-4.3$, i.e., far from the saturation level of coronal sources. The X-ray emission thus is compatible with the properties of main-sequence low-mass stars (outside flaring times).

\begin{figure*}
\includegraphics[width=9cm]{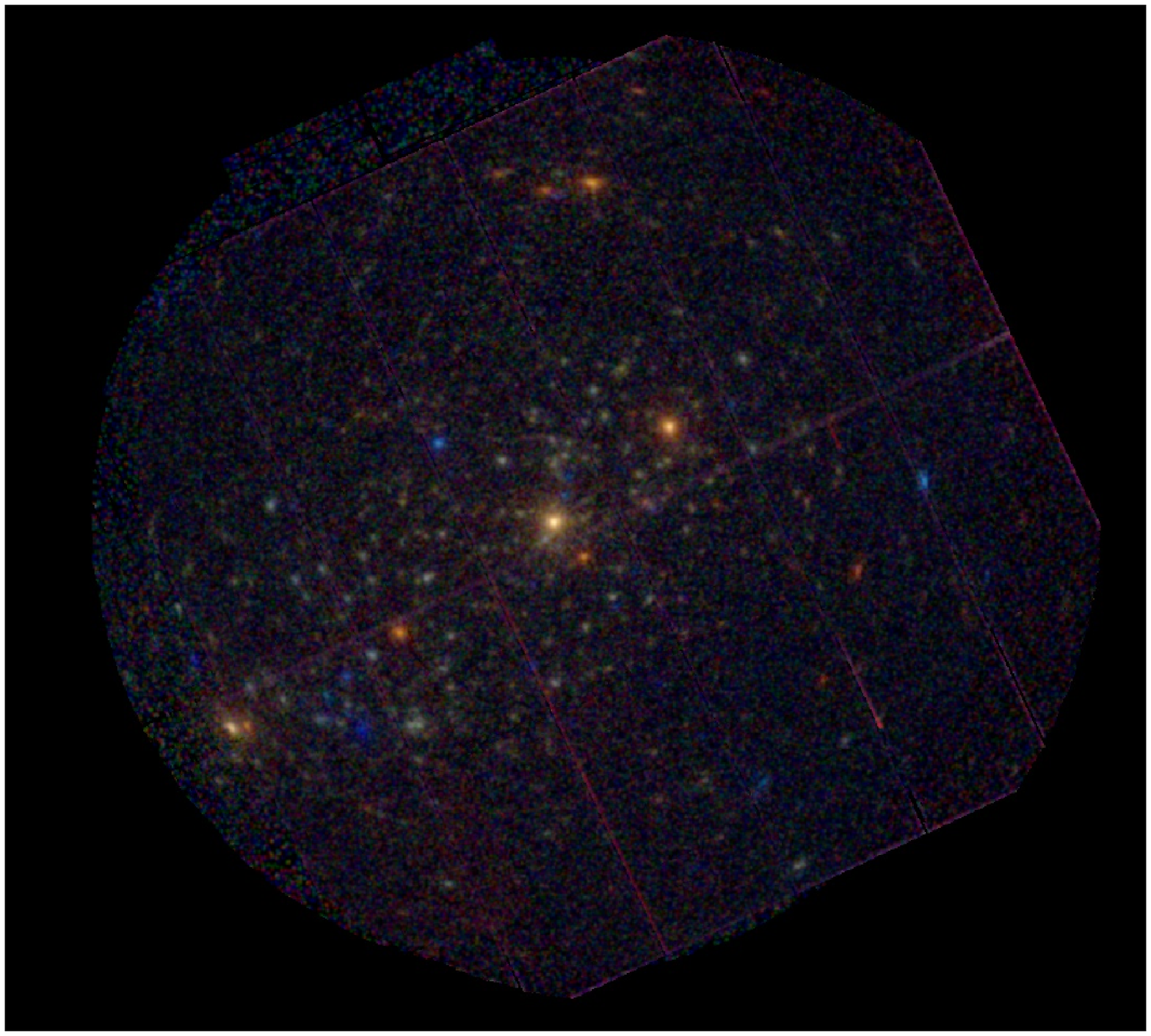}
\includegraphics[width=9cm,bb=40 170 540 675, clip]{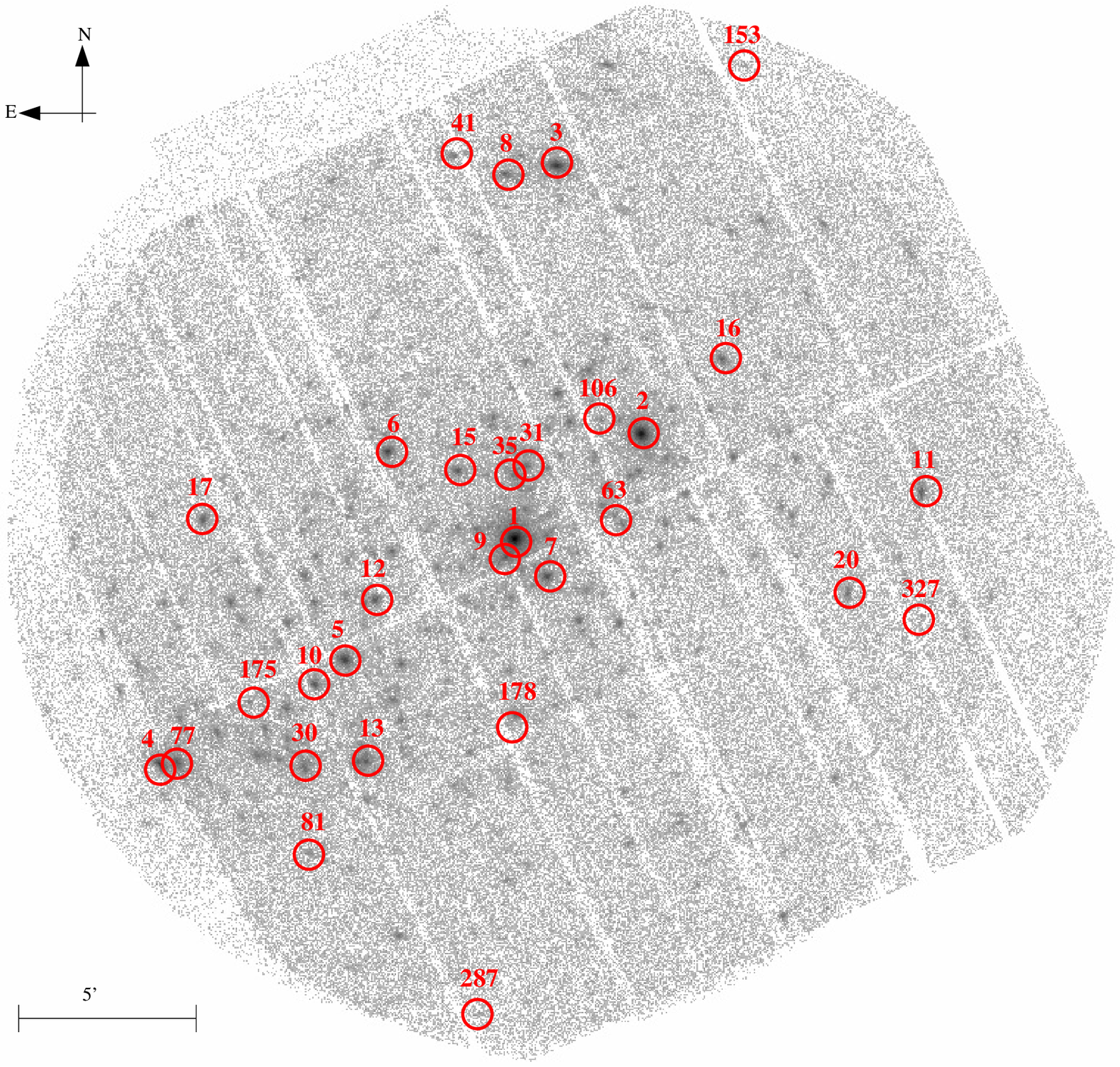}
\caption{Same as Fig. \ref{hm1col}, but for IC\,2944/2948. The yellow source at the field's center is HD\,101205. The color version of this figure is available online.}
\label{iccol}
\end{figure*}

\subsubsection{XID 8}
This off-axis source does not possess any visible or IR counterpart. The MOS data (source in gap for pn) indicate a high extinction ($N_{\rm H}> 2\times 10^{22}$\,cm$^{-2}$) and a hard emission (Fig. \ref{hm1HR} and Table \ref{hm1fit}). Therefore, XID 8 displays all properties of a background accreting object (X-ray binary or extragalactic active galactic nuclei, AGN).

\subsubsection{XID 9}
This source is not as hard as XID 8 (see Fig. \ref{hm1HR}) and has an IR counterpart. Its absorption is close to the interstellar value of the cluster, and the associated plasma temperature, about 3\,keV, is higher than for normal massive stars. Without more information, it is difficult to ascertain the nature of this source.

\subsection{Faint X-ray sources}
To learn more about the faint X-ray sources, we focused on the sources that display NIR counterparts. We considered only the 29 objects with reliable 2MASS photometry (i.e., best 2MASS quality flag Qflag=AAA), and Fig. \ref{hm12mass} shows the associated color-magnitude and color-color diagrams. 

As expected, the massive stars are at the top of the main sequence, although the two WRs, VB4, and VB5 are slightly too bright: for the latter two, this could be explained by binarity. The remaining 18 stars can be split into two classes. The first one contains foreground objects, which are less distant and extincted than cluster stars. The second one corresponds to a population of reddened objects at the distance of the cluster. Their age, derived from the \citet{sie00} isochrones, is about 0.5--2\,Myr, i.e., they are younger or of similar age as the main stars of the cluster, and they do not display high IR-excesses. They may thus correspond to weak-line TTs, the first ones detected in this cluster. Note that the position of the most reddened objects is not entirely compatible with the extinction law previously attributed to the cluster \citep[$R_V=3.3$, ][]{vaz01}, which could indicate anomalous extinction.

\section{IC\,2944/2948}
IC\,2944 and IC\,2948 belong to the Cen\,OB2 association. The existence of these two clusters has been questioned in the past: while some authors advocate that these are one (or several) true physical cluster(s) \citep[e.g.][]{wal87}, others instead see the groupings as a chance superposition of isolated hot stars scattered along the Carina arm \citep[e.g.][]{per86}. Moreover, even the identification of the two groupings is often confused in the literature \citep{rei08}. In the X-ray image (Fig. \ref{iccol}), a tight cluster of X-ray sources is detected, suggesting at first sight that IC\,2944/2948 is indeed a true cluster.

Few recent (i.e., CCD) photometric studies of the cluster exist. Searching for Be stars, \citet{mac05} provided $b$, $y$, and H$\alpha$ measurements for the cluster core. They evaluated the distance to be 1.8\,kpc, the age to be 6.6\,Myr, and the reddening to be $E(B-V)\sim0.32$\,mag. \citet{kar05} found similar distance and reddening, but \citet{san11} instead favored a larger distance of 2.3\,kpc for the O-stars, in agreement with the results of \citet[2.2\,kpc]{tov98}. In this paper, we use the higher value.

\begin{figure*}
\includegraphics[width=6.cm]{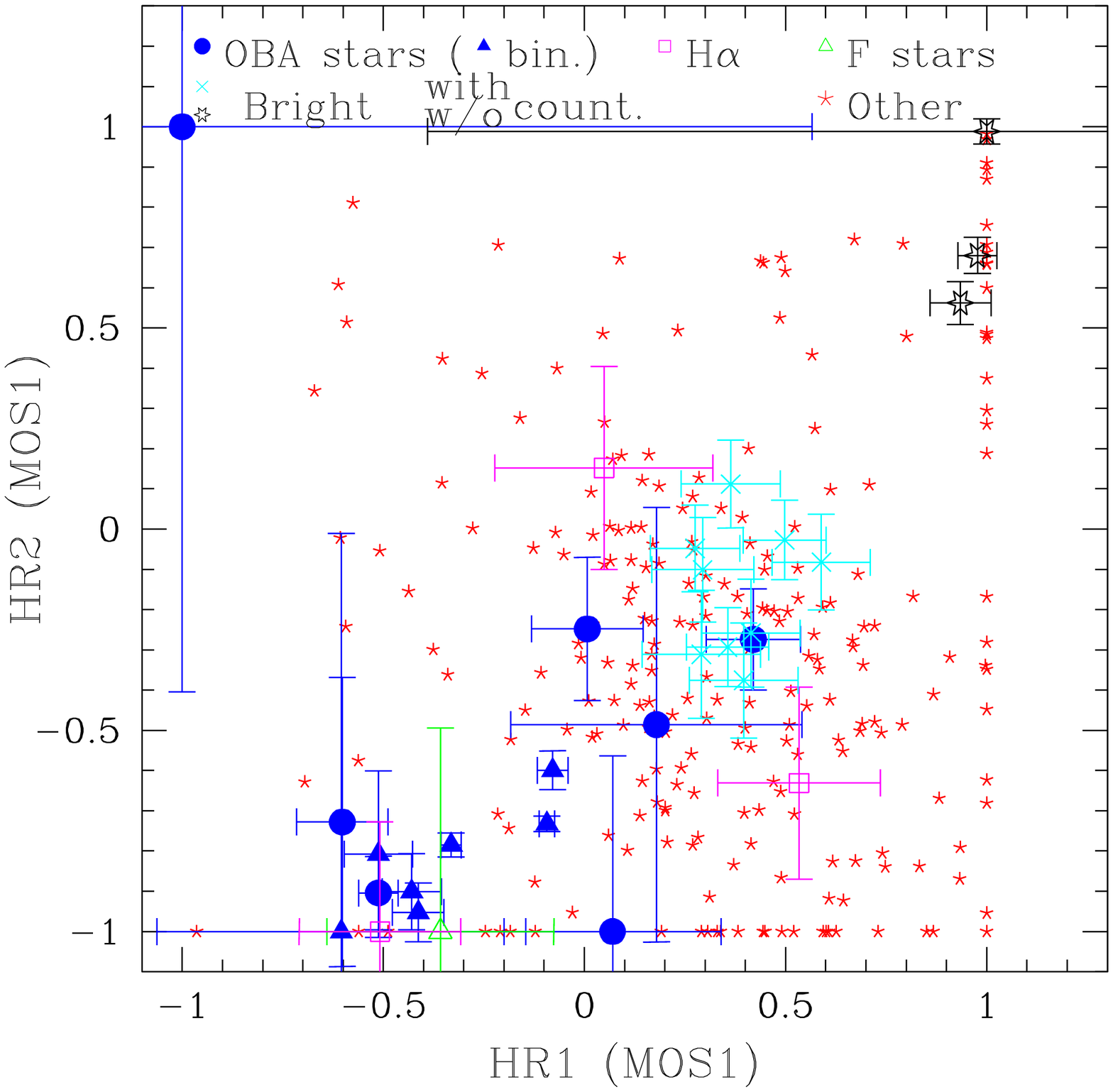}
\includegraphics[width=6.cm]{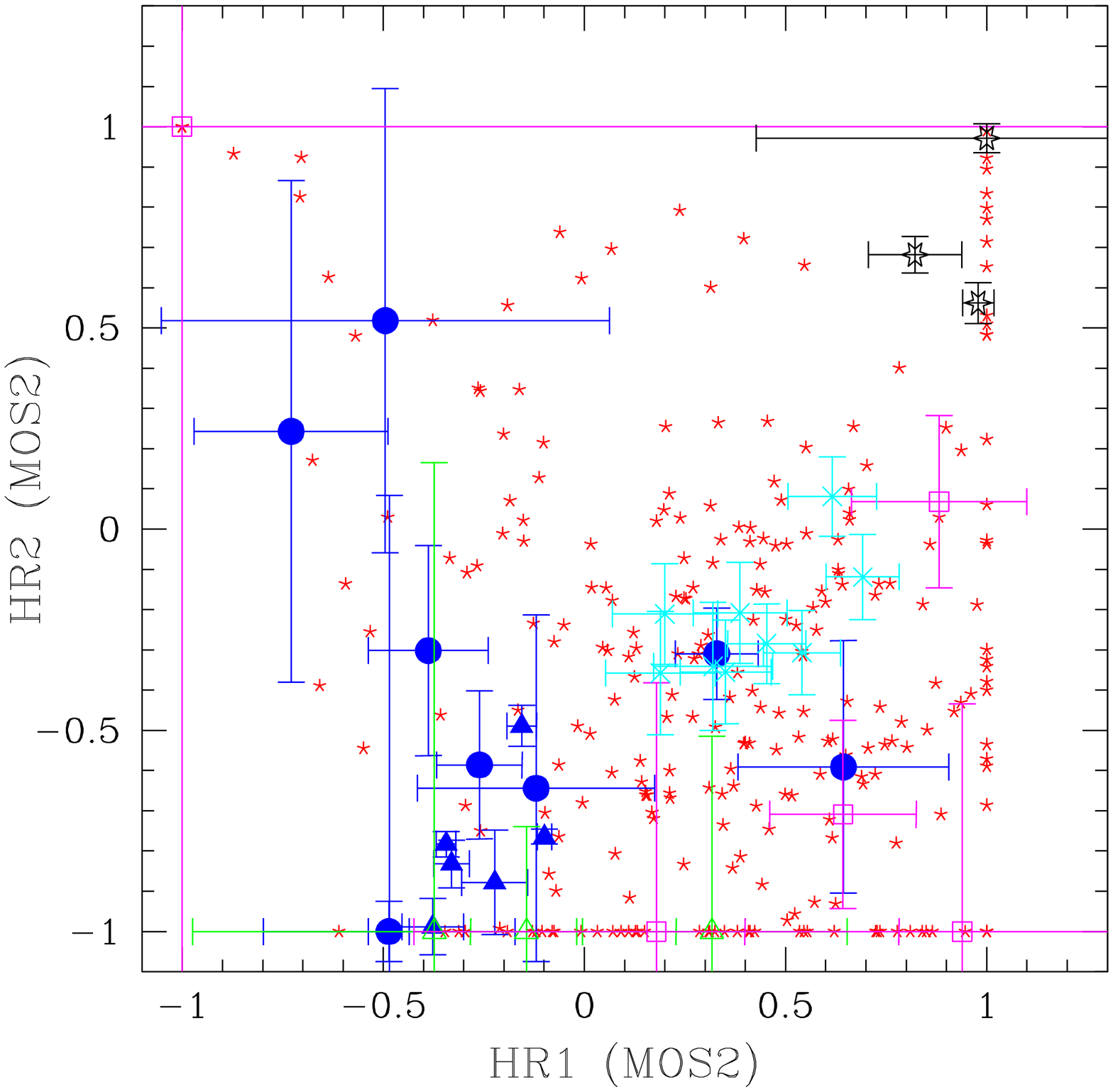}
\includegraphics[width=6.cm]{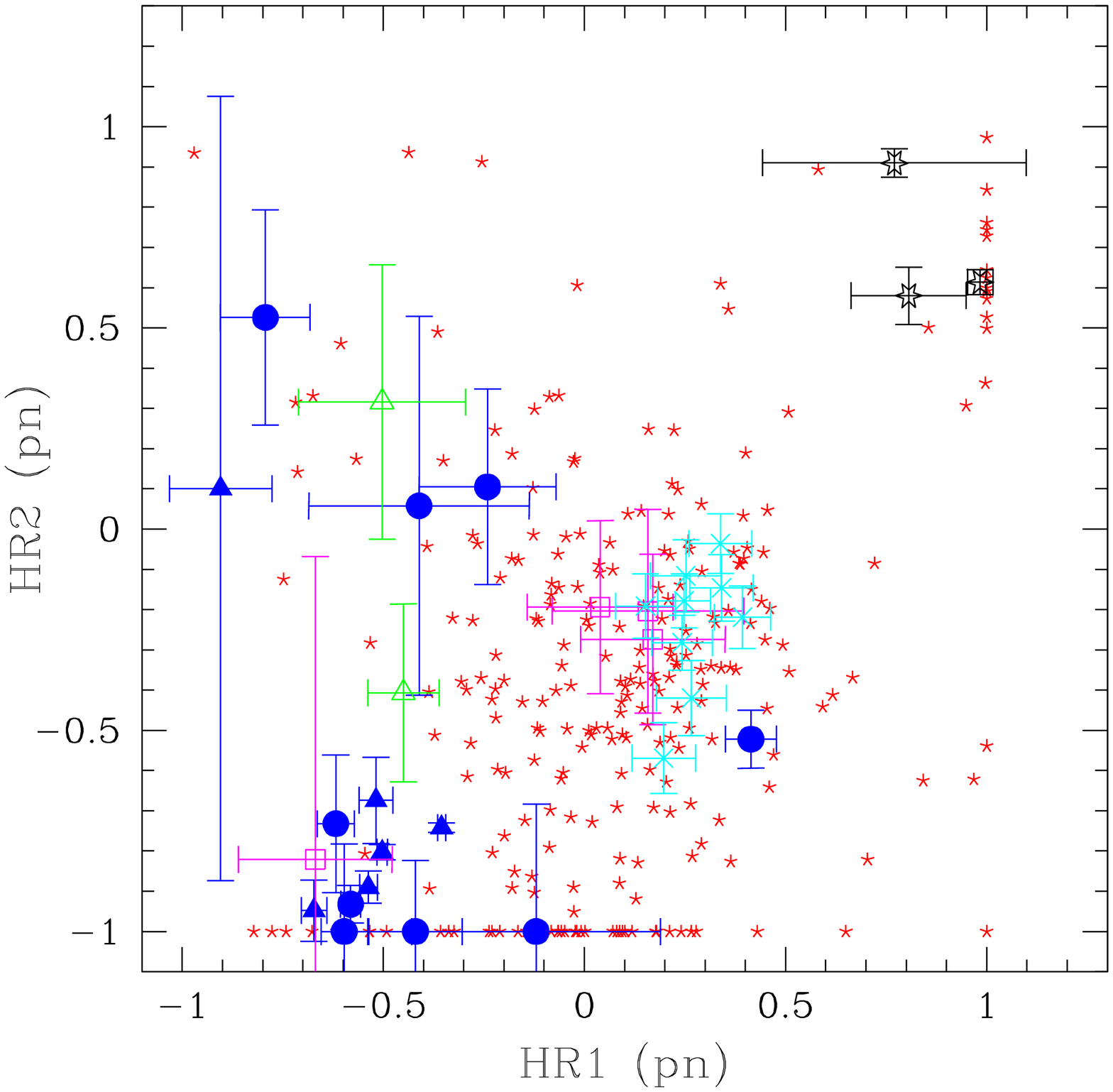}
\caption{Same as Fig. \ref{hm1HR} for sources in IC\,2944/2948. The magenta squared symbols are added to represent H$\alpha$ emitters. The color version of this figure is available online.}
\label{HR_ic}
\end{figure*}

\subsection{Detection}

In X-rays, the field of IC\,2944/2948 is much more crowded than that of HM1, with several sources overlapping due to the broad PSF. The XMM detection algorithm does not easily account for both source crowding (i.e., overlapping PSFs) and source extension (i.e., sources with PSFs larger than that of a single source), however. Therefore, we performed the source detection in two steps. First, we allowed the detection of extended sources, but not simultaneous fitting in dense groups of objects: 267 sources were then found, with 6 extended sources (all with a low extension, $<$6''), 16 probably spurious objects (not detected by eye) and many obvious sources missing. Clearly, crowding impairs source detection in this case. Second, we allowed the simultaneous fitting of up to four objects, to detect and separate sources in dense groups, but without the possibility of extension. This indeed solves the crowding problem, resulting in a much more complete list of 368 sources, with 53 probably spurious cases (Table \ref{icdet}). Note that we kept the results from the first trial in two cases: (1) XID 30, since it is a hard source that is larger than nearby point sources - its extension, detected in the first run, is thus not spurious (see bottom left of Fig. \ref{iccol}) - and (2) XID 77, since the position of this off-axis source associated with HD\,101413 seemed better fitted in the first trial. 

When the position of the massive O-stars of the field is superposed on the X-ray image, a shift is clearly seen. Comparing the Hipparcos positions of the massive O-stars with their associated X-ray sources (excluding HD\,101413 and HD\,101436, which are heavily distorted because of their off-axis position), we found an average shift of $RA_{Hip}-RA_{XMM}=0.45$s and $DEC_{Hip}-DEC_{XMM}=0.41$'' in DEC. We thus corrected the coordinates of the X-ray sources by adding these values to those found by the detection algorithm (Table \ref{icdet}), and only then correlated with other existing catalogs. 

\subsection{Cross-correlation}

A cross-correlation with 2MASS yields 260 counterparts to the 368 X-ray sources within 3'' (this radius seems the best to avoid spurious identifications, see above). Amongst these, a second object is found within the same radius in only 11 cases: Table \ref{iccorrel} therefore only lists the closest counterpart. In total, 75\% of these 260 counterparts have the best quality for 2MASS photometry ($Q$ flag='AAA') and can be used for further photometric studies (see below).

A cross-correlation with the catalog of \citet{mac05} was also performed. Within 3'' radius, only 13 of the 368 X-ray sources have a counterpart. We derived their H$\alpha$ equivalent widths using formula (4) of \citet{mac05a}. In six cases, H$\alpha$ can be considered to be in emission because the absolute equivalent widths exceed 10\,\AA. 

Finally, a last cross-correlation was performed with Simbad, again within 3'': 28 of the 368 X-ray sources have a counterpart. Ten of these objects are massive O-stars recently studied by \citet{san11}, six other objects display a B-type, one an A-type, and three an F-type. The latter ones are probably foreground sources unrelated to the cluster. The remaining eight objects have no known spectral type.

Table \ref{iccorrel} gives the results of these correlations.

\subsection{Additional information}
Fig. \ref{HR_ic} shows the hardness ratios of the detected objects. The massive stars, which are soft, form a group with low hardness ratios, whereas the X-ray bright objects without counterparts are at the other extreme, i.e., their X-ray emission is very hard. In between these two groups are the other, most probably low-mass pre-main sequence objects. 

The detection algorithm finds 21 sources with at least 500 EPIC counts. We analyzed these objects in more detail but, because the field is crowded, extracting their spectra and light curves is difficult. We therefore took the background in nearby source-free regions on the same CCD as the source under consideration and limited the extraction radius for sources to 37'' at maximum (9'' at minimum). Even with this restriction, XID 13 and 14 are so close to each other that their properties cannot be studied individually: we therefore extracted one spectrum and one lightcurve for both sources.  Moreover, because of their distorted PSF (due to their large off-axis angle) and the presence of close neighbours, the spectra and light curves of XID 3, 4 and 77 had to be extracted in elliptical regions. Light Curves are shown in Fig. \ref{iclc} and spectral fittings are provided in Table \ref{icfit}.

As for HM1, the light curves were tested using $\chi^2$ tests: XID 1, 9, 16 and 30 are found to be marginally variable (between 1 and 10\% significance level), while XID 7 (for 1\,ks bins), 10, 12, 15, 31 (for 1\,ks bins), and 35 (for 1\,ks bins) are definitely variable, with a significance level $<$1\%.

  \begin{sidewaystable*}[htb]
  \tiny
  \centering
  \caption{Parameters of the detected X-ray sources in IC\,2944/2948.}
  \label{icdet} 
\begin{tabular}{>{\footnotesize}l >{\footnotesize}c >{\footnotesize}c >{\footnotesize}c >{\footnotesize}c | >{\footnotesize}c >{\footnotesize}c >{\footnotesize}c >{\footnotesize}c >{\footnotesize}c }
  \hline
XID & RA & DEC & RA(cor) & DEC(cor) &CR(MOS1) & CR(MOS2) & CR(pn) & $HR_1$(pn) & $HR_2$(pn)  \\
  \hline
1    & 11:38:20.023  & -63:22:22.04   & 11:38:20.473 & -63:22:22.45  & $107.288 \pm  1.936$ & $105.828 \pm  1.914$ & $398.641 \pm  4.006$ & $-0.355 \pm  0.010$ & $-0.742 \pm  0.012$  \\
2    & 11:37:47.998  & -63:19:23.55   & 11:37:48.448 & -63:19:23.96  & $ 65.630 \pm  1.697$ & $ 57.319 \pm  1.527$ & $228.961 \pm  3.346$ & $-0.503 \pm  0.013$ & $-0.803 \pm  0.019$  \\
3    & 11:38:09.387  & -63:11:47.36   & 11:38:09.837 & -63:11:47.77  & $ 35.349 \pm  2.487$ & $ 32.971 \pm  1.530$ & $123.280 \pm  3.266$ & $-0.537 \pm  0.022$ & $-0.890 \pm  0.040$  \\
4    & 11:39:49.856  & -63:28:43.21   & 11:39:50.306 & -63:28:43.62  & $ 52.690 \pm  1.966$ & $ 61.910 \pm  2.238$ & $                  $ & $                 $ & $                 $  \\
5    & 11:39:03.011  & -63:25:46.18   & 11:39:03.461 & -63:25:46.59  & $ 13.636 \pm  0.767$ & $ 13.394 \pm  0.777$ & $ 53.026 \pm  1.616$ & $-0.581 \pm  0.025$ & $-0.932 \pm  0.047$  \\
6    & 11:38:51.967  & -63:19:53.04   & 11:38:52.417 & -63:19:53.45  & $ 10.122 \pm  0.662$ & $ 11.064 \pm  0.682$ & $ 31.628 \pm  1.259$ & $ 0.984 \pm  0.032$ & $ 0.614 \pm  0.031$  \\
7    & 11:38:11.822  & -63:23:26.51   & 11:38:12.272 & -63:23:26.92  & $  6.764 \pm  0.539$ & $  5.843 \pm  0.478$ & $ 26.483 \pm  1.103$ & $-0.672 \pm  0.032$ & $-0.948 \pm  0.076$  \\
\hline
  \end{tabular}
  \end{sidewaystable*}

  \addtocounter{table}{-1}

  \begin{sidewaystable*}[htb]
  \tiny
  \centering
  \caption{Continued.}
  \begin{tabular}{>{\footnotesize}l >{\footnotesize}c >{\footnotesize}c >{\footnotesize}c >{\footnotesize}c | >{\footnotesize}c >{\footnotesize}c >{\footnotesize}c >{\footnotesize}c >{\footnotesize}c }
  \hline
XID & RA & DEC & RA(cor) & DEC(cor) &CR(MOS1) & CR(MOS2) & CR(pn) & $HR_1$(pn) & $HR_2$(pn)  \\
  \hline
56   & 11:39:44.751  & -63:27:28.25   & 11:39:45.201 & -63:27:28.66  & $  4.917 \pm  0.657$ & $  5.006 \pm  0.697$ & $ 14.656 \pm  2.183$ & $ 0.363 \pm  0.155$ & $-0.343 \pm  0.179$  \\
57   & 11:39:17.384  & -63:24:43.43   & 11:39:17.834 & -63:24:43.84  & $  1.380 \pm  0.384$ & $  2.070 \pm  0.381$ & $  7.085 \pm  0.748$ & $ 0.388 \pm  0.122$ & $-0.087 \pm  0.118$  \\
58   & 11:39:25.240  & -63:28:29.56   & 11:39:25.690 & -63:28:29.97  & $  3.473 \pm  0.574$ & $  4.042 \pm  0.646$ & $  9.022 \pm  1.066$ & $ 0.455 \pm  0.142$ & $ 0.047 \pm  0.126$  \\
59   & 11:39:23.700  & -63:28:27.27   & 11:39:24.150 & -63:28:27.68  & $  1.577 \pm  0.480$ & $  1.932 \pm  0.543$ & $  5.782 \pm  0.987$ & $ 0.384 \pm  0.177$ & $-0.085 \pm  0.199$  \\
60?  & 11:39:21.053  & -63:28:30.09   & 11:39:21.503 & -63:28:30.50  & $  2.114 \pm  0.579$ & $  1.575 \pm  0.541$ & $  5.000 \pm  0.928$ & $ 0.100 \pm  0.250$ & $ 0.283 \pm  0.200$  \\
61   & 11:38:57.915  & -63:24:10.33   & 11:38:58.365 & -63:24:10.74  & $  0.633 \pm  0.209$ & $  0.597 \pm  0.226$ & $  3.230 \pm  0.511$ & $ 0.212 \pm  0.147$ & $-1.000 \pm  0.195$  \\
62   & 11:38:50.831  & -63:26:54.73   & 11:38:51.281 & -63:26:55.14  & $  1.969 \pm  0.359$ & $  2.315 \pm  0.391$ & $  6.944 \pm  0.736$ & $ 0.137 \pm  0.112$ & $-0.344 \pm  0.136$  \\
\hline
  \end{tabular}
  \end{sidewaystable*}

  \addtocounter{table}{-1}

  \begin{sidewaystable*}[htb]
  \tiny
  \centering
  \caption{Continued.}
  \begin{tabular}{>{\footnotesize}l >{\footnotesize}c >{\footnotesize}c >{\footnotesize}c >{\footnotesize}c | >{\footnotesize}c >{\footnotesize}c >{\footnotesize}c >{\footnotesize}c >{\footnotesize}c }
  \hline
XID & RA & DEC & RA(cor) & DEC(cor) &CR(MOS1) & CR(MOS2) & CR(pn) & $HR_1$(pn) & $HR_2$(pn)  \\
  \hline
111  & 11:38:06.927  & -63:17:29.33  :& 11:38:07.377 & -63:17:29.74  & $  1.051 \pm  0.282$ & $  0.574 \pm  0.208$ & $  2.755 \pm  0.596$ & $-0.180 \pm  0.252$ & $ 0.187 \pm  0.273$  \\
112  & 11:38:19.110  & -63:12:13.79   & 11:38:19.560 & -63:12:14.20  & $                  $ & $  2.622 \pm  0.544$ & $  7.241 \pm  1.109$ & $ 0.338 \pm  0.369$ & $ 0.610 \pm  0.125$  \\
113  & 11:37:52.239  & -63:29:31.00   & 11:37:52.689 & -63:29:31.41  & $  0.710 \pm  0.248$ & $  1.887 \pm  0.456$ & $  3.636 \pm  0.678$ & $ 0.204 \pm  0.173$ & $-0.628 \pm  0.277$  \\
114  & 11:38:17.961  & -63:20:58.08   & 11:38:18.411:& -63:20:58.49  & $  1.115 \pm  0.306$ & $  1.426 \pm  0.317$ & $  3.261 \pm  0.546$ & $ 0.721 \pm  0.215$ & $-0.085 \pm  0.165$  \\
115  & 11:37:34.888  & -63:21:47.72   & 11:37:35.338 & -63:21:48.13  & $  1.264 \pm  0.273$ & $  0.296 \pm  0.188$ & $  1.957 \pm  0.426$ & $-0.669 \pm  0.191$ & $-0.821 \pm  0.753$  \\
116  & 11:37:25.504  & -63:13:38.18   & 11:37:25.954 & -63:13:38.59  & $  1.466 \pm  0.497$ & $  2.274 \pm  0.513$ & $  5.485 \pm  0.971$ & $ 0.401 \pm  0.212$ & $ 0.189 \pm  0.179$  \\
117  & 11:38:15.840  & -63:20:57.49   & 11:38:16.290 & -63:20:57.90  & $  1.582 \pm  0.309$ & $  0.916 \pm  0.262$ & $  1.910 \pm  0.435$ & $ 1.000 \pm  0.323$ & $ 0.591 \pm  0.206$  \\
\hline
  \end{tabular}
  \end{sidewaystable*}

  \addtocounter{table}{-1}

  \begin{sidewaystable*}[htb]
  \tiny
  \centering
  \caption{Continued.}
  \begin{tabular}{>{\footnotesize}l >{\footnotesize}c >{\footnotesize}c >{\footnotesize}c >{\footnotesize}c | >{\footnotesize}c >{\footnotesize}c >{\footnotesize}c >{\footnotesize}c >{\footnotesize}c }
  \hline
XID & RA & DEC & RA(cor) & DEC(cor) &CR(MOS1) & CR(MOS2) & CR(pn) & $HR_1$(pn) & $HR_2$(pn)  \\
  \hline
166  & 11:37:36.678  & -63:20:07.31   & 11:37:37.128 & -63:20:07.72  & $  1.190 \pm  0.819$ & $  0.567 \pm  0.277$ & $  2.453 \pm  0.498$ & $ 0.364 \pm  0.197$ & $-0.826 \pm  0.231$  \\
167  & 11:36:40.208  & -63:14:16.18   & 11:36:40.658 & -63:14:16.59  & $  3.707 \pm  1.204$ & $  3.462 \pm  0.835$ & $  9.210 \pm  1.557$ & $ 0.232 \pm  0.198$ & $ 0.099 \pm  0.185$  \\
168? & 11:39:12.799  & -63:28:39.13   & 11:39:13.249 & -63:28:39.54  & $  1.538 \pm  0.457$ & $  2.202 \pm  0.546$ & $  5.996 \pm  0.921$ & $ 0.651 \pm  1.940$ & $ 0.966 \pm  0.056$  \\
169  & 11:37:43.688  & -63:25:46.97   & 11:37:44.138 & -63:25:47.38  & $  0.923 \pm  0.278$ & $  0.543 \pm  0.221$ & $                  $ & $                 $ & $                 $  \\
170  & 11:38:04.260  & -63:33:45.89   & 11:38:04.710 & -63:33:46.30  & $  0.514 \pm  0.385$ & $  0.428 \pm  0.354$ & $  3.438 \pm  0.684$ & $-0.492 \pm  0.174$ & $-1.000 \pm  0.445$  \\
171  & 11:39:37.741  & -63:29:13.73   & 11:39:38.191 & -63:29:14.14  & $  1.121 \pm  0.412$ & $  0.969 \pm  0.395$ & $  2.897 \pm  0.779$ & $ 0.144 \pm  0.251$ & $-0.445 \pm  0.409$  \\
172  & 11:38:08.646  & -63:12:15.40   & 11:38:09.096 & -63:12:15.81  & $                  $ & $  0.910 \pm  0.373$ & $  1.709 \pm  0.635$ & $-0.777 \pm  0.270$ & $-1.000 \pm  1.815$  \\
\hline
  \end{tabular}
  \end{sidewaystable*}

  \addtocounter{table}{-1}

  \begin{sidewaystable*}[htb]
  \tiny
  \centering
  \caption{Continued.}
  \begin{tabular}{>{\footnotesize}l >{\footnotesize}c >{\footnotesize}c >{\footnotesize}c >{\footnotesize}c | >{\footnotesize}c >{\footnotesize}c >{\footnotesize}c >{\footnotesize}c >{\footnotesize}c }
  \hline
XID & RA & DEC & RA(cor) & DEC(cor) &CR(MOS1) & CR(MOS2) & CR(pn) & $HR_1$(pn) & $HR_2$(pn)  \\
  \hline
221  & 11:39:02.103  & -63:31:41.19   & 11:39:02.553 & -63:31:41.60  & $  1.151 \pm  0.375$ & $  0.714 \pm  0.296$ & $  3.095 \pm  0.764$ & $ 0.159 \pm  0.301$ & $ 0.248 \pm  0.260$  \\
222  & 11:39:29.209  & -63:30:13.84   & 11:39:29.659 & -63:30:14.25  & $  1.297 \pm  0.429$ & $  1.354 \pm  0.449$ & $  2.366 \pm  0.640$ & $-0.027 \pm  0.246$ & $-0.890 \pm  0.477$  \\
223  & 11:38:33.504  & -63:23:09.92   & 11:38:33.954 & -63:23:10.33  & $  0.569 \pm  0.205$ & $  0.384 \pm  0.186$ & $  1.709 \pm  0.422$ & $-0.120 \pm  0.257$ & $-0.223 \pm  0.354$  \\
224  & 11:38:09.867  & -63:27:04.90   & 11:38:10.317 & -63:27:05.31  & $  0.793 \pm  0.255$ & $  0.828 \pm  0.253$ & $  2.049 \pm  0.445$ & $-0.104 \pm  0.215$ & $-0.428 \pm  0.339$  \\
225  & 11:37:46.716  & -63:16:42.04   & 11:37:47.166 & -63:16:42.45  & $  0.290 \pm  0.242$ & $  0.574 \pm  0.239$ & $  2.415 \pm  0.542$ & $ 1.000 \pm  0.315$ & $ 0.599 \pm  0.191$  \\
226  & 11:38:00.471  & -63:18:27.03   & 11:38:00.921 & -63:18:27.44  & $  0.455 \pm  0.232$ & $  0.500 \pm  0.212$ & $  1.424 \pm  0.433$ & $ 0.592 \pm  0.344$ & $-0.441 \pm  0.321$  \\
227  & 11:37:43.057  & -63:28:37.36   & 11:37:43.507 & -63:28:37.77  & $  0.177 \pm  0.185$ & $  0.506 \pm  0.235$ & $  1.826 \pm  0.507$ & $ 0.101 \pm  0.235$ & $-1.000 \pm  0.546$  \\
\hline
  \end{tabular}
  \end{sidewaystable*}

  \addtocounter{table}{-1}

  \begin{sidewaystable*}[htb]
  \tiny
  \centering
  \caption{Continued.}
  \begin{tabular}{>{\footnotesize}l >{\footnotesize}c >{\footnotesize}c >{\footnotesize}c >{\footnotesize}c | >{\footnotesize}c >{\footnotesize}c >{\footnotesize}c >{\footnotesize}c >{\footnotesize}c }
  \hline
XID & RA & DEC & RA(cor) & DEC(cor) &CR(MOS1) & CR(MOS2) & CR(pn) & $HR_1$(pn) & $HR_2$(pn)  \\
  \hline
276  & 11:39:51.943  & -63:21:39.93   & 11:39:52.393 & -63:21:40.34  & $  0.212 \pm  0.212$ & $  1.012 \pm  0.329$ & $  2.100 \pm  0.622$ & $ 0.106 \pm  0.268$ & $-0.413 \pm  0.468$  \\
277  & 11:39:27.304  & -63:29:35.52   & 11:39:27.754 & -63:29:35.93  & $  0.705 \pm  0.354$ & $  1.172 \pm  0.394$ & $  1.841 \pm  0.685$ & $ 0.223 \pm  0.444$ & $ 0.246 \pm  0.376$  \\
278? & 11:39:10.883  & -63:21:08.06   & 11:39:11.333 & -63:21:08.47  & $  0.446 \pm  0.189$ & $  0.434 \pm  0.222$ & $  1.573 \pm  0.429$ & $-0.106 \pm  0.272$ & $-0.383 \pm  0.422$  \\
279  & 11:38:18.898  & -63:09:37.49   & 11:38:19.348 & -63:09:37.90  & $                  $ & $  0.664 \pm  0.349$ & $  2.798 \pm  0.894$ & $-0.718 \pm  0.261$ & $ 0.315 \pm  0.655$  \\
280  & 11:36:45.189  & -63:11:23.84   & 11:36:45.639 & -63:11:24.25  & $                  $ & $                  $ & $  3.027 \pm  0.866$ & $ 0.080 \pm  0.273$ & $-1.000 \pm  0.335$  \\
281? & 11:39:31.166  & -63:17:01.09   & 11:39:31.616 & -63:17:01.50  & $                  $ & $  0.840 \pm  0.322$ & $  2.590 \pm  0.645$ & $ 0.698 \pm  0.250$ & $-0.177 \pm  0.270$  \\
282  & 11:38:02.751  & -63:09:46.23   & 11:38:03.201 & -63:09:46.64  & $  0.613 \pm  0.476$ & $  0.758 \pm  0.366$ & $  3.689 \pm  0.999$ & $-0.128 \pm  0.284$ & $ 0.103 \pm  0.355$  \\
\hline
  \end{tabular}
  \end{sidewaystable*}

  \addtocounter{table}{-1}

  \begin{sidewaystable*}[htb]
  \tiny
  \centering
  \caption{Continued.}
  \begin{tabular}{>{\footnotesize}l >{\footnotesize}c >{\footnotesize}c >{\footnotesize}c >{\footnotesize}c | >{\footnotesize}c >{\footnotesize}c >{\footnotesize}c >{\footnotesize}c >{\footnotesize}c }
  \hline
XID & RA & DEC & RA(cor) & DEC(cor) &CR(MOS1) & CR(MOS2) & CR(pn) & $HR_1$(pn) & $HR_2$(pn)  \\
  \hline
331  & 11:36:49.297  & -63:28:27.88   & 11:36:49.747 & -63:28:28.29  & $  0.885 \pm  0.481$ & $  1.224 \pm  0.463$ & $                  $ & $                 $ & $                 $  \\
332  & 11:38:53.872  & -63:11:38.84   & 11:38:54.322 & -63:11:39.25  & $                  $ & $  0.400 \pm  0.307$ & $  4.019 \pm  0.876$ & $ 0.454 \pm  0.196$ & $-0.446 \pm  0.289$  \\
333? & 11:38:46.526  & -63:15:37.77   & 11:38:46.976 & -63:15:38.18  & $                  $ & $  0.000 \pm  0.143$ & $  1.641 \pm  0.483$ & $-1.000 \pm  0.272$ & $ 1.000 \pm  0.486$  \\
334  & 11:37:42.749  & -63:12:28.11   & 11:37:43.199 & -63:12:28.52  & $  1.042 \pm  0.381$ & $  0.296 \pm  0.253$ & $  2.699 \pm  0.793$ & $-0.064 \pm  0.359$ & $ 0.332 \pm  0.319$  \\
335  & 11:38:58.264  & -63:24:45.74   & 11:38:58.714 & -63:24:46.15  & $  0.370 \pm  0.193$ & $  0.672 \pm  0.232$ & $  1.369 \pm  0.434$ & $-0.266 \pm  0.335$ & $-0.036 \pm  0.469$  \\
336  & 11:40:05.958  & -63:21:38.28   & 11:40:06.408 & -63:21:38.69  & $  0.947 \pm  0.431$ & $  1.393 \pm  0.544$ & $  2.087 \pm  0.690$ & $                 $ & $ 1.000 \pm  0.138$  \\
337  & 11:40:11.053  & -63:25:54.15   & 11:40:11.503 & -63:25:54.56  & $  0.925 \pm  0.532$ & $  1.767 \pm  0.539$ & $                  $ & $                 $ & $                 $  \\
\hline
  \end{tabular}
\tablefoot{A `?' indicates probably spurious sources, whereas `ext' refers to an extended source. The count rates are given in cts\,ks$^{-1}$ for the total band (0.3--10.\,keV energy band) ; uncertain values are indicated by a ``:'' (source partially in a gap) ; missing values could not be calculated (e.g., the source is in a gap, over a bad column, or out of the field-of-view for that instrument). The hardness ratios $HR_1$ and $HR_2$ are computed as $(M-S)/(M+S)$ and $(H-M)/(H+M)$, respectively, where $S$, $M$, and $H$ are the count rates recorded in the soft (0.3--1.0\,keV) energy band, medium (1.0--2.0\,keV) energy band, and hard (2.0--10.0\,keV) energy band, respectively.}
  \end{sidewaystable*}

\begin{table*}
  \centering
  \caption{Counterparts of the X-ray sources of IC\,2944/2948 in the 2MASS, \citet{mac05}, and Simbad catalogs.}
  \label{iccorrel}
  \begin{tabular}{l | cc | ccc | ccc}
  \hline
XID & d('') & 2MASS & d('') & MG & $W_{\rm H\alpha}$ (\AA) & d('') & Simbad  & Info \\
  \hline
 1  & 0.782 & 11382038-6322219 &       &    &       & 0.94 & HD\,101205 & O7IIIn((f))+?, $P\sim2$\,d \\
 2  & 0.427 & 11374844-6319235 &       &    &       & 0.45 & HD\,101131 & O6.5V((f))+O8.5V, $P\sim10$\,d \\
 3  & 1.043 & 11380992-6311486 &       &    &       & 0.99 & HD\,101190 & O4V((f))+O7V, $P\sim6$\,d \\
 4  & 2.269 & 11394996-6328435 &       &    &       & 2.33 & HD\,101436 & O6.5V+O7V, $P\sim37$\,d \\
 5  & 1.307 & 11390329-6325472 &       &    &       & 1.32 & HD\,101298 & O6III((f)) \\
 7  & 0.713 & 11381216-6323268 & 0.78  &  72&   2.9 & 0.77 & HD\,101191 & O8V+?, $P>100$\,d \\
 8  & 1.810 & 11382277-6312028 &       &    &       & 1.78 & HD\,101223 & O8V \\
 9  & 1.224 & 11382269-6322530 &       &    &       & & \\
10  & 1.745 & 11391084-6326288 &       &    &       & & & \\
12  & 1.517 & 11385556-6324056 &       &    &       & & & \\
13  & 0.953 & 11385797-6328390 &       &    &       & 0.99 & HD\,308829 & B5 \\
14  & 1.736 & 11390012-6328402 &       &    &       & & & \\
15  & 0.368 & 11383475-6320264 &       &    &       & & & \\
16  & 1.129 & 11372812-6317175 &       &    &       & & & \\
17+ & 0.928 & 11393894-6321483 &       &    &       & & & \\
18  & 0.955 & 11382006-6327213 &       &    &       & & & \\
19  & 0.338 & 11380947-6318157 &       &    &       & & & \\
20  & 1.518 & 11365618-6323525 &       &    &       & 1.55 & HD\,101008 & B1II/III \\
21  & 0.706 & 11380641-6319035 &       &    &       & & & \\
24  & 1.417 & 11375723-6308573 &       &    &       & & & \\
26  & 0.624 & 11372525-6320033 &       &    &       & & & \\
28  & 0.752 & 11375384-6324459 &       &    &       & & & \\
29  & 0.796 & 11384910-6325545 &       &    &       & & & \\
31  & 0.520 & 11381731-6320180 &       &    &       & & & \\
32  & 1.719 & 11371208-6333010 &       &    &       & & & \\
33  & 0.794 & 11380411-6324419 &       &    &       & & & \\
34  & 1.862 & 11384974-6333355 &       &    &       & & & \\
35  & 2.758 & 11382068-6320344 & 2.71  &  35&   2.9 & & & \\
36  & 1.529 & 11382101-6319463 &       &    &       & & & \\
37  & 0.544 & 11382986-6322057 &       &    &       & & & \\
38  & 1.132 & 11380185-6317354 &       &    &       & & & \\
39  & 1.718 & 11371812-6313222 &       &    &       & 1.71 & GSC\,08976-03815 & \\
40  & 1.006 & 11391102-6324103 &       &    &       & & & \\
41  & 1.188 & 11383542-6311321 &       &    &       & 1.12 & HD\,308806 & F5V\\
43  & 0.566 & 11375679-6320522 &       &    &       & & & \\
44  & 0.441 & 11381244-6322164 &       &    &       & & & \\
45  & 0.655 & 11381635-6319530 &       &    &       & & & \\
46  & 1.152 & 11381791-6321318 &       &    &       & 1.15 & 2MASS\,J11381791-6321318 & \\
47  & 2.106 & 11374896-6320258 &       &    &       & & & \\
48  & 1.459 & 11391348-6323244 &       &    &       & & & \\
49  & 2.340 & 11400472-6324584 &       &    &       & & & \\
50  & 1.235 & 11393568-6327510 &       &    &       & & & \\
51  & 1.314 & 11382551-6318565 &       &    &       & & & \\
52  & 2.074 & 11382426-6323026 &       &    &       & & & \\
56  & 2.271 & 11394498-6327268 &       &    &       & & & \\
57  & 0.201 & 11391786-6324438 &       &    &       & & & \\
59+ & 1.238 & 11392398-6328281 &       &    &       & & & \\
61  & 1.945 & 11385807-6324105 &       &    &       & & & \\
62  & 0.730 & 11385123-6326557 &       &    &       & & & \\
63  & 2.739 & 11375518-6321453 & 2.82  &  47&   4.2 & 2.68 & HD\,308818 & B0V \\
64  & 1.237 & 11384924-6327308 &       &    &       & & & \\
65  & 0.285 & 11381877-6311312 &       &    &       & & & \\
66  & 1.519 & 11374212-6321565 &       &    &       & & & \\
67  & 1.655 & 11392351-6327445 &       &    &       & & & \\
68  & 0.932 & 11371030-6321429 &       &    &       & & & \\
69  & 1.121 & 11373723-6321049 &       &    &       & & & \\
70  & 1.738 & 11394905-6324094 &       &    &       & & & \\
71  & 2.007 & 11392016-6328346 &       &    &       & & & \\
72  & 0.457 & 11370951-6320055 &       &    &       & & & \\
74  & 1.223 & 11370265-6313279 &       &    &       & & & \\
75  & 2.154 & 11390231-6321443 &       &    &       & & & \\
77  & 1.053 & 11394585-6328401 &       &    &       & 1.21 & HD\,101413 & O8V+B3:V, $P\sim150$\,d \\
79  & 1.799 & 11383473-6322012 & 1.86  &  54&   2.6 & 1.75 & 2MASS\,J11383473-6322012 & \\
80  & 2.039 & 11392911-6324384 &       &    &       & & & \\
81  & 1.445 & 11391277-6331135 &       &    &       & 1.41 & HD\,101333 & O9.5V \\
82  & 1.736 & 11381556-6323425 &       &    &       & & & \\
83  & 1.377 & 11382126-6324033 &       &    &       & & & \\
84  & 1.331 & 11374099-6318422 &       &    &       & & & \\
85  & 0.960 & 11381762-6317211 & 0.85  &   8& {\bf $-$14.5} & & & \\
86  & 2.011 & 11374173-6320067 &       &    &       & & & \\
87  & 0.374 & 11385818-6318340 &       &    &       & & & \\
89  & 0.873 & 11383982-6323106 &       &    &       & & & \\
90  & 1.686 & 11390604-6312121 &       &    &       & & & \\
  \hline
  \end{tabular}
  \end{table*}

  \addtocounter{table}{-1}

\begin{table*}
  \centering
  \caption{Continued.}
  \begin{tabular}{l | cc | ccc | ccc}
  \hline
XID & d('') & 2MASS & d('') & MG & $W_{\rm H\alpha}$ (\AA) & d('') & Simbad  & Info \\
  \hline
91  & 1.392 & 11365924-6329067 &       &    &       & & & \\
92  & 0.843 & 11375818-6321251 &       &    &       & & & \\
93  & 1.854 & 11391037-6322514 &       &    &       & & & \\
95  & 1.069 & 11381191-6320226 &       &    &       & & & \\
96  & 1.252 & 11382230-6332021 &       &    &       & & & \\
97  & 1.622 & 11370142-6320439 &       &    &       & & & \\
98  & 2.828 & 11371107-6319198 &       &    &       & & & \\
99  & 2.421 & 11375989-6319594 &       &    &       & & & \\
100 & 0.922 & 11381200-6319148 & 1.34  &  22& {\bf $-$11.1} & & & \\
101 & 0.896 & 11381153-6325376 &       &    &       & & & \\
102 & 0.815 & 11374375-6322348 &       &    &       & & & \\
103 & 2.606 & 11380361-6325001 &       &    &       & & & \\
105 & 0.827 & 11390727-6319005 &       &    &       & & & \\
106 & 0.327 & 11375844-6318594 & 0.30  &  19&   3.5 & 0.38 & HD308813 & O9.5V+? \\
108 & 1.208 & 11371335-6321260 &       &    &       & & & \\
109 & 2.003 & 11370561-6327151 &       &    &       & & & \\
110 & 0.671 & 11385610-6322474 &       &    &       & & & \\
111 & 2.953 & 11380714-6317322 &       &    &       & & & \\
113 & 2.162 & 11375241-6329302 &       &    &       & & & \\
115 & 1.431 & 11373512-6321479 & 1.46  &  48& {\bf $-$29.4} & & & \\
117 & 0.521 & 11381622-6320581 &       &    &       & & & \\
118 & 2.101 & 11373384-6332083 &       &    &       & & & \\
119 & 1.647 & 11382187-6326057 &       &    &       & & & \\
120 & 1.825 & 11385865-6329469 &       &    &       & & & \\
121 & 1.293 & 11373871-6330240 &       &    &       & 1.48 & TYC\,8976-2364-1 & \\
122 & 1.909 & 11391878-6319304 &       &    &       & & & \\
123 & 2.348 & 11362645-6325049 &       &    &       & & & \\
125 & 1.163 & 11382846-6325135 &       &    &       & & & \\
126 & 1.956 & 11372189-6330245 &       &    &       & & & \\
127 & 2.233 & 11385903-6328109 &       &    &       & & & \\
130 & 0.996 & 11391193-6314295 &       &    &       & & & \\
131 & 2.261 & 11391965-6325162 &       &    &       & & & \\
132 & 1.228 & 11375151-6330594 &       &    &       & & & \\
133 & 0.515 & 11383591-6329206 &       &    &       & & & \\
135+& 2.193 & 11395595-6325528 &       &    &       & & & \\
136 & 1.737 & 11383158-6328256 &       &    &       & & & \\
137 & 2.075 & 11375390-6318596 & 2.61  &  18&$-$1.9 & 2.11 & 2MASS\,J11375390-6318596 & \\
139+& 1.937 & 11383979-6314145 &       &    &       & & & \\
140 & 1.590 & 11380451-6322477 &       &    &       & & & \\
141 & 1.981 & 11383189-6327271 &       &    &       & & & \\
142 & 2.490 & 11400147-6323432 &       &    &       & & & \\
143 & 0.113 & 11375033-6323583 &       &    &       & & & \\
144 & 1.975 & 11380921-6310517 &       &    &       & & & \\
148 & 1.042 & 11375723-6316433 &       &    &       & & & \\
149 & 1.253 & 11391206-6317571 &       &    &       & & & \\
150 & 0.998 & 11374677-6315505 &       &    &       & & & \\
153 & 1.333 & 11372270-6308585 &       &    &       & 1.37 & HD\,101070 & B0.5IV \\
154 & 2.871 & 11394528-6324512 &       &    &       & & & \\
156 & 2.416 & 11380063-6317212 &       &    &       & & & \\
157 & 2.735 & 11381152-6321443 &       &    &       & & & \\
160 & 1.223 & 11382910-6314156 &       &    &       & & & \\
161 & 1.817 & 11384249-6322082 &       &    &       & & & \\
162 & 0.390 & 11385311-6319241 &       &    &       & & & \\
163 & 0.477 & 11385607-6328589 &       &    &       & & & \\
164 & 1.206 & 11382734-6320326 &       &    &       & & & \\
165 & 1.913 & 11384248-6324578 &       &    &       & & & \\
166 & 2.783 & 11373732-6320052 &       &    &       & & & \\
167 & 0.704 & 11364070-6314159 &       &    &       & & & \\
169 & 1.019 & 11374406-6325464 &       &    &       & & & \\
170 & 1.690 & 11380456-6333449 &       &    &       & & & \\
171 & 2.816 & 11393786-6329123 &       &    &       & & & \\
174 & 2.547 & 11374014-6319062 & 2.56  &  20& {\bf $-$10.4} & 2.56 & 2MASS\,J11374014-6319062 & \\
175 & 1.109 & 11392755-6326590 &       &    &       & 1.01 & HD\,308831 & B1.5V \\
176 & 2.571 & 11390303-6322565 &       &    &       & & & \\
177 & 0.340 & 11380521-6326551 &       &    &       & & & \\
178 & 1.876 & 11382214-6327380 & 1.92  & 107& {\bf $-$28.6} & 1.67 & HD\,308828 & F2 \\
180 & 1.885 & 11370392-6319066 &       &    &       & 1.81 & GSC\,08976-01143 & \\
181 & 1.700 & 11394164-6327003 &       &    &       & & & \\
182 & 1.953 & 11381514-6323091 &       &    &       & & & \\
  \hline
  \end{tabular}
  \end{table*}

  \addtocounter{table}{-1}

\begin{table*}
  \centering
  \caption{Continued.}
  \begin{tabular}{l | cc | ccc | ccc}
  \hline
XID & d('') & 2MASS & d('') & MG & $W_{\rm H\alpha}$ (\AA) & d('') & Simbad  & Info \\
  \hline
184+& 1.088 & 11363007-6317309 &       &    &       & & & \\
187 & 1.912 & 11370695-6315275 &       &    &       & & & \\
189 & 1.443 & 11375071-6312287 &       &    &       & & & \\
190 & 1.507 & 11371996-6322488 &       &    &       & & & \\
191 & 0.407 & 11372795-6323157 &       &    &       & & & \\
192 & 0.840 & 11371355-6317061 &       &    &       & & & \\
193 & 1.222 & 11382067-6318051 &       &    &       & & & \\
195 & 0.935 & 11371374-6319283 &       &    &       & & & \\
197 & 2.300 & 11394205-6316323 &       &    &       & & & \\
198 & 0.549 & 11383488-6314432 &       &    &       & & & \\
199 & 1.198 & 11401217-6322204 &       &    &       & & & \\
202 & 0.621 & 11381753-6331352 &       &    &       & & & \\
205 & 1.123 & 11393042-6321110 &       &    &       & & & \\
207 & 1.353 & 11383886-6324582 &       &    &       & & & \\
209 & 2.531 & 11395445-6330245 &       &    &       & & & \\
211 & 1.519 & 11361403-6319273 &       &    &       & & & \\
212 & 2.104 & 11384886-6328169 &       &    &       & & & \\
215+& 2.131 & 11383289-6329310 &       &    &       & & & \\
217 & 1.768 & 11391404-6328077 &       &    &       & & & \\
219 & 0.748 & 11395181-6322464 &       &    &       & & & \\
220 & 0.748 & 11374910-6321185 &       &    &       & & & \\
222 & 1.336 & 11392970-6330129 &       &    &       & & & \\
223 & 1.948 & 11383383-6323120 &       &    &       & & & \\
224 & 2.005 & 11381003-6327046 &       &    &       & & & \\
226 & 0.686 & 11380102-6318273 &       &    &       & & & \\
227 & 2.786 & 11374317-6328394 &       &    &       & & & \\
228 & 1.557 & 11393033-6322596 &       &    &       & & & \\
229 & 0.438 & 11375636-6309400 &       &    &       & & & \\
230 & 2.993 & 11390595-6321180 &       &    &       & & & \\
231 & 1.777 & 11390277-6323580 &       &    &       & & & \\
232 & 1.614 & 11383029-6327489 &       &    &       & & & \\
234 & 2.109 & 11391325-6313196 &       &    &       & & & \\
235 & 0.842 & 11372370-6330448 &       &    &       & & & \\
238 & 1.556 & 11382206-6319148 &       &    &       & & & \\
240 & 0.434 & 11384703-6316562 &       &    &       & & & \\
241 & 1.652 & 11370236-6321099 &       &    &       & & & \\
242 & 2.567 & 11393525-6323153 &       &    &       & & & \\
243 & 1.999 & 11391757-6328198 &       &    &       & & & \\
245 & 0.897 & 11394470-6324118 &       &    &       & & & \\
246 & 1.431 & 11401369-6325004 &       &    &       & & & \\
247 & 2.105 & 11373389-6325297 &       &    &       & & & \\
249 & 2.451 & 11391489-6314592 &       &    &       & & & \\
250 & 1.475 & 11365017-6316233 &       &    &       & & & \\
252 & 2.901 & 11392288-6319199 &       &    &       & & & \\
253 & 2.045 & 11384737-6327565 &       &    &       & & & \\
254 & 1.003 & 11385658-6317112 &       &    &       & & & \\
255 & 2.293 & 11393388-6329347 &       &    &       & & & \\
256 & 0.861 & 11384423-6320549 &       &    &       & & & \\
258 & 0.855 & 11375328-6313024 &       &    &       & & & \\
259 & 2.160 & 11393397-6320390 &       &    &       & & & \\
260 & 2.116 & 11401877-6327086 &       &    &       & & & \\
262 & 0.964 & 11385360-6318406 &       &    &       & & & \\
263 & 2.422 & 11374496-6318149 &       &    &       & & & \\
264 & 0.950 & 11375810-6323559 &       &    &       & & & \\
266 & 0.756 & 11390545-6328118 &       &    &       & & & \\
268 & 0.494 & 11364324-6322498 &       &    &       & & & \\
269 & 0.896 & 11391338-6321331 &       &    &       & & & \\
271 & 1.697 & 11375010-6321385 &       &    &       & & & \\
272 & 2.292 & 11375972-6326484 &       &    &       & & & \\
277 & 1.390 & 11392781-6329346 &       &    &       & & & \\
279 & 2.798 & 11381969-6309394 &       &    &       & & & \\
283 & 0.964 & 11391178-6321538 &       &    &       & & & \\
284 & 1.622 & 11375337-6326077 &       &    &       & & & \\
  \hline
  \end{tabular}
  \end{table*}

  \addtocounter{table}{-1}

\begin{table*}
  \centering
  \caption{Continued.}
  \begin{tabular}{l | cc | ccc | ccc}
  \hline
XID & d('') & 2MASS & d('') & MG & $W_{\rm H\alpha}$ (\AA) & d('') & Simbad  & Info \\
  \hline
286 & 1.618 & 11383228-6324097 &       &    &       & & & \\
287 & 2.692 & 11382937-6335480 &       &    &       & 2.85 & HD\,308839 & F8 \\
290 & 0.710 & 11372112-6315557 &       &    &       & & & \\
291 & 1.628 & 11391005-6319087 &       &    &       & & & \\
294 & 2.496 & 11372666-6328266 &       &    &       & & & \\
297 & 1.569 & 11381736-6316337 &       &    &       & & & \\
299+& 0.404 & 11371723-6317482 &       &    &       & & & \\
304 & 2.418 & 11375358-6311511 &       &    &       & & & \\
305 & 1.067 & 11381597-6328317 &       &    &       & & & \\
310 & 2.238 & 11365566-6326533 &       &    &       & & & \\
311 & 1.313 & 11373652-6330384 &       &    &       & & & \\
316 & 0.695 & 11374561-6309509 &       &    &       & & & \\
319 & 2.053 & 11373173-6313248 &       &    &       & & & \\
325 & 1.206 & 11373263-6309119 &       &    &       & & & \\
326 & 2.689 & 11373114-6321022 &       &    &       & & & \\
327 & 0.798 & 11363768-6324324 &       &    &       & 0.84 & HD\,308713 & A3V \\
330 & 2.889 & 11395790-6321503 &       &    &       & & & \\
332 & 2.758 & 11385465-6311408 &       &    &       & & & \\
334 & 1.628 & 11374312-6312269 &       &    &       & & & \\
335 & 0.759 & 11385873-6324454 &       &    &       & & & \\
336 & 1.273 & 11400652-6321377 &       &    &       & & & \\
337 & 0.919 & 11401138-6325541 &       &    &       & & & \\
339 & 1.864 & 11384942-6320473 &       &    &       & & & \\
341 & 1.600 & 11373115-6331256 &       &    &       & & & \\
343 & 2.794 & 11382002-6320249 &       &    &       & & & \\
344 & 0.640 & 11380080-6325230 &       &    &       & & & \\
345 & 0.543 & 11375123-6320345 &       &    &       & & & \\
347 & 1.848 & 11380991-6327559 & 1.71  & 110& {\bf $-$22.0} & & & \\
349 & 0.899 & 11390185-6321292 &       &    &       & & & \\
350 & 2.387 & 11385483-6325303 &       &    &       & & & \\
356+& 0.506 & 11374127-6314340 &       &    &       & & & \\
359+& 1.965 & 11400265-6323555 &       &    &       & & & \\
363 & 2.482 & 11385824-6322416 &       &    &       & & & \\
367 & 1.743 & 11382628-6319097 &       &    &       & & & \\
\multicolumn{9}{l}{\it Potentially spurious sources}\\
128 & 1.061 & 11385931-6328546 &       &    &       & & & \\
134 & 0.933 & 11382455-6322009 &       &    &       & & & \\
146 & 2.292 & 11394677-6328344 &       &    &       & & & \\
183 & 2.806 & 11381269-6312059 &       &    &       & 2.83 & GSC\,08976-05121 & \\
186 & 1.124 & 11380507-6308112 &       &    &       & & & \\
188 & 1.264 & 11395348-6328004 &       &    &       & & & \\
214 & 0.916 & 11382154-6311121 &       &    &       & & & \\
261 & 2.516 & 11385234-6315141 &       &    &       & & & \\
267 & 2.158 & 11383524-6319485 &       &    &       & & & \\
270+& 2.152 & 11390069-6327372 &       &    &       & & & \\
273 & 2.153 & 11362430-6329483 &       &    &       & & & \\
278 & 2.166 & 11391112-6321101 &       &    &       & & & \\
298 & 1.473 & 11380303-6317451 &       &    &       & & & \\
303+& 1.194 & 11382946-6313488 &       &    &       & & & \\
324 & 2.842 & 11382700-6314415 &       &    &       & & & \\
328 & 1.444 & 11383079-6333324 &       &    &       & & & \\
333 & 1.606 & 11384718-6315389 &       &    &       & & & \\
353 & 2.286 & 11380955-6308221 &       &    &       & & & \\
354 & 1.855 & 11383024-6324174 &       &    &       & & & \\
357 & 2.193 & 11390803-6312569 &       &    &       & & & \\
358 & 2.634 & 11374623-6320040 & 2.56  &  32&   4.1 & 2.59 & CPD$-$62$^{\circ}$2153 & B1V \\
  \hline
  \end{tabular}
\tablefoot{A `+' indicates the presence of a second 2MASS counterpart within 3'', at a larger distance than the first one listed here. Equivalent widths $W$ are positive for H$\alpha$ absorption, boldface indicating strong H$\alpha$ emitters. Note that for XID 100, there is a second counterpart in the \citet{mac05} catalog: MG 23, at 2.38'', which has an equivalent width $W= -13.0$\,\AA. Spectral types come from  \citet{san11} for O-stars, from Simbad otherwise. }
\end{table*}

  \begin{sidewaystable*}[htb]
  \tiny
  \centering
  \caption{Spectral parameters of the best-fit thermal models to EPIC data of IC\,2944/2948 (see Table \ref{hm1fit} for definition of errors).}
  \label{icfit}
  \begin{tabular}{l c c c c c c c c c c c c }
  \hline
XID & $N_{\rm H}$ & $kT_1$ & $norm_1$ & $kT_2$ & $norm_2$ & $kT_3$ & $norm_3$ & $\chi^2$/dof (dof) & $F_{\rm X}^{obs}$ & $L_{\rm X}^{abscor}$ & $\log(L_{\rm BOL})$ & $\log(L_{\rm X}^{abscor}/L_{\rm BOL})$ \\
& $10^{22}$\,cm$^{-2}$ & keV & $10^{-3}$\,cm$^{-5}$ & keV & $10^{-3}$\,cm$^{-5}$ & keV & $10^{-3}$\,cm$^{-5}$ & & 10$^{-13}$\,erg\,cm$^{-2}$\,s$^{-1}$ & 10$^{32}$\,erg\,s$^{-1}$ & (L$_{\odot}$) & \\
  \hline
 1 & $0.19 + (0.32\pm0.03)$ & $0.153\pm0.004$ & $5.83\pm1.50$ & $0.601\pm0.015$ & $0.50\pm0.04$ & $1.23\pm0.04$ & $0.188\pm0.014$  & 1.57 (274) & 6.73 & 8.10$\pm$0.06 &  5.63 & $-6.305\pm0.003$ \\
 2 & $0.19 + (0.20\pm0.03)$ & $0.188\pm0.004$ & $1.39\pm0.32$ & $0.77\pm0.02$ & $0.188\pm0.010$ &  &   & 1.45 (191) & 3.64 & 4.93$\pm$0.06 &  5.43 & $-6.320\pm0.005$ \\
 3 & $0.19 + (0.06\pm0.05)$ & $0.154\pm0.012$ & $0.32\pm0.18$ & $0.73\pm0.02$ & $0.102\pm0.013$ &  &   & 1.50 (81)  & 1.97 & 2.72$\pm$0.06 &  5.53 & $-6.679\pm0.010$ \\
 4 & $0.19 + (0.10\pm0.08)$ & $0.148\pm0.017$ & $0.61\pm0.77$ & $0.77\pm0.03$ & $0.144\pm0.042$ & $2.38\pm0.50$ & $0.115\pm0.018$  & 1.38 (99)  & 3.73 & 4.36$\pm$0.11 & 5.31 & $-6.254\pm0.011$ \\
 5 & $0.19 + (0.39\pm0.06)$ & $0.045\pm0.011$ & $1902\pm32335$ & $0.24\pm0.02$ & $0.58\pm0.30$ &  &   & 0.96 (89) & 0.79 & 1.19$\pm$0.03 & 5.14 & $-6.647\pm0.011$\\
 6 & $0.19 + (1.42\pm0.14)$ & $21.7\pm11.7$ & $0.156\pm0.008$ &  &  &  &   & 0.88 (111) & 2.17 &  &  & \\
 6-& $2.15\pm0.27$ & $1.73\pm0.13$ & $0.059\pm0.014$ &  &  &  &   & 0.84 (111) & 2.07 &  &  & \\
 7 & $0.19 + (0.\pm0.34)$ & $0.24\pm0.06$ & $0.047\pm0.440$ & $0.99\pm0.30$ & $0.0086\pm0.0020$ &  &   & 0.94 (62) & 0.41 & 0.62$\pm$0.02 & 4.85 & $-6.641\pm0.015$ \\
 8 & $0.19 + (0.23\pm0.11)$ & $0.155\pm0.024$ & $0.30\pm0.50$ & $0.61\pm0.10$ & $0.033\pm0.011$ &  &   & 0.95 (22) & 0.44 & 0.62$\pm$0.03 & 4.77 & $-6.561\pm0.022$\\
 9 & $0.19 + (0.38\pm0.16)$ & $0.22\pm0.04$ & $0.12\pm0.32$ & $2.45\pm0.33$ & $0.043\pm0.005$ & &  & 0.86 (53) & 0.50 & &  & \\
 9-& $0.34\pm0.04$ & $2.75\pm0.18$ & $0.027\pm0.004$ &  &  &  &   & 1.08 (55) & 0.47 &  &  & \\
10 & $0.19 + (0.06\pm0.06)$ & $3.97\pm0.71$ & $0.046\pm0.004$ &  &  &  &   & 1.06 (38) & 0.63 & &  & \\
10-& $0.39\pm0.08$ & $2.18\pm0.19$ & $0.023\pm0.005$ &  &  &  &   & 1.07 (38) & 0.63 &  &  & \\
11 & $0.19 + (1.22\pm0.24)$ & $>14.2$ & $0.37\pm0.09$ &  &  &  &   & 1.31 (44) & 5.28 & &  & \\
11-& $1.52\pm0.39$ & $1.43\pm0.26$ & $0.088\pm0.040$ &  &  &  &   & 1.30 (44) & 5.27 &  &  & \\
12 & $0.19 + (0.\pm0.04)$ & $3.37\pm0.42$ & $0.0371\pm0.0019$ &  &  &  &   & 1.17 (44) & 0.49 & &  & \\
12-& $0.34\pm0.07$ & $2.44\pm0.19$ & $0.021\pm0.004$ &  &  &  &   & 1.12 (44) & 0.47 &  &  & \\
13+14 & $0.36 + (0.96\pm0.17)$ & $0.25\pm0.07$ & $1.97\pm1.27$ & $3.17\pm0.92$ & $0.0090\pm0.0014$ &  &   & 1.48 (84) & 1.12 & 1.02$\pm$0.05 & 3.40 & $-4.975\pm0.021$ \\
15 & $0.19 + (0.\pm0.05)$ & $0.34\pm0.30$ & $0.0021\pm0.0030$ & $3.97\pm0.59$ & $0.0261\pm0.0017$ &  &   & 1.39 (35) & 0.39 & &  & \\
15-& $0.23\pm0.06$ & $2.17\pm0.21$ & $0.012\pm0.002$ &  &  &  &   & 1.25 (37) & 0.38 &  &  & \\
16 & $0.19 + (0.\pm0.59)$ & $0.58\pm0.47$ & $0.003\pm0.502$ & $4.46\pm1.10$ & $0.039\pm0.003$ &  &   & 0.59 (30) & 0.63 & &  & \\
17 & $0.19 + (0.\pm0.04)$ & $4.59\pm0.91$ & $0.041\pm0.003$ &  &  &  &   & 1.14 (36) & 0.61 & &  & \\
17-& $0.33\pm0.09)$ & $2.14\pm0.25$ & $0.019\pm0.005$ &  &  &  &   & 1.08 (36) & 0.58 &  &  & \\
30 & $0.19 + (3.27\pm0.45)$ & $>13.0$ & $0.180\pm0.011$ &  &  &  &   & 0.97 (51) & 2.16 & &  & \\
30-& $5.99\pm1.32$ & $2.30\pm0.38$ & $0.19\pm0.17$ &  &  &  &   & 0.83 (51) & 1.90 &  &  & \\
31 & $0.19 + (0.85\pm0.09)$ & $0.85\pm0.09$ & $0.047\pm0.012$ &  &  &  &   & 1.47 (21) & 0.15 & &  & \\
35 & $0.19 + (0.61\pm0.16)$ & $0.11\pm0.02$ & $4.88\pm45.2$ & $1.55\pm0.16$ & $0.028\pm0.005$ &  &   & 1.06 (26) & 0.23 & &  & \\
77 & $0.19 + (0.32\pm0.07)$ & $0.14\pm0.02$ & $1.52\pm3.83$ & $0.56\pm0.10$ & $0.083\pm0.046$ &  &   & 1.05 (62) & 0.82 &  1.16$\pm$0.05& 4.89 & $-6.409\pm0.017$\\
  \hline
  \end{tabular}
\tablefoot{ The fitted model has the form $wabs\times wabs\times \sum apec$, where the first, interstellar absorption was fixed to $1.9\times 10^{21}$\,cm$^{-2}$ except for XID\,13 (since HD\,308829 displays a stronger absorption, with $E(B-V)\sim 0.62$). A symbol `-' gives the results of the alternative fitting by a power law, with no absorption fixed and the third column corresponding in these cases to the photon index. Fluxes refer to the 0.5--10.\,keV  energy band, luminosities were corrected for the interstellar absorption only, and abundances of the thermal emission component and additional absorption are solar \citep{and89}. When the 1$\sigma$ confidence interval is asymmetric, the largest value of the 1$\sigma$ error is here given.}
  \end{sidewaystable*}

\onlfig{9}{
\begin{figure*}
\includegraphics[width=9.cm]{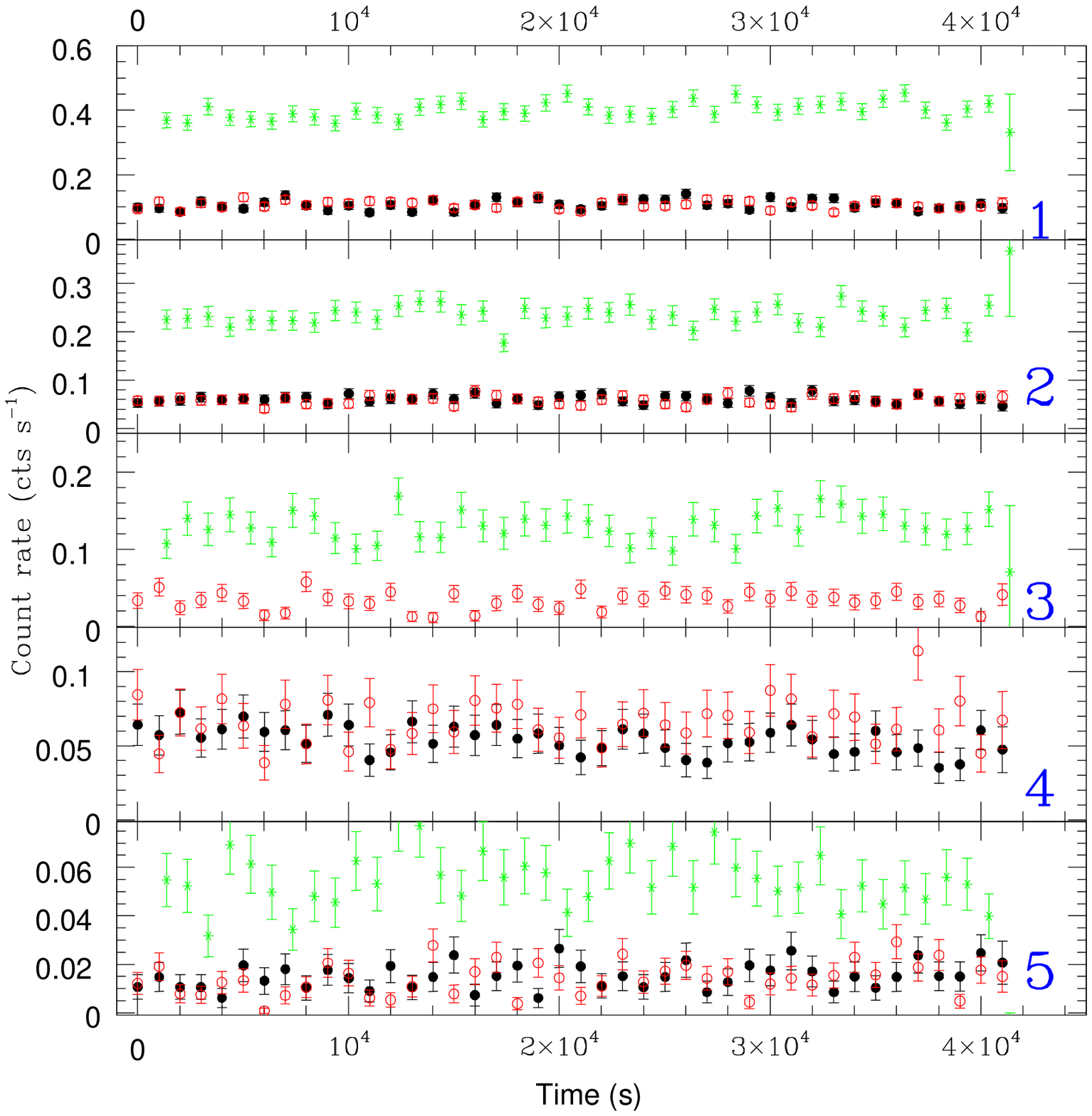}
\includegraphics[width=9.cm]{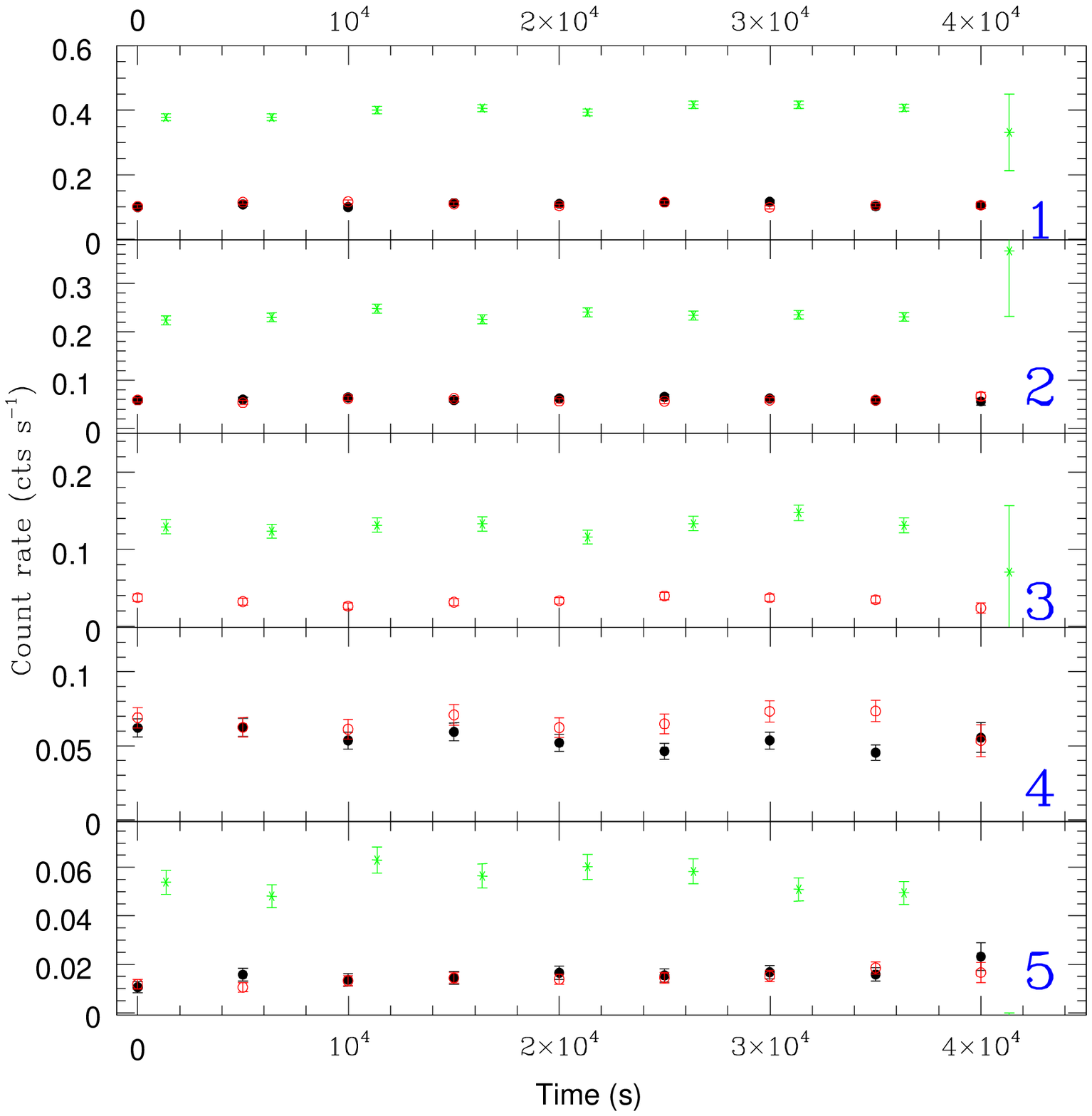}
\includegraphics[width=9.cm]{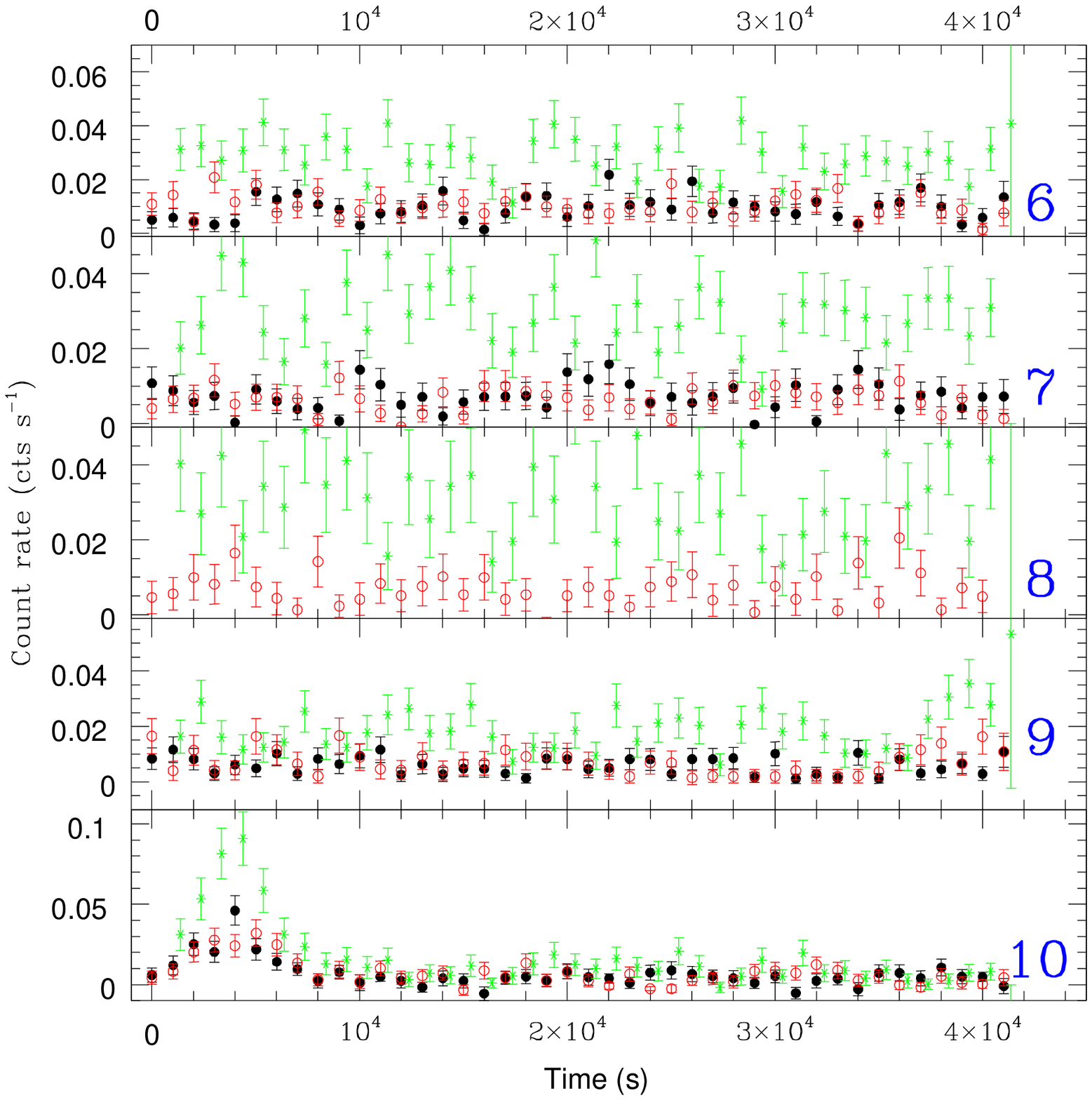}
\includegraphics[width=9.cm]{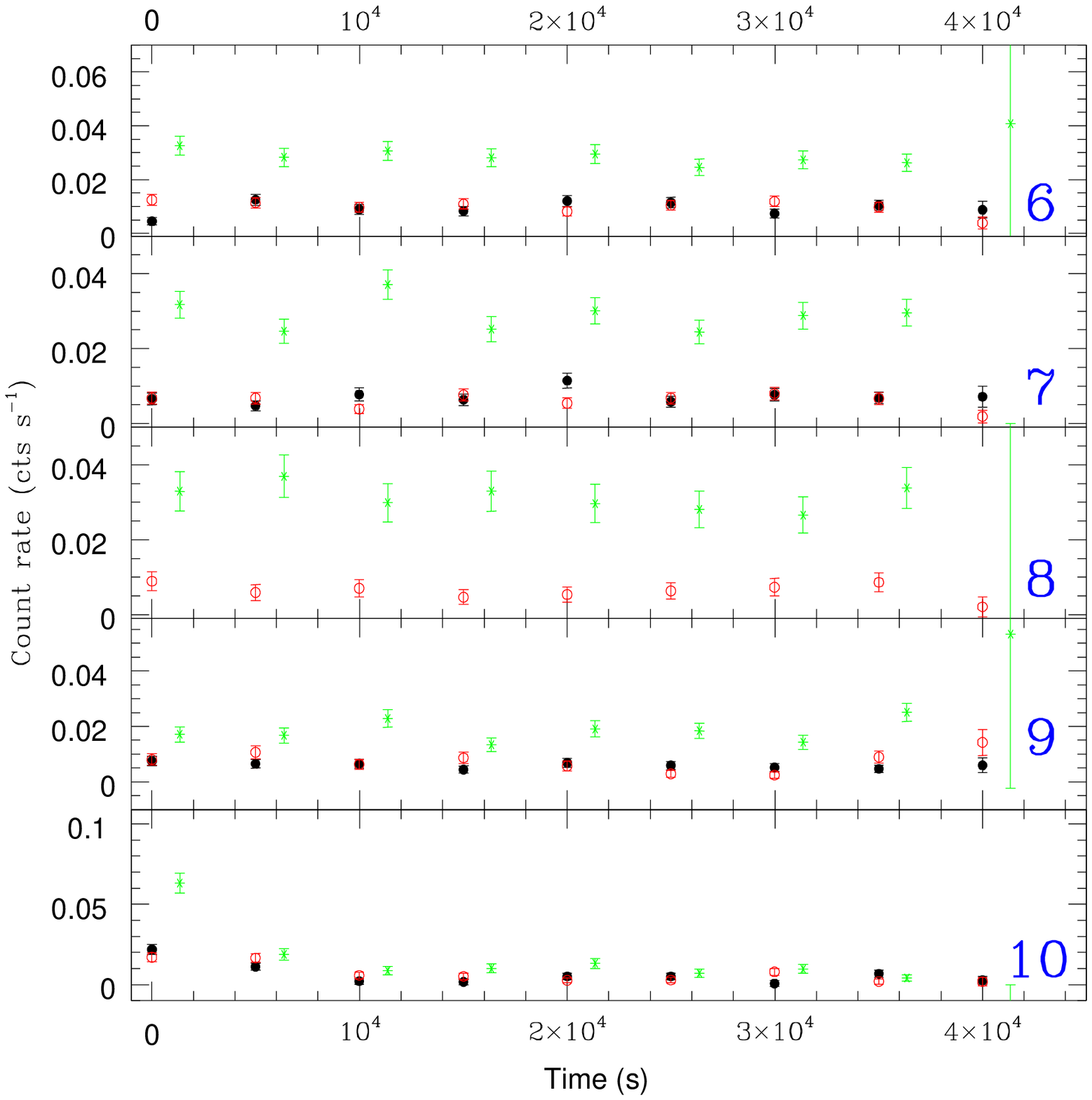}
\caption{Same as Fig. \ref{hm1lc} for the brightest X-ray sources of IC\,2944/2948. XID 1, 9, 16 and 30 are marginally variable (significance level between 1 and 10\%,  whereas XID 7, 10, 12, 15, 31, 35 are significantly variable (significance level $<$1\%). The color version of this figure is available online.}
\label{iclc}
\end{figure*}
\setcounter{figure}{8}
\begin{figure*}
\includegraphics[width=9.cm]{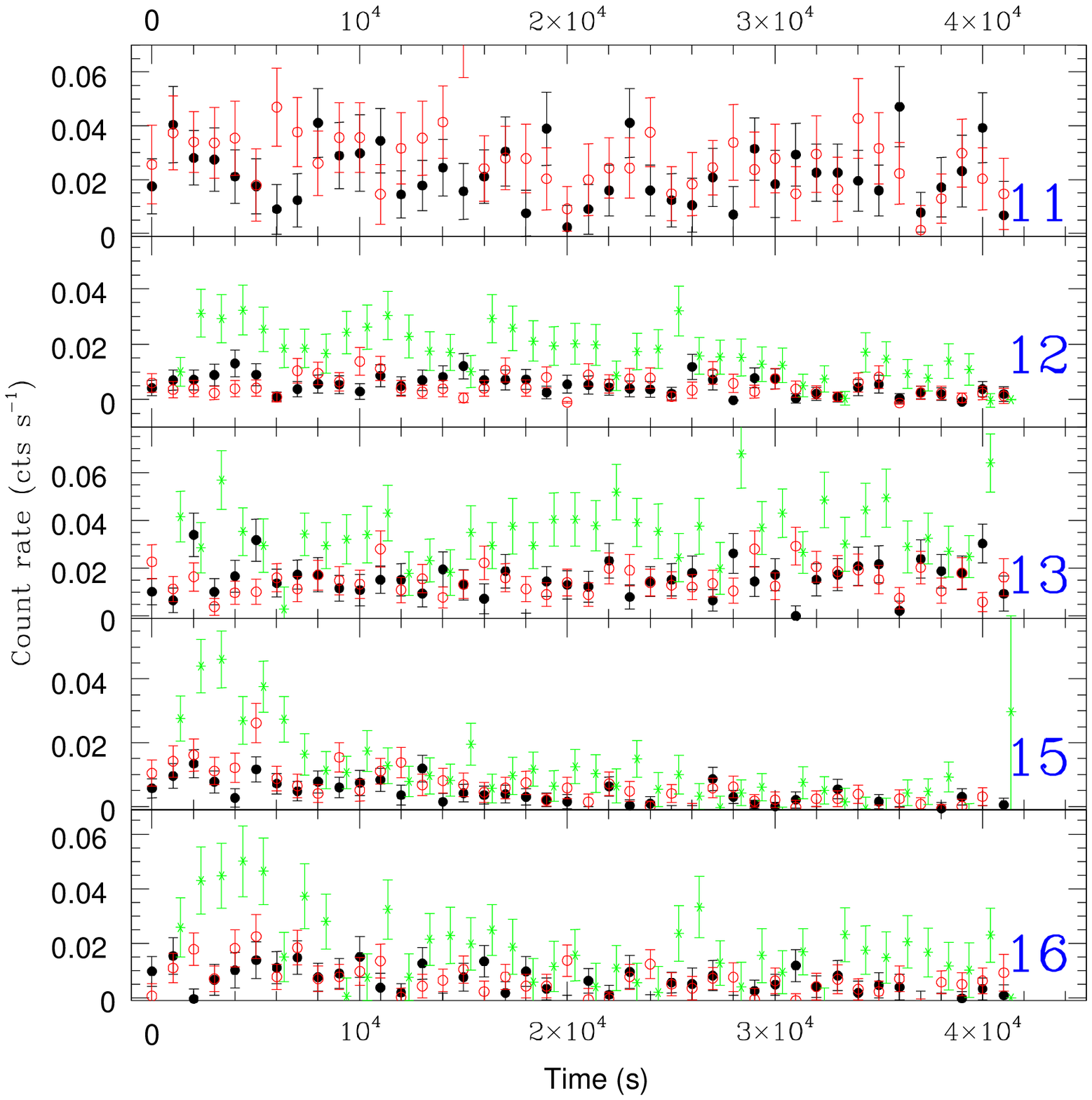}
\includegraphics[width=9.cm]{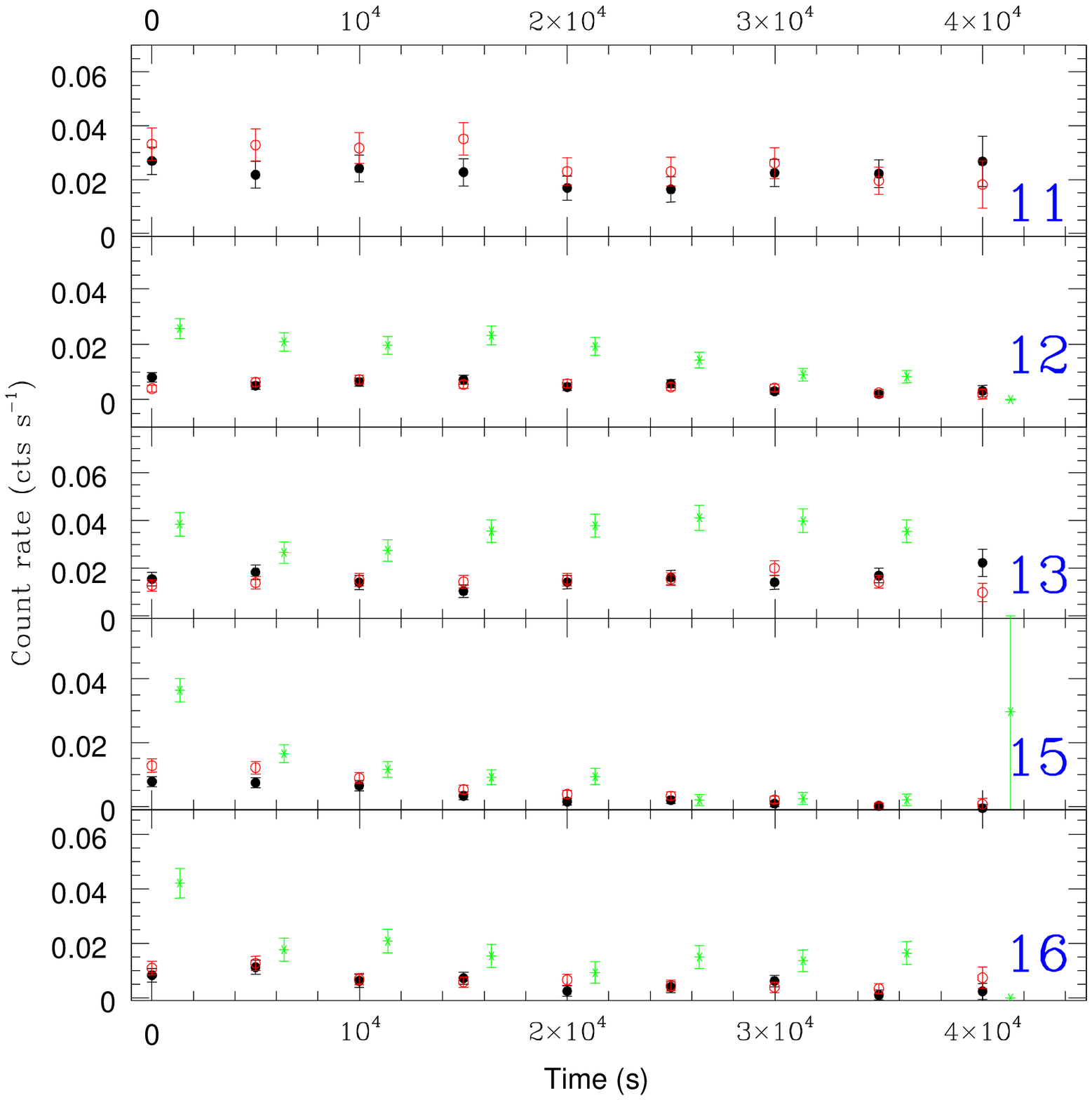}
\includegraphics[width=9.cm]{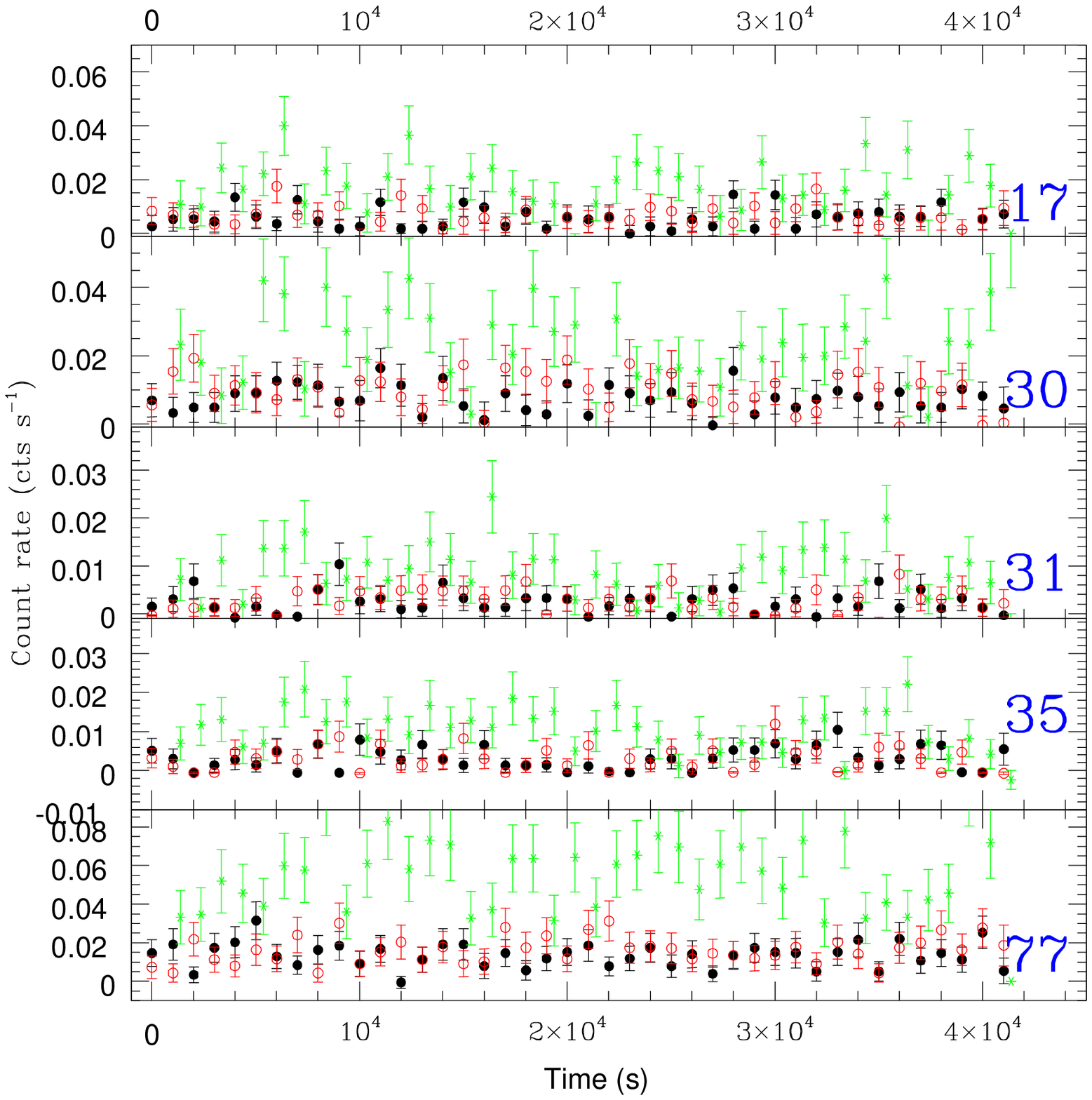}
\includegraphics[width=9.cm]{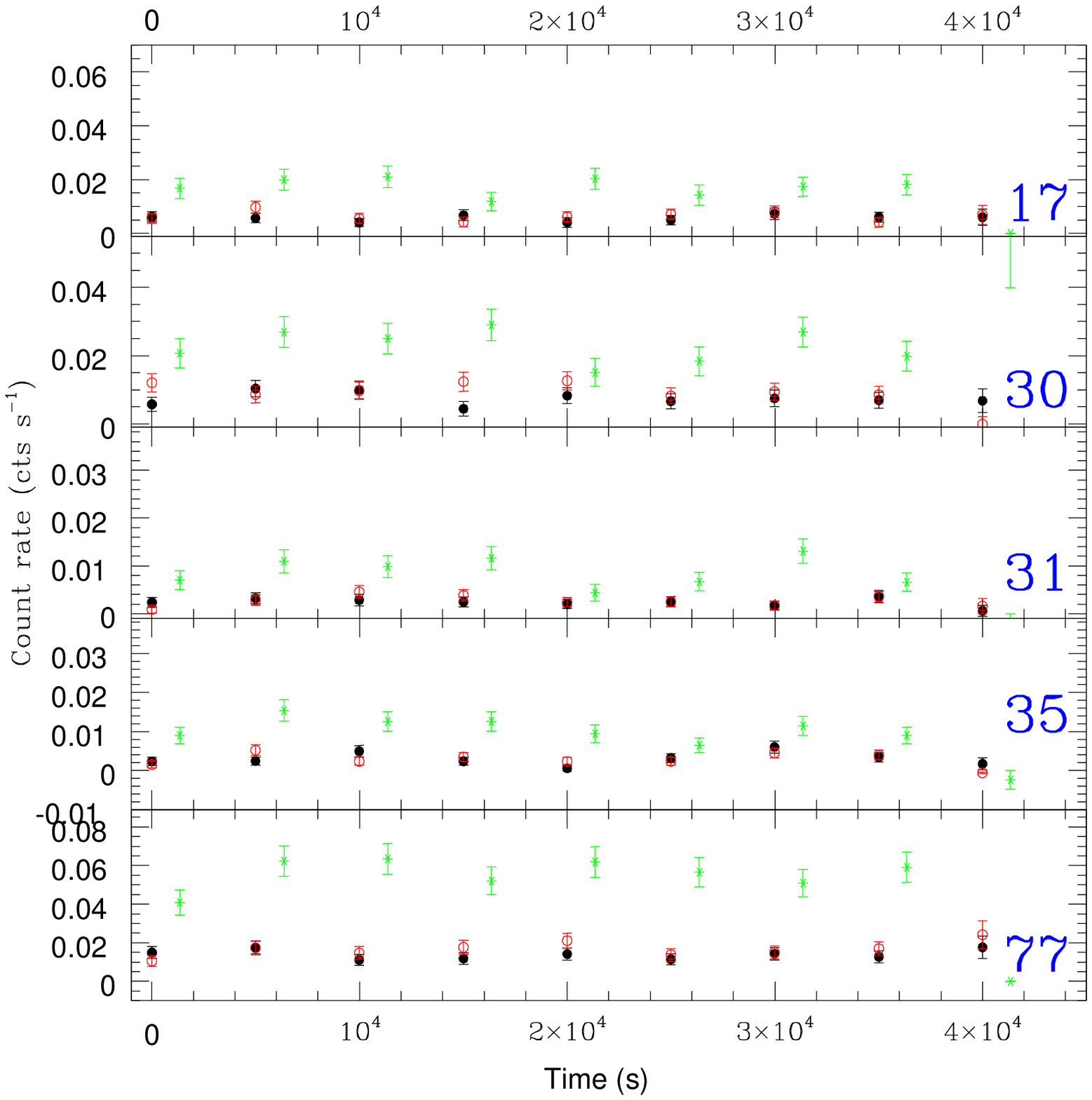}
\caption{Continued.}
\end{figure*}
}
\subsection{Individual objects}
\subsubsection{Massive stars}
The massive star population was recently studied in depth by \citet{san11}. Of the objects listed in that paper, two are located outside of the \xmm\ field-of-view (HD\,100099 and HD\,101545), and two remain undetected (CPD\,-62$^{\circ}$2198 and HD\,308804) - but both are of late-type (O9.7 and B3, respectively) and are located at the edge of the field (CPD\,-62$^{\circ}$2198 even being in the field-of-view of MOS2 only!), which explains their non-detection.

The remaining ten massive O-stars are detected (HD\,101131, HD\,101333, HD\,101190, HD\,101191, HD\,101205, HD\,101223, HD\,101298, HD\,101413, HD\,101436, and HD\,308813). The Be star and two Be candidates of \citet{mac05} are not detected, but five additional AB stars, not studied by \citet{san11} but listed in Simbad, are observed (Table \ref{iccorrel}).

Most massive stars are amongst the brightest X-ray sources of the field: eight have enough counts to be analyzed in more detail (Fig. \ref{iclc} and Table \ref{icfit}). As already indicated by their hardness ratios (see Fig. \ref{HR_ic}), their spectral properties are similar, with a moderate additional absorption and two plasma temperatures of about 0.1 and 0.7\,keV needed for a good fitting. 

XID 1 (HD\,101205, a multiple system containing an eclipsing binary) is marginally variable (significance level between 1 and 10\%), with a linear trend being significantly better (significance level $<$1\%) to fit the pn lightcurve than a simple constant. The link between this increasing trend seen in pn and the eclipsing nature of the binary in the system is not straightforward, though, since the \xmm\ observation probes a large part of the (circular) orbit between $\phi=0.46$ and 0.70, i.e., before, during, and after one of the eclipses \citep[using the ephemeris of ][]{ote07}, so that a monotonic trend is not expected. XID 7 (HD\,101191) is definitely variable at the $<$1\% significance level, but only with the shortest bin (1\,ks); no significant and coherent variability is found for the six other objects. The short-term variations are thus limited, as expected for massive stars. 

To study long-term variations, we need additional X-ray observations of the cluster. IC\,2944/2948 was observed by the {\it ROSAT} PSPC instrument in January 1993 (ObsID=200706). These archival data reveal only four sources: HD\,101131, HD\,101190, HD\,101205, and the pair HD\,101413+HD\,101436. This pair is not only entirely blended, since it looks like a single source, but its position is also problematic, because it is in the shadow of the inner ring that supports the entrance window of the PSPC. Therefore, the individual properties of the two sources forming the pair cannot be assessed with precision, and we thus discarded the {\it ROSAT} data for them. The spectra of the other three sources were extracted with circular regions of 65'' radii, with a nearby background region, of 80'' radius, located between HD\,101190 and HD\,101131 in an area as devoid of sources as possible (as seen in the \xmm\ images). The recorded count rates were compared with those predicted using the best-fit models to \xmm\ spectra, folded through the {\it ROSAT} response matrices. In addition, the spectra were compared with the best-fit \xmm\ models, and also fitted individually. Taking into account the low signal-to-noise ratio of the {\it ROSAT} data, the derived count rates and fluxes agree well with the results from \xmm\ fits: potential variations remain within one or two sigma error bars. No significant long-term variation of the X-ray properties of these three objects are therefore detected.

The X-ray luminosities, corrected for interstellar absorption, were calculated for all massive stars in the 0.5--10.\,keV energy band. For the X-ray bright ones, these luminosities were derived from the spectral fits (Table \ref{icfit}). For the fainter objects, we converted the count rates into fluxes using (1) the interstellar absorption and a thermal plasma with temperature of 0.3, 0.6, or 1.\,keV, and (2) the interstellar absorption and the typical O-star spectrum found in \citet{naz09}. This resulted in similar conversion factors: $\sim 1.2\times 10^{-11}$\,erg\,cm$^{-2}$\,s$^{-1}$ for 1 cts\,s$^{-1}$ recorded with the MOS camera equipped with the thick filter, and  $\sim 3.1\times 10^{-12}$\,erg\,cm$^{-2}$\,s$^{-1}$ for 1 cts\,s$^{-1}$ recorded with the pn camera equipped with the same filter. The derived X-ray luminosities are listed in Table \ref{icfaint}. The bolometric luminosities were derived using the visual magnitude from Simbad, an interstellar reddening of $E(B-V)=0.32$\,mag (with $R_V=3.1$), a distance of 2.3\,kpc, and bolometric corrections from \citet{mar05} for O-stars and from Schmidt-Kaler for other spectral types. When an object is a known binary, the correction factor is a weighted mean of the individual correction factors. Fig. \ref{hm1lxlb} graphically shows the results. 

The average \loglxlb\ for the ten O-stars is $-6.62$ with a dispersion of 0.35\,dex (see also Fig. \ref{hm1lxlb}), it is $-6.48$ with a dispersion of 0.18\,dex for the eight O-stars with more than 500 EPIC counts, $-6.75$ with a dispersion of 0.26\,dex for the three single O-stars and $-6.57$ with a dispersion of 0.39\,dex for the seven known binaries (see discussion below). Two massive objects are clearly overluminous: XID\,13 and 327. The former forms a blend with XID \,14, of unknown type and bolometric luminosity - it is therefore uncertain whether the full X-ray luminosity should be attributed to XID\,13 (indeed, the two sources display similar count rates, see Table \ref{icdet}). The latter object, XID\,327, is an A-star: such stars rarely emit X-rays, hence it is possible that a companion to that A-star is the actual X-ray emitter. On the other hand, one object is underluminous: XID\,106, associated with the late O-type binary HD\,308813 (Table \ref{iccorrel}), has a luminosity that is 0.9\,dex below the value derived from its bolometric luminosity and the average \lxlb\ of the eight O-stars. This difference corresponds to about five times the cluster dispersion and ten times the formal error on the X-ray luminosity of this object (Fig. \ref{hm1lxlb}). Only a few O-type stars are underluminous in X-rays compared with other O-stars of the same cluster, e.g., the late-type binary FO15 in the Carina Nebula \citep{naz11} - the reason for their low luminosity remains unknown to this day, however. 

  \begin{table}[htb]
  \tiny
  \centering
  \caption{Luminosities of the faint OBA stars in the 0.5--10.\,keV energy band (see Table \ref{hm1fit} for definition of errors).}
  \label{icfaint}
  \begin{tabular}{l c c c }
  \hline
XID & $L_{\rm X}^{abscor}$ & $\log(L_{\rm BOL})$ & $\log(L_{\rm X}^{abscor}/L_{\rm BOL})$ \\
& 10$^{32}$\,erg\,s$^{-1}$ & (L$_{\odot}$) & \\
  \hline
 20 & 0.32$\pm$0.02 & 4.38 & $-$6.46$\pm$0.03 \\
 63 & 0.26$\pm$0.03 & 4.23 & $-$6.40$\pm$0.06 \\
 81 & 0.12$\pm$0.01 & 4.55 & $-$7.05$\pm$0.05 \\
106 &0.044$\pm$0.007& 4.42 & $-$7.36$\pm$0.07 \\
153 & 0.17$\pm$0.03 & 4.51 & $-$6.86$\pm$0.07 \\
175 &0.064$\pm$0.012& 3.68 & $-$6.46$\pm$0.08 \\
327 &0.052$\pm$0.013& 2.52 & $-$5.39$\pm$0.11 \\
  \hline
  \end{tabular}
  \end{table}

\subsubsection{Foreground objects}

\begin{figure*}
\includegraphics[width=6cm]{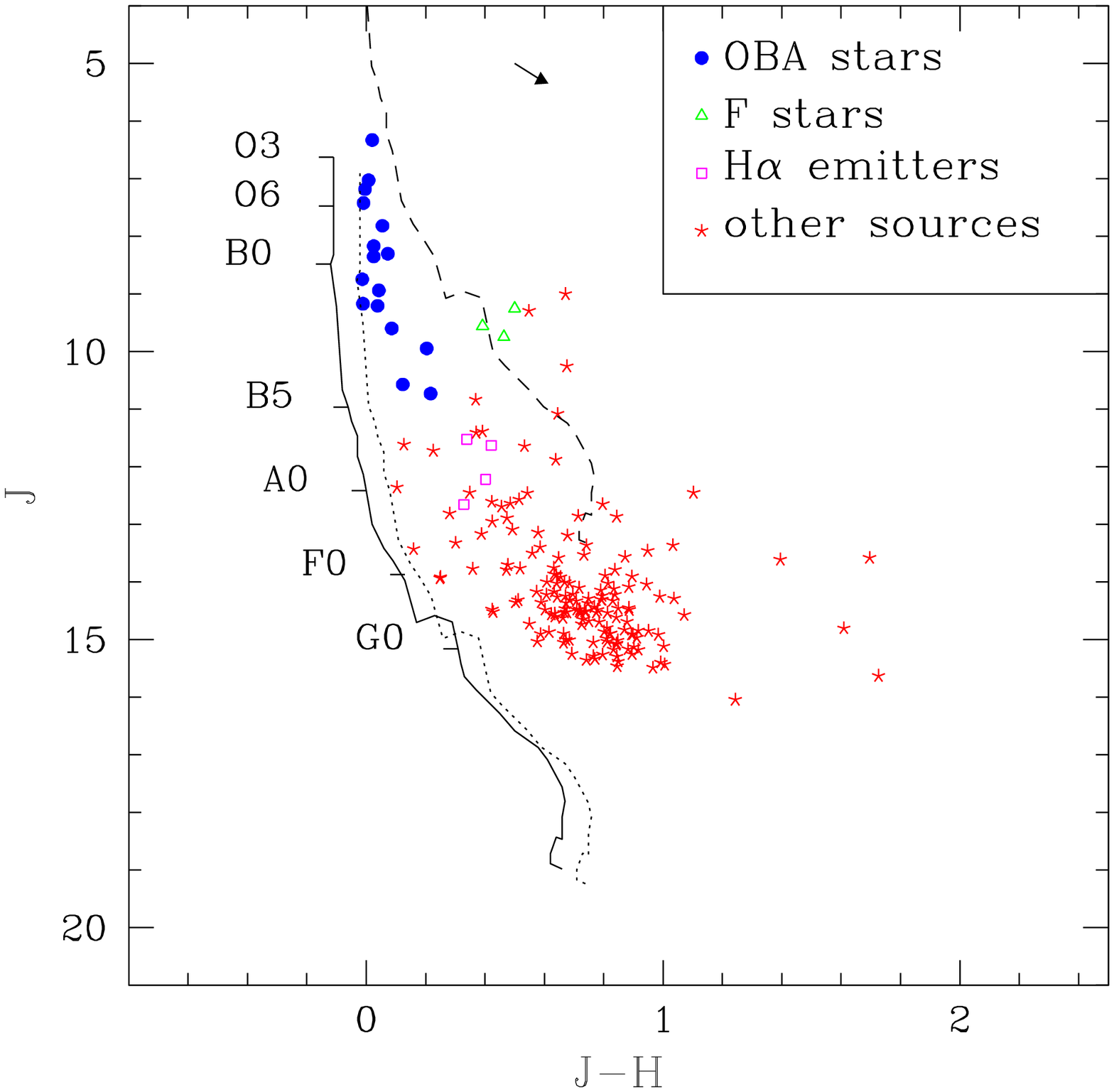}
\includegraphics[width=6cm]{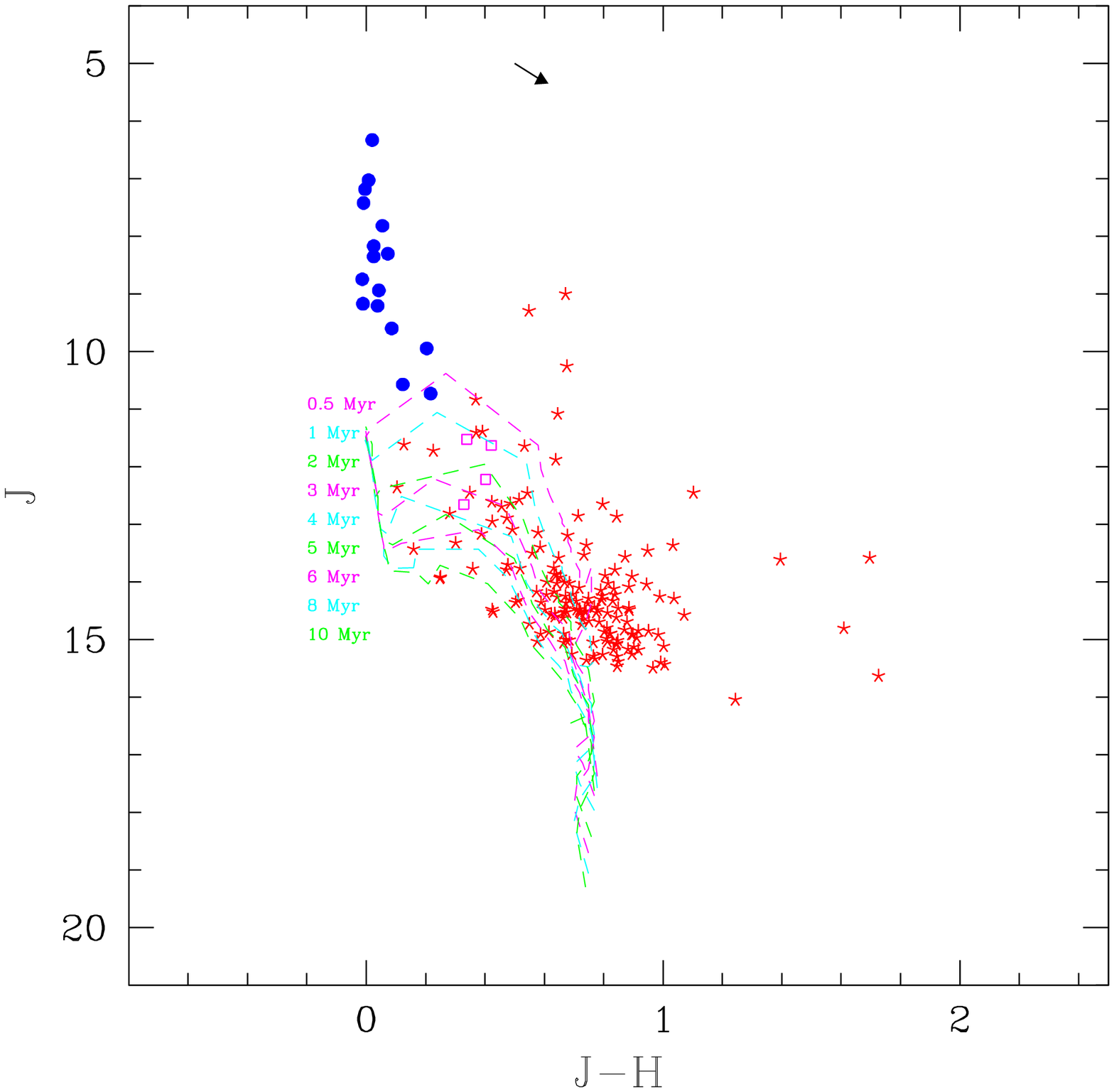}
\includegraphics[width=6cm]{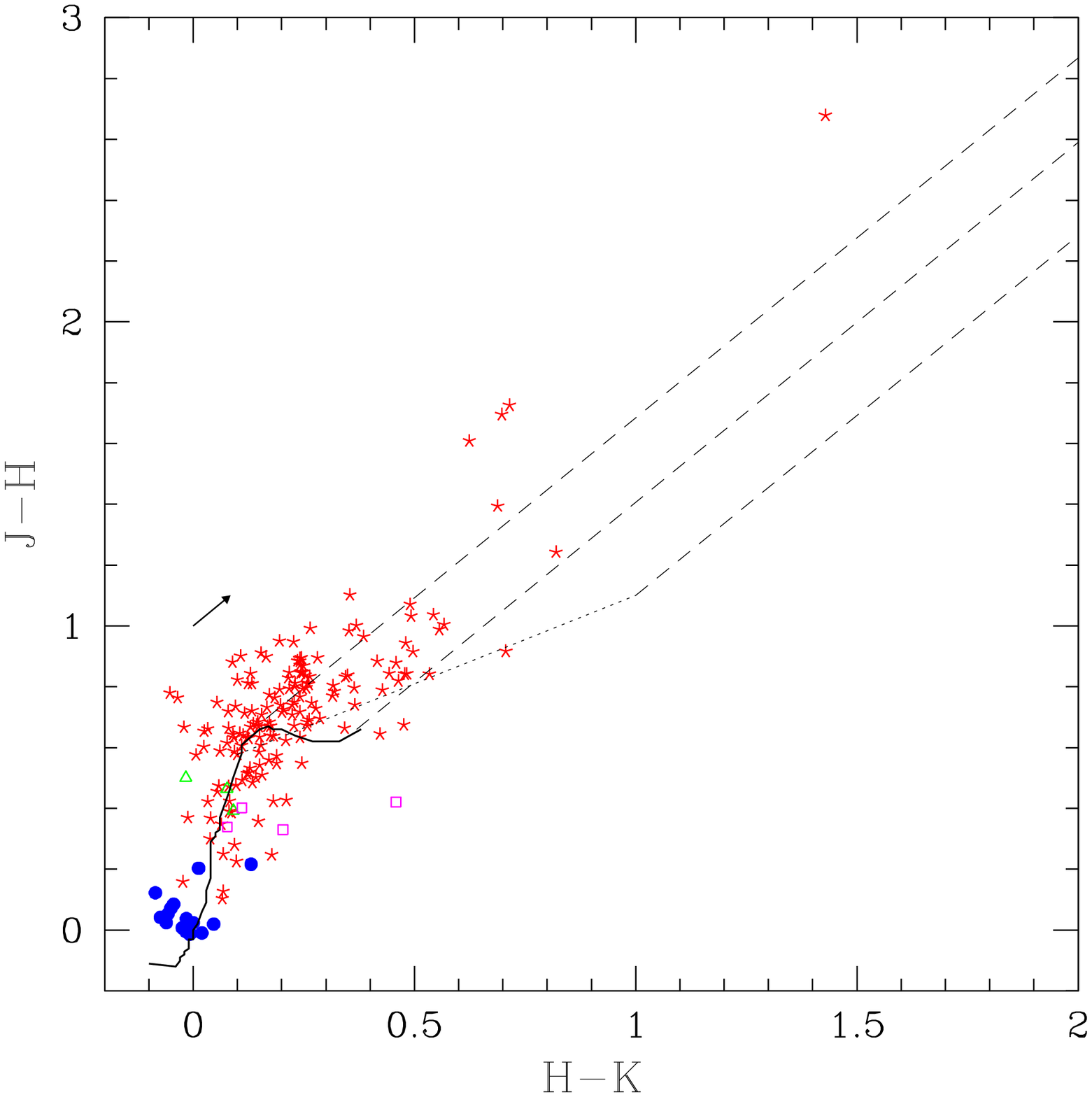}
\caption{Same as Fig. \ref{hm12mass} for IC\,2944/2948. {\it Left:} The dashed line corresponds to the main sequence shifted for a distance of 150\,pc and for a reddening of $E(B-V)=0.35$\,mag, the solid and dotted lines correspond to the main sequences for a distance of 2.3\,kpc, typical of the cluster \citep{san11}, without and with correction for the cluster's absorption \citep[$E(B-V)=0.32$\,mag, $R_V=3.1$, see ][]{mac05}, respectively. {\it Middle:} The same color-magnitude diagram with the \citet{sie00} isochrones, shifted for a distance of 2.3\,kpc and $E(B-V)=0.32$\,mag, superimposed as dashed lines. {\it Right:} Color-color diagram for the same objects (using $R_V=3.1$). The color version of this figure is available online.}
\label{ic2mass}
\end{figure*}

The three detected F-type stars were not the target of in-depth studies, and only basic information (spectral type, BV magnitudes) is available in Simbad. Compared with typical values for main-sequence stars of the same subclass, their bolometric luminosities are about three times that of the Sun, their reddenings amount to about $E(B-V)\sim0.35$\,mag, and their distances to about 150\,pc, in agreement with their 2MASS photometry (Fig. \ref{ic2mass}). Because of their faint X-ray emission, we converted their count rates into fluxes corrected for the interstellar absorption using a single hot plasma component (two temperatures, 0.3\,keV and 1.\,keV were tried and yielded similar results). The resulting X-ray luminosities amount to about $2\times10^{29}$\,erg\,cm$^{-2}$\,s$^{-1}$ for HD\,308806 (XID 41), $2\times10^{28}$\,erg\,cm$^{-2}$\,s$^{-1}$ for HD\,308828 (XID 178), $4\times10^{29}$\,erg\,cm$^{-2}$\,s$^{-1}$ for HD\,308839 (XID 237), corresponding to \loglxlb\ between $-5$ and $-6$, i.e., far from the saturation level of coronal sources. 

\subsubsection{Accreting objects}

Several very hard X-ray sources exist in the FOV, e.g., XID 6, 11, and 30 (Figs. \ref{iccol} and \ref{HR_ic}). For the brightest among them, a spectral fitting was attempted, which results in high absorption ($N_{\rm H}>10^{22}$\,cm$^{-2}$) and unrealistically high plasma temperature ($kT>10.$\,keV), or high absorption and a power law with photon index of about 2 (see Table \ref{icfit}). These are good candidates for background accreting sources (X-ray binaries or accreting AGNs). 

\subsubsection{Young stars}

The star formation in IC\,2944/2948 area has been studied several times in the past 15 years. \citet{rei97} found seven H$\alpha$ emitters, hence candidate TTs: only two are positioned in our field-of-view. IRAS sources embedded in clouds have also been reported by \citet{sug94} and \citet{yam99}: none of the SFO sources of \citet{sug94} are located in our FOV, and only the clouds [YSM99] 44 and 45, corresponding to nine IRAS sources, are located in our field-of-view. The X-ray emission of these sources, if any, is too weak to be detected by our observation, however.

\citet{mac05} provided $y$ and H$\alpha$ photometry, which enables us to find the H$\alpha$ equivalent width (see formula in \citealt{mac05a} and Table \ref{iccorrel}): six objects, including one B-star and one F-star, have a clear emission in this line, i.e., a very small fraction of our sources. The lack of simultaneous H$\alpha$ and X-ray emissions is not surprising and has been remarked on several times \citep[e.g., NGC6383, ][]{rau10}: indeed, H$\alpha$ traces the accreting gas, whereas strong X-ray emission is mostly a marker of coronal activity; weak-line TTs often show the latter but not the former, while classical TTs display the opposite behavior. 

Using the best 2MASS photometry (i.e., only AAA quality flag), we can draw color-mag and color-color diagrams (Fig. \ref{ic2mass}). While there are only few highly reddened objects and red classical TTs, there is a population of slightly reddened, low-mass stars. Their age is between 0.5 and 10\,Myr, which is compatible with the age derived from the presence of massive stars. 

Regarding variability, only the brighter sources can be analyzed (Fig. \ref{iclc}). None shows a clear, big flare, but small-amplitude flare-like activity can be seen in the light curves of XID 10, 15, and 16. In addition, sources XID 12, 31, and 35 are significantly variable (with a significance level $<$1\%). 

Most X-ray sources are found in the vicinity of the O-type stars HD\,101205 and HD\,101298. In contrast, the massive stars to the north (HD\,101190, HD\,101223 and HD\,308804, CPD\,--62$^{\circ}$2198) are isolated from the main source clustering area, with no concentration of sources in their neighborhood. It is true that \xmm\ sensitivity decreases with off-axis angle, but faint objects are seen near the pair formed by HD\,101413 and HD\,101436, to the east of the field-of-view, whereas nothing similar exists for the northern group, despite their similar off-axis angle. Ejection of several massive stars in the same direction and at similar apparent distances is improbable, though not impossible. The X-ray data therefore suggest that this northern group may have formed separately from the main cluster, though the few X-ray sources detected in the area do not occupy an obviously different locus in the HR diagram. 

\section{\lxlb\ relation}

The relation between X-ray luminosity and bolometric luminosity has been known since the first X-ray observations of massive stars. The first homogeneous study of that relation for a large sample of objects (more than 200 OB stars) was conducted by \citet{ber97} using {\it ROSAT} count rates. Today, spectral fits are used instead for deriving the X-ray luminosities. For example, the analysis of twelve O-stars in NGC6231 yields \loglxlb $=-6.91$ with a dispersion of 0.15\,dex \citep{Sana}. A larger study was made for the Carina nebula: \loglxlb\ was first found to be $-6.66$ with a dispersion of 0.26\,dex \citep[\xmm, for ten presumably single O-stars, derived by ][]{Igor} and then $-7.24$ with a dispersion of 0.20\,dex \citep[$Chandra$, for 27 presumably single O-stars,][]{naz11}. The difference in the two latter values may appear puzzling because they concern the same region. But it is mostly due to changes in the treatment of absorption and derivation of the bolometric luminosities (one could derive a logarithmic ratio of $-6.99$ with $Chandra$, simply by using $R_V=3.1$ instead of 4), with a minor role played by the choice of energy bands (0.4--10.0\,keV energy band for the former, 0.5--10.0\,keV  energy band for the latter) and instrumental cross-calibration problems. Similar variations are also found in other clusters, e.g., M17, where the bolometric luminosities are highly uncertain \citep[Fig. 7 in ][]{rau13}. Therefore, deriving one single value for the \lxlb\ ratio, or simply comparing values found in different studies, is a difficult, if not impossible, task. That is why homogeneous studies (same method, same instrument) have also been performed: the 2XMM survey data yield \loglxlb $=-6.45$ with a dispersion of 0.51\,dex for 78 O-stars \citep{naz09}. The dispersion is higher in these overall studies than in analyses of a single cluster (0.5\,dex vs 0.2\,dex), and different subgroups in the overall sample corresponding to clusters yield values at a few 0.1\,dex of each other, two facts that indicate possible cluster-to-cluster differences.

In that framework, our results bring some more information. Indeed, the \lxlb\ value in IC\,2944/2948 is higher than in HM1: this is particularly obvious when comparing the two sets of points in Fig. \ref{hm1lxlb}). However, as shown in Sect. 3.1.3, this could be due to the high extinction (and therefore low signal-to-noise ratio at low energies) in HM1. In fact, the Cyg OB2 association, which is also strongly reddened ($E(B-V)\sim2$), shows similar features: \loglxlb $< -7$ for 3 out of the 4 O-stars studied by \citet{Rauw}, high plasma temperatures derived for 1T fits (0.7--0.9\,keV for Cyg\,OB2\,7 and 1.5--2.0\,keV for CPR2002\,A11). A true cluster-to-cluster difference in \lxlb\ thus remains to be securely established.

Another, popular aspect of the \lxlb\ ratio is its ``strong'' link with binarity: wind-wind collisions are expected in massive binaries, and should lead to additional hard X-ray emission. However, whilst X-ray overluminosities often indicate wind-wind collisions in binaries (or magnetically confined winds in single stars), the opposite is certainly not observed, as found from {\it ROSAT} data \citep[e.g., 29\,CMa, see ][]{ber94} and in more recent \xmm\ and $Chandra$ observations \citep{pit00,osk05, naz09, naz11}. Except in a few exceptional cases (such as WR89 here), the differences in luminosity are small: overluminosities of $\sim$0.3\,dex (=1.5 times the dispersion) for two overluminous binaries in  NGC6231, of 0.14\,dex ($\sim$ dispersion) for O-type binaries in Carina, and of 0.25\,dex (=half the dispersion) for O-type binaries in the 2XMM sample. Remarkably, however, these differences, though small compared to the dispersion, are systematic (i.e., binaries are, on average, always brighter than single stars). The situation is similar for our two clusters: low overluminosity, at best, without any link with period (as in \citealt{naz09} and \citealt{naz11}).

Is this non-detection of overluminosities a result of observing the systems at a particular phase? Detailed ephemeris is available only for four of the eight O-type binaries studied here \citep[and references therein]{san11}: the short-period eclipsing binary in HD\,101205 is circular and was observed before, during, and after an eclipse without phase-locked variation in the X-ray domain (see 4.1.1); the 10\,d, slightly eccentric binary HD\,101131 was observed at $\phi=0.45$ (with an observation length of $\Delta\phi\sim0.05$), i.e., close to apastron; the 6\,d, slightly eccentric binary HD\,101190 was observed at $\phi=0.35-40$ (with an observation length of $\Delta\phi\sim0.08$) and the 37\,d, slightly eccentric binary HD\,101436 was observed at $\phi=0.2-0.25$ (with an observation length of $\Delta\phi\sim0.01$), i.e., between periastron and apastron in both cases. None of the systems appears to be observed exactly at periastron or apastron, to the best of current knowledge, thus preventing us from testing the hypothesis of an overluminosity at these phases (for adiabatic and radiative cases, respectively). Simple snapshots such as our observations are therefore inadequate for studies of the colliding-wind phenomenon. Moreover, colliding-wind systems display changes in plasma temperatures and/or variations of their X-ray flux (as shown for, e.g., Cyg\,OB2\,\#8A, \citealt{deb}, and Cyg\,OB2\,\#9, \citealt{naz12}): a monitoring of the two clusters would certainly yield more constraining information.

\section{Conclusion}
We have obtained \xmm\ data of two clusters, never analyzed before in the high-energy range: HM1 and IC\,2944/2948. The two datasets enabled us to study the X-ray emission of massive stars and surrounding low-mass stars.

For HM1, both WR stars are detected: WR89 is extremely overluminous, but much softer than WR87. The high luminosity of the former and its apparent variability (found when comparing the new data with older {\it ROSAT} data) point toward a possible colliding-wind origin (although no sign of radial velocity variation is detected in the available visible spectra), whereas the hardness of the latter, without any strong overluminosity, remains a puzzle because it is neither compatible with wind-wind collisions nor with magnetically confined winds. The X-ray luminosities of the other massive stars are proportional to the bolometric luminosities, but with a \loglxlb\ below the canonical value of $-$7. Whereas variations of \loglxlb\ amongst clusters are possible, the most probable cause for this low value is an underestimate of the X-ray luminosities, due to the high interstellar absorption of the cluster, which reduces the signal-to-noise ratio at low energies, impairing a correct determination of the X-ray luminosities. We also detected a population of faint X-ray sources associated with reddened objects at the distance of the cluster, most probably weak-line TTs, as well as a few foreground coronal sources and background accreting objects.

The lower extinction of IC\,2944/2948 results in the detection of many more X-ray sources. Most are grouped around HD\,101205, HD\,101298, and HD\,101413-436, but the massive stars to the north appear isolated, probably forming a separate (sub)cluster. The X-ray emissions of O-type stars are normal: they are soft, not significantly variable compared to archival {\it ROSAT} data, without overluminosities, and following \loglxlb = $-$6.6. Known massive O-type binaries, whatever their period, are not significantly brighter than single objects. Many low-mass, pre-main-sequence objects are detected, with an age of a few Myr ; only few of them are strong H$\alpha$ emitters, suggesting a weak-line TTs nature for most of these X-ray emitters. 

\begin{acknowledgements}
We acknowledge support from the Fonds National de la Recherche Scientifique (Belgium), and the Communaut\'e Fran\c caise de Belgique, the PRODEX XMM and Integral contracts. ADS and CDS were used for preparing this document. YN acknowledges E. Gosset for careful reading.
\end{acknowledgements}

\end{document}